\newcommand{\figures}{./}
\documentclass[aps,prl,showpacs,twocolumn,superscriptaddress,letterpaper,longbibliography]{revtex4-1}
\usepackage{graphicx}
\usepackage{dcolumn}
\usepackage{bm}
\usepackage{amsmath, amssymb}
\usepackage{epstopdf}
\usepackage{color}
\usepackage[normalem]{ulem}
\usepackage{multirow}
\usepackage{enumitem}
\usepackage{comment}
\usepackage{hyperref}
\usepackage{cleveref}
\usepackage{lineno}

\hypersetup{breaklinks = true, colorlinks = true, citecolor = blue, linkcolor = blue, urlcolor = blue}

\newcommand{\uB}{MicroBooNE}
\newcommand{\CCIpOpi}{CC1p0$\pi$}

\begin{document}

\title{First double-differential measurement of kinematic imbalance\texorpdfstring{\\}{}in neutrino interactions with the MicroBooNE detector}

\date{\today}



\newcommand{\ANL}{Argonne National Laboratory (ANL), Lemont, IL, 60439, USA}
\newcommand{\Bern}{Universit{\"a}t Bern, Bern CH-3012, Switzerland}
\newcommand{\BNL}{Brookhaven National Laboratory (BNL), Upton, NY, 11973, USA}
\newcommand{\UCSB}{University of California, Santa Barbara, CA, 93106, USA}
\newcommand{\Cambridge}{University of Cambridge, Cambridge CB3 0HE, United Kingdom}
\newcommand{\CIEMAT}{Centro de Investigaciones Energ\'{e}ticas, Medioambientales y Tecnol\'{o}gicas (CIEMAT), Madrid E-28040, Spain}
\newcommand{\Chicago}{University of Chicago, Chicago, IL, 60637, USA}
\newcommand{\Cincinnati}{University of Cincinnati, Cincinnati, OH, 45221, USA}
\newcommand{\CSU}{Colorado State University, Fort Collins, CO, 80523, USA}
\newcommand{\Columbia}{Columbia University, New York, NY, 10027, USA}
\newcommand{\Edinburgh}{University of Edinburgh, Edinburgh EH9 3FD, United Kingdom}
\newcommand{\FNAL}{Fermi National Accelerator Laboratory (FNAL), Batavia, IL 60510, USA}
\newcommand{\Granada}{Universidad de Granada, Granada E-18071, Spain}
\newcommand{\Harvard}{Harvard University, Cambridge, MA 02138, USA}
\newcommand{\IIT}{Illinois Institute of Technology (IIT), Chicago, IL 60616, USA}
\newcommand{\KSU}{Kansas State University (KSU), Manhattan, KS, 66506, USA}
\newcommand{\Lancaster}{Lancaster University, Lancaster LA1 4YW, United Kingdom}
\newcommand{\LANL}{Los Alamos National Laboratory (LANL), Los Alamos, NM, 87545, USA}
\newcommand{\Louisiana}{Louisiana State University, Baton Rouge, LA, 70803, USA}
\newcommand{\Manchester}{The University of Manchester, Manchester M13 9PL, United Kingdom}
\newcommand{\MIT}{Massachusetts Institute of Technology (MIT), Cambridge, MA, 02139, USA}
\newcommand{\Michigan}{University of Michigan, Ann Arbor, MI, 48109, USA}
\newcommand{\Minnesota}{University of Minnesota, Minneapolis, MN, 55455, USA}
\newcommand{\NMSU}{New Mexico State University (NMSU), Las Cruces, NM, 88003, USA}
\newcommand{\Oxford}{University of Oxford, Oxford OX1 3RH, United Kingdom}
\newcommand{\Pitt}{University of Pittsburgh, Pittsburgh, PA, 15260, USA}
\newcommand{\Rutgers}{Rutgers University, Piscataway, NJ, 08854, USA}
\newcommand{\SLAC}{SLAC National Accelerator Laboratory, Menlo Park, CA, 94025, USA}
\newcommand{\SDSMT}{South Dakota School of Mines and Technology (SDSMT), Rapid City, SD, 57701, USA}
\newcommand{\Maine}{University of Southern Maine, Portland, ME, 04104, USA}
\newcommand{\Syracuse}{Syracuse University, Syracuse, NY, 13244, USA}
\newcommand{\TelAviv}{Tel Aviv University, Tel Aviv, Israel, 69978}
\newcommand{\Tennessee}{University of Tennessee, Knoxville, TN, 37996, USA}
\newcommand{\UTA}{University of Texas, Arlington, TX, 76019, USA}
\newcommand{\Tufts}{Tufts University, Medford, MA, 02155, USA}
\newcommand{\UCL}{University College London, London WC1E 6BT, United Kingdom}
\newcommand{\VTech}{Center for Neutrino Physics, Virginia Tech, Blacksburg, VA, 24061, USA}
\newcommand{\Warwick}{University of Warwick, Coventry CV4 7AL, United Kingdom}
\newcommand{\Yale}{Wright Laboratory, Department of Physics, Yale University, New Haven, CT, 06520, USA}

\affiliation{\ANL}
\affiliation{\Bern}
\affiliation{\BNL}
\affiliation{\UCSB}
\affiliation{\Cambridge}
\affiliation{\CIEMAT}
\affiliation{\Chicago}
\affiliation{\Cincinnati}
\affiliation{\CSU}
\affiliation{\Columbia}
\affiliation{\Edinburgh}
\affiliation{\FNAL}
\affiliation{\Granada}
\affiliation{\Harvard}
\affiliation{\IIT}
\affiliation{\KSU}
\affiliation{\Lancaster}
\affiliation{\LANL}
\affiliation{\Louisiana}
\affiliation{\Manchester}
\affiliation{\MIT}
\affiliation{\Michigan}
\affiliation{\Minnesota}
\affiliation{\NMSU}
\affiliation{\Oxford}
\affiliation{\Pitt}
\affiliation{\Rutgers}
\affiliation{\SLAC}
\affiliation{\SDSMT}
\affiliation{\Maine}
\affiliation{\Syracuse}
\affiliation{\TelAviv}
\affiliation{\Tennessee}
\affiliation{\UTA}
\affiliation{\Tufts}
\affiliation{\UCL}
\affiliation{\VTech}
\affiliation{\Warwick}
\affiliation{\Yale}

\author{P.~Abratenko} \affiliation{\Tufts}
\author{O.~Alterkait} \affiliation{\Tufts}
\author{D.~Andrade~Aldana} \affiliation{\IIT}
\author{J.~Anthony} \affiliation{\Cambridge}
\author{L.~Arellano} \affiliation{\Manchester}
\author{J.~Asaadi} \affiliation{\UTA}
\author{A.~Ashkenazi}\affiliation{\TelAviv}
\author{S.~Balasubramanian}\affiliation{\FNAL}
\author{B.~Baller} \affiliation{\FNAL}
\author{G.~Barr} \affiliation{\Oxford}
% ...

\collaboration{The MicroBooNE Collaboration}
\thanks{microboone\_info@fnal.gov}\noaffiliation


\begin{abstract}
We report the first measurement of flux-integrated double-differential quasielastic-like neutrino-argon cross sections, which have been made using the Booster Neutrino Beam and the \uB\ detector at Fermi National Accelerator Laboratory.
The data are presented as a function of kinematic imbalance variables which are sensitive to nuclear ground state distributions and hadronic reinteraction processes.
We find that the measured cross sections in different phase-space regions are sensitive to different nuclear effects.
Therefore, they enable the impact of specific nuclear effects on the neutrino-nucleus interaction to be isolated more completely than was possible using previous single-differential cross section measurements.
Our results provide precision data to help test and improve neutrino-nucleus interaction models.
They further support ongoing neutrino-oscillation studies by establishing phase-space regions where precise reaction modeling has already been achieved.
\end{abstract}

\maketitle


Neutrino oscillation measurements aim to extract neutrino mixing angles, mass differences, and the charge-parity violating phase, and to search for new physics beyond the Standard Model~\cite{pdg2018,T2KNature20,PhysRevLett.123.151803}.
The analysis of such measurements traditionally relies on detailed comparisons of measured and theoretically-expected neutrino interaction rates in the corresponding detectors. 
Therefore, a precise understanding of neutrino-nucleus interactions is required to fully exploit the discovery potential of current and next-generation experiments.

With a growing number of neutrino-oscillation experiments employing liquid argon time projection chamber (LArTPC) neutrino detectors~\cite{DUNE1:2016oaz,DUNE2:2016oaz,DUNE3:2016oaz,Antonello:2015lea,Tortorici:2018yns,Abi:2018dnh}, high-accuracy modeling of neutrino-argon interactions is becoming of paramount importance~\cite{Dolan:2018sbb,PhysRevLett.116.192501,Rocco2020}. 
The overarching goal of these efforts is both to achieve few-percent-level modeling of neutrino-argon interaction rates and to provide a detailed understanding of the final-state kinematics of emitted particles that are used to reconstruct the energies of the interacting neutrinos~\cite{HK,Abi:2020evt}.

This Letter reports the first measurement of flux-integrated double-differential cross sections for muon-neutrino-argon ($\nu_{\mu}$-Ar) charged-current (CC) quasielastic (QE)-like scattering reactions as a function of transverse kinematic imbalance variables.
Building upon a previous analysis of neutrino-argon cross sections with a similar signal event topology~\cite{PhysRevLett.125.201803}, we focus on reactions where the neutrino removes a single intact proton from the nucleus without producing any additional detected particles.
The results reported here are obtained using the Booster Neutrino Beam (BNB) and the \uB\ detector at Fermi National Accelerator Laboratory with an exposure of $6.79 \times 10^{20}$ protons on target.

Transverse kinematic imbalance variables were previously shown to be sensitive to the modeling of the nuclear ground-state distribution and to nuclear medium effects, such as hadronic final-state interactions (FSI)~\cite{Bodek2019,PhysRevC.94.015503,Abe:2018pwo,PhysRevLett.121.022504,PhysRevD.101.092001,Bathe-Peters:2022kkj}.
By measuring the components of the muon and proton momenta perpendicular to the neutrino direction, $\vec{p}_{T}\,^{\mu}$ and $\vec{p}_{T}\,^{p}$ respectively, we construct the transverse missing momentum, $\delta\vec{p}_{T} = \vec{p}_{T}\,^{\mu} + \vec{p}_{T}\,^{p}$, and its angular orientation with respect to $\vec{p}_{T}\,^{\mu}$, $\delta \alpha_{T} = \arccos\left(\cfrac{- \vec{p}_{T}\,^{\mu} \cdot \delta \vec{p}_{T}}{p_{T}\,^{\mu} \,\, \delta p_{T}}\right)$.
Due to the isotropic nature of Fermi motion, $\delta\alpha_{T}$ is expected to be uniformly distributed in the absence of any FSI.
In the presence of FSI, the proton momentum is generally reduced and the $\delta \alpha_{T}$ distribution becomes enhanced towards 180$^{\circ}$.
Similarly, the shape of the $\delta p_{T}$ distribution encapsulates information related to Fermi motion and is further smeared due to FSI and multi-nucleon effects.
Given the sensitivity of $\delta\alpha_{T}$ to FSI and of $\delta p_{T}$ to both FSI and Fermi motion, a simultaneous measurement of these two observables can help to disentangle the individual impact of each nuclear effect on the neutrino-nucleus interaction.
Similarly, the muon-proton momentum imbalance components transverse and parallel to the transverse lepton momentum, $\delta p_{T,x} = \delta p_{T} \cdot \sin\delta\alpha_{T}$ and $\delta p_{T,y} =\delta p_{T} \cdot \cos\delta\alpha_{T}$, provide further handles on Fermi motion and FSI processes, respectively.

\begin{figure*}[htb!]
\centering 
\includegraphics[width=0.32\linewidth]{\figures 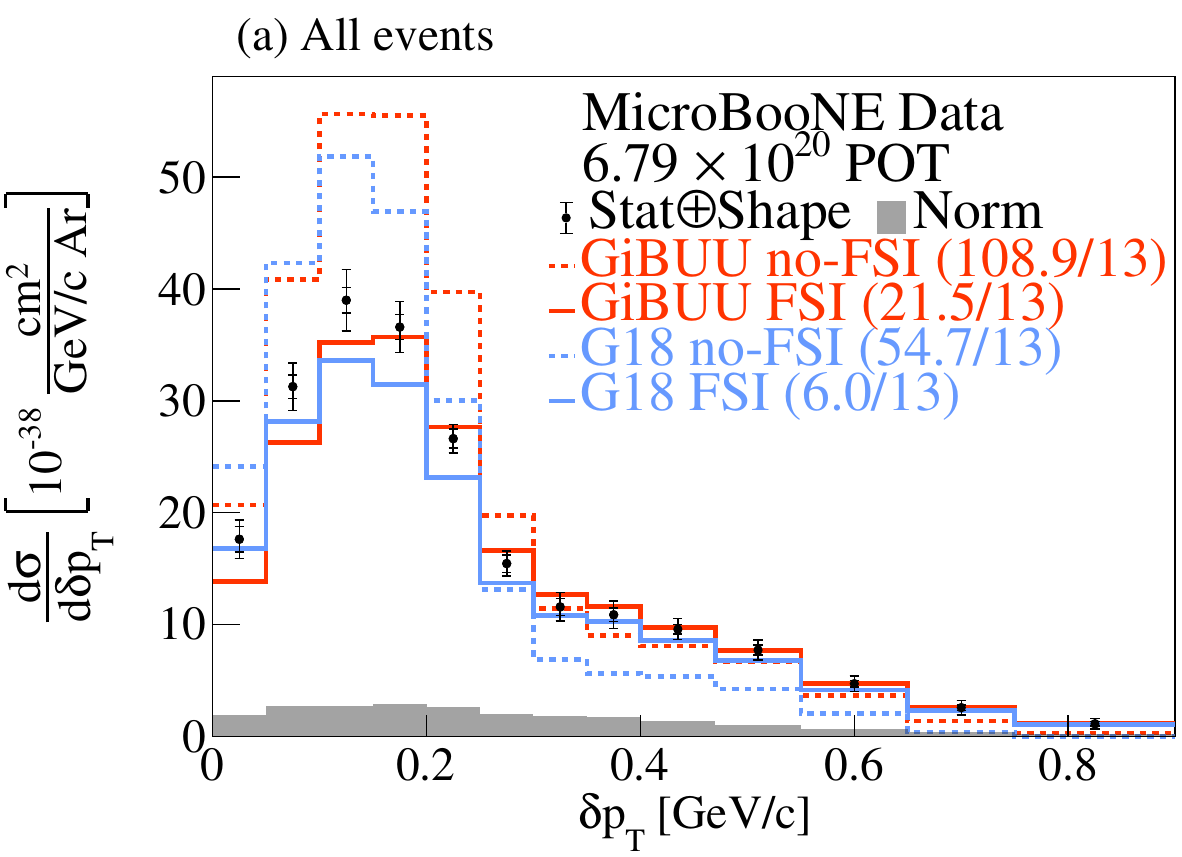}
\includegraphics[width=0.32\linewidth]{\figures 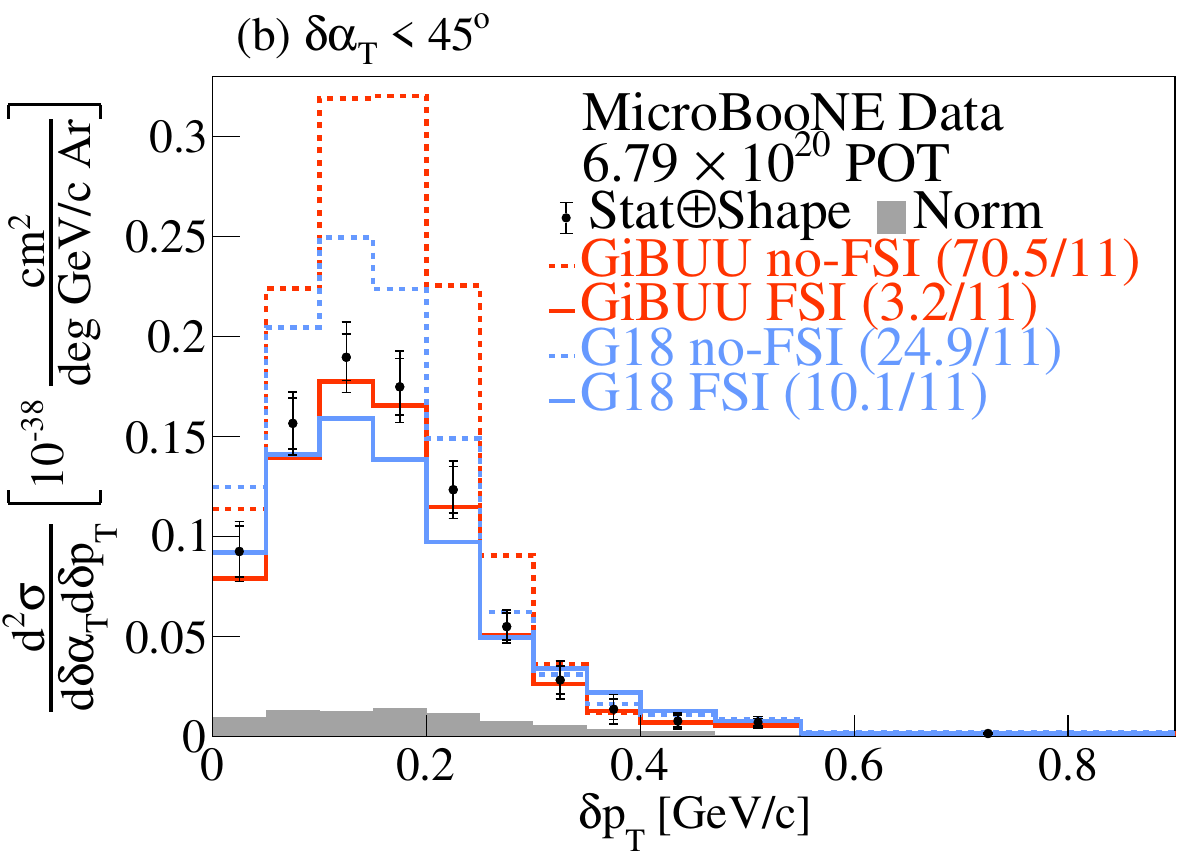}
\includegraphics[width=0.32\linewidth]{\figures 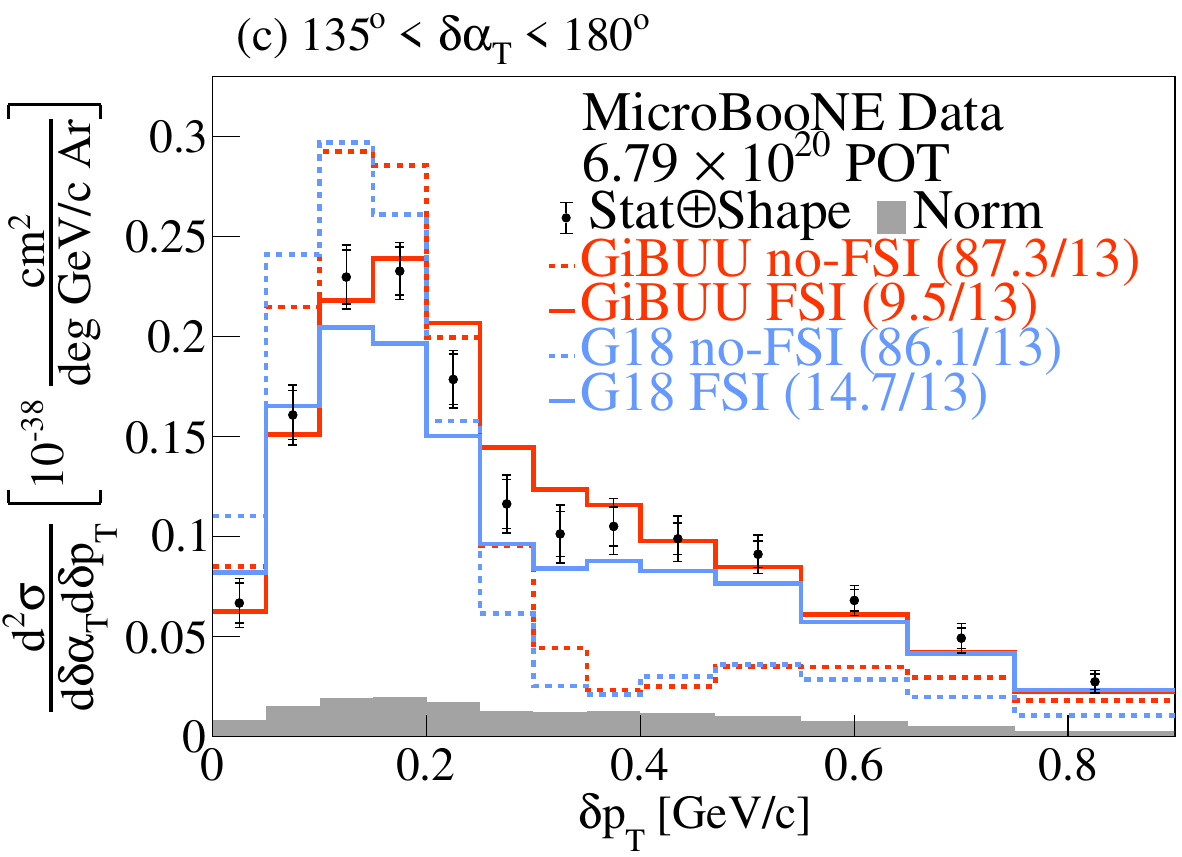}
\caption{
The flux-integrated (a) single- and (b-c) double- (in $\delta\alpha_{T}$ bins) differential \CCIpOpi\ cross sections as a function of the transverse missing momentum $\delta p_{T}$. 
Inner and outer error bars show the statistical and total (statistical and shape systematic) uncertainty at the 1$\sigma$, or 68\%, confidence level. 
The gray band shows the separate normalization systematic uncertainty.
Colored lines show the results of theoretical cross section calculations with (solid line) and without (dashed line) FSI based on the $\texttt{\scriptsize{GENIE}}$ (blue) and $\texttt{GiBUU}$ (orange) event generators.
}
\label{DeltaPTInDeltaAlphaT}
\end{figure*}


The active volume of the \uB\ LArTPC contains 85 tonnes of argon~\cite{Acciarri:2016smi}. 
It is exposed to the BNB neutrino energy spectrum that peaks around 0.8\,GeV and extends to about 2\,GeV.

Neutrinos are detected by measuring the charged particles produced following their interactions with argon nuclei in the LArTPC active volume.
These charged particles travel through the liquid argon, producing both scintillation light and trails of ionization electrons. 
In the presence of a uniform 273\,V/cm electric field, the ionization electrons drift through the argon and are detected by a system of three anode wire planes that are perpendicular to the field.
The scintillation light is measured by photomultiplier tubes (PMTs).
Events are recorded if the PMT signals are in time coincidence with the beam arrival time. 
Trigger hardware and software selection cuts reject background events, mostly from cosmic muons, providing enriched data samples in which a neutrino interaction occurs in $\approx$ 15\% of selected beam spills~\cite{Kaleko:2013eda}. 

The $\texttt{Pandora}$ reconstruction package~\cite{Acciarri:2017hat} is used to form individual tracks from the measured ionization signals in the enriched data samples.
Particle identification and momentum determination are performed using the measured track energy-deposition profile and track length~\cite{PDG_spline_table,stoppingpowersource}.

Candidate muon-proton pairs are identified by requiring exactly two track-like objects and no shower-like objects based on a track-score variable from $\texttt{Pandora}$~\cite{VanDePontseele:2020tqz,PhysRevD.105.112004}.
The discriminant described in~\cite{NicoloPID} is used to distinguish muon and proton candidates.
We further apply quality cuts to avoid mis-reconstructed tracks.
Details are given in~\cite{RefPRD}.

To reduce contributions from cosmic tracks and to minimize bin-migration effects, the event selection considers only muon and proton track pairs that are fully contained within a fiducial volume of 10\,cm from the edge of the detector active volume.

The signal definition used in this analysis includes all $\nu_{\mu}$-Ar scattering events with a final-state muon with momentum 0.1 $< p_{\mu} <$ 1.2\,GeV/$c$ and exactly one final-state proton with 0.3 $< p_{p} <$ 1\,GeV/$c$.
Events with final-state neutral pions at any momentum are excluded.
Signal events may contain additional protons with momentum less than 300 MeV/$c$ or greater than 1\,GeV/$c$, neutrons at any momentum, and charged pions with momentum lower than 70 MeV/$c$.
We refer to the signal events as CC1p0$\pi$.
Due to the requirement for a single proton and no pions in the final state, the CC1p0$\pi$ topology of interest is dominated by QE events.
Yet, more complex interactions, namely meson exchange currents (MEC), resonance interactions (RES) and deep inelastic scattering events (DIS), can still produce the CC1p0$\pi$ experimental signature.
Events that do not satisfy the CC1p0$\pi$ signal definition at a truth level are treated as background.
Such events are referred to as non-CC1p0$\pi$ and are dominated by interactions with two protons in the momentum range of interest, where the second proton was not reconstructed.
This topology is studied in~\cite{MicroBooNE:2022emb}, where a good data-simulation agreement is observed. 

After the application of the event selection, we retain 9051 data events that satisfy all criteria.
Event distributions for all the aforementioned variables of interest and details on the \CCIpOpi\ event selection can be found in~\cite{RefPRD}.

\begin{figure*}[htb!]
\centering 
\includegraphics[width=0.32\linewidth]{\figures 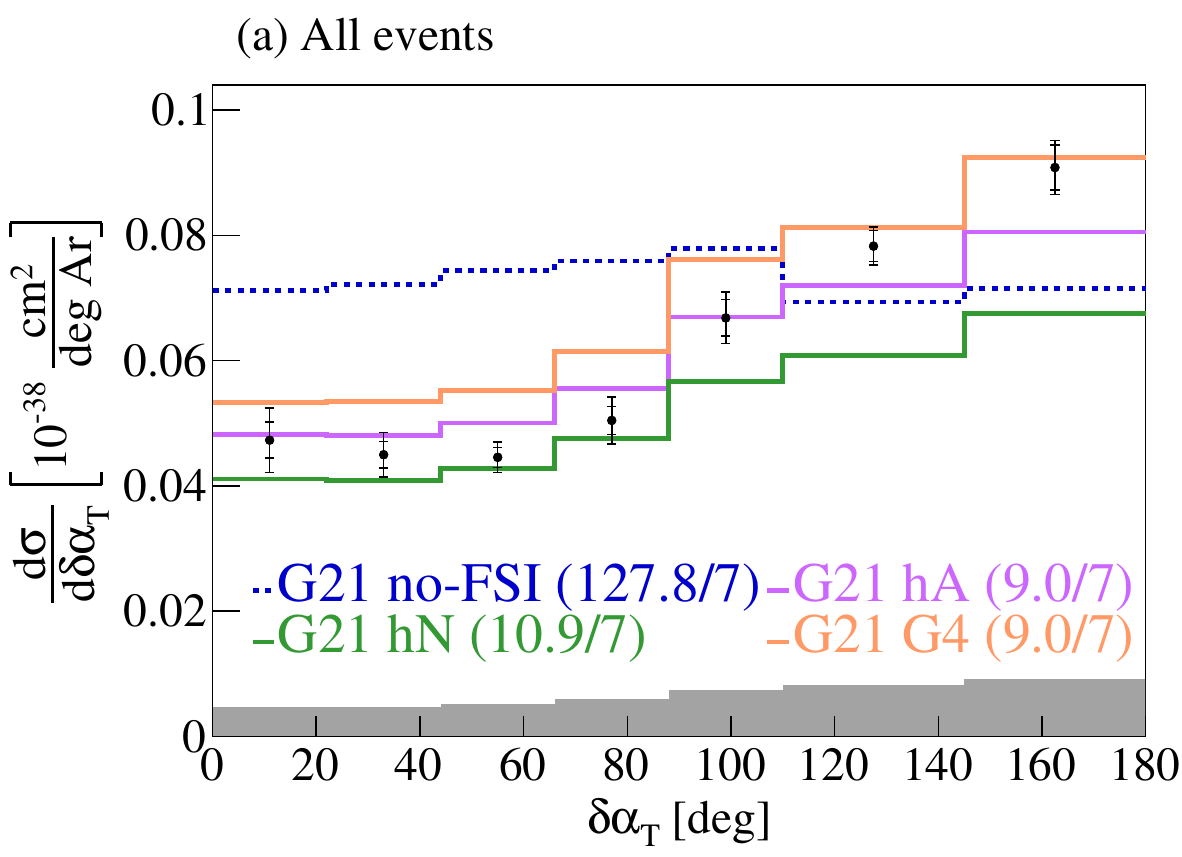}
\includegraphics[width=0.32\linewidth]{\figures 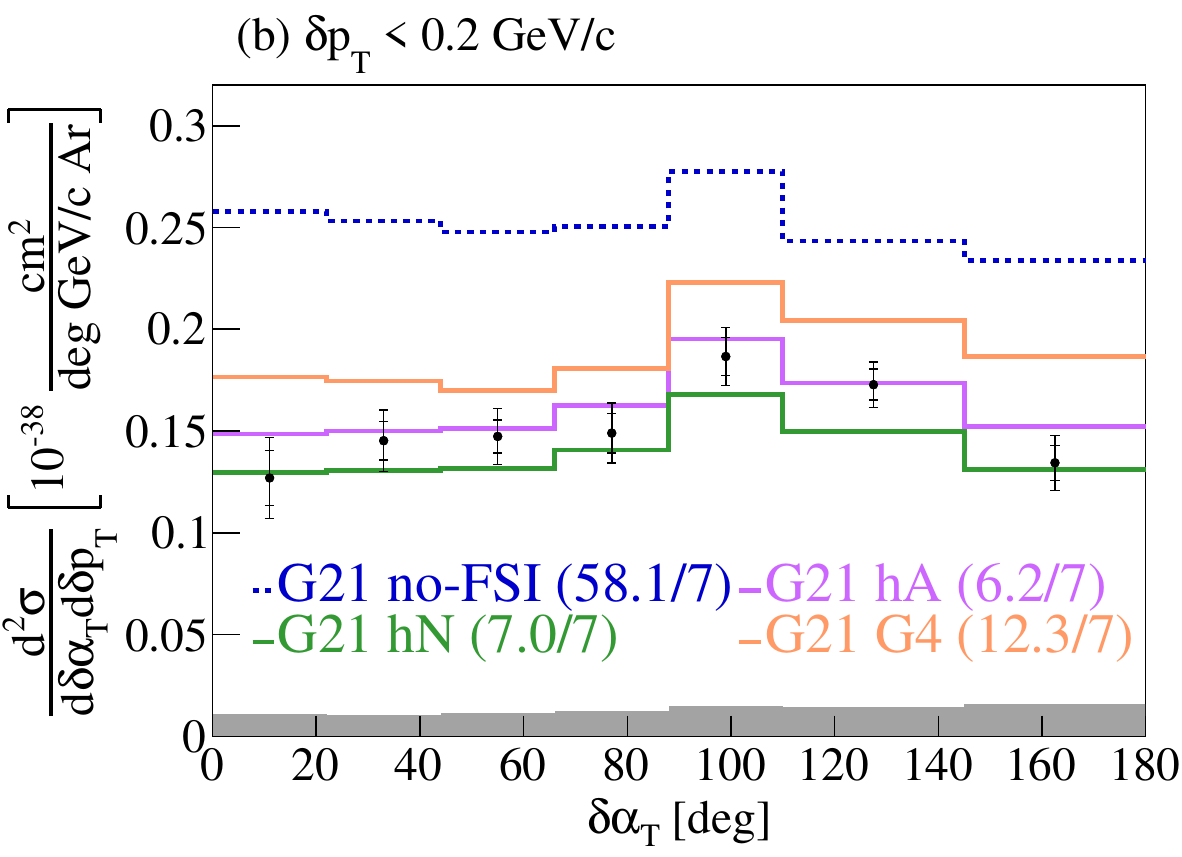}
\includegraphics[width=0.32\linewidth]{\figures 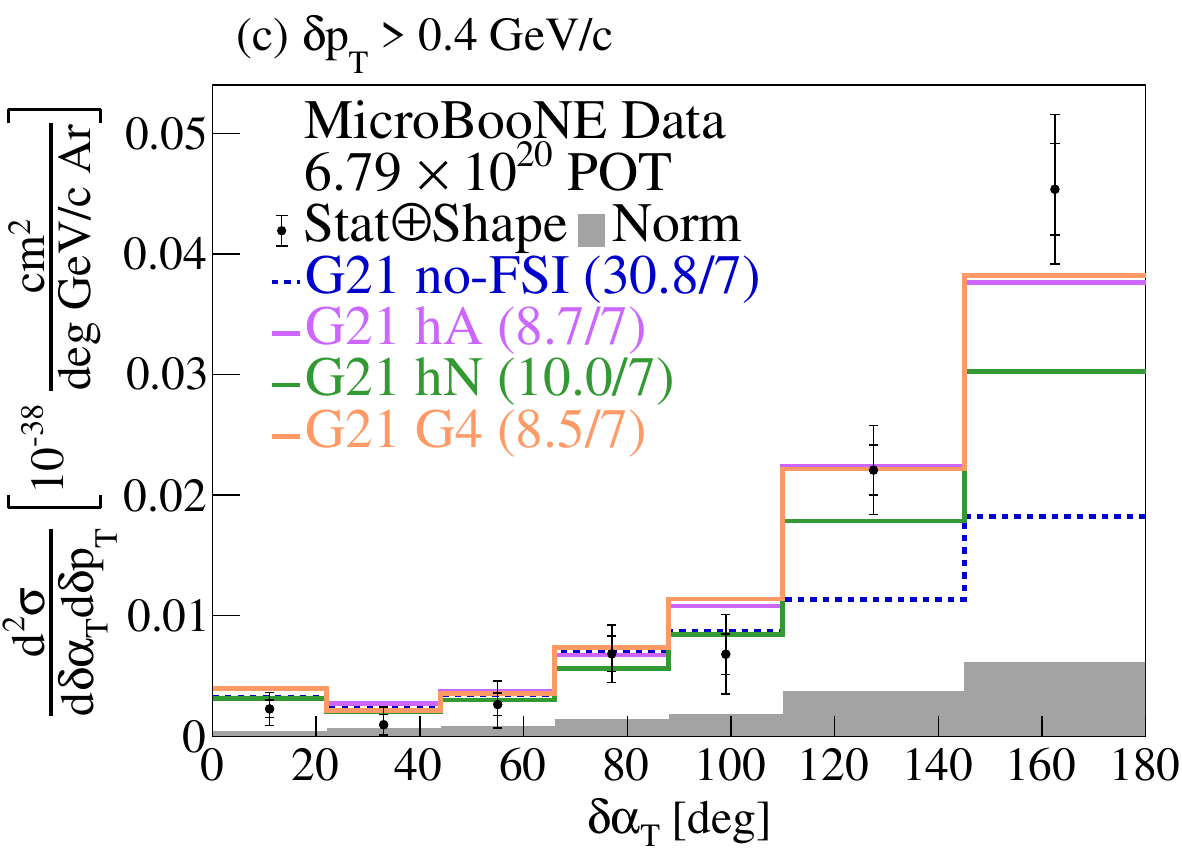}
\caption{The flux-integrated (a) single- and (b-c) double- (in $\delta p_{T}$ bins) differential \CCIpOpi\ cross sections as a function of the angle $\delta\alpha_{T}$. 
Inner and outer error bars show the statistical and total (statistical and shape systematic) uncertainty at the 1$\sigma$, or 68\%, confidence level. 
The gray band shows the separate normalization systematic uncertainty.
Colored lines show the results of theoretical cross section calculations with a number of FSI-modeling choices based on the $\texttt{GENIE}$ event generator.}
\label{DeltaAlphaTInDeltaPT}
\end{figure*}


The flux-averaged differential event rate as a function of a given variable $x$ in bin $i$ is obtained by
\begin{equation}
\frac{dR}{dx_{i}} = \frac{N_{i} - B_{i}}{T \cdot \Phi_{\nu} \cdot \Delta_{i}}
\label{fluxnormrate}
\end{equation}
where $N_{i}$ and $B_{i}$ are the number of measured events and the expected background events, respectively.
$T$ is the number of target argon nuclei in the fiducial volume of interest.
$\Phi_{\nu}$ corresponds to the total BNB flux and, finally, $\Delta_{i}$ corresponds to the $i$-th bin width or area for the single- and double-differential results, respectively.

We report the extracted cross sections for the measured interaction using the Wiener singular value decomposition (Wiener-SVD) unfolding technique as a function of unfolded kinematic variables~\cite{Tang_2017}.
More details on the unfolding procedure can be found in~\cite{RefPRD}. 
The unfolding machinery returns the unfolded differential cross section and the corresponding uncertainties.
Apart from the unfolded result, an additional smearing matrix $A_{C}$ is obtained, which accounts for the regularization and bias of the measurement.
When a comparison to the unfolded data is performed, the corresponding $A_{C}$ matrices must be applied to the true cross section predictions.
See Supplemental Material for the data release, the unfolded covariance matrices, and the additional matrices $A_{C}$.

As in previous MicroBooNE measurements~\cite{Adams:2018lzd,PhysRevLett.125.201803,PhysRevLett.128.151801,PhysRevD.105.L051102}, the full Monte Carlo (MC) simulation used in the unfolding procedure consists of a combination of simulated neutrino interactions overlaid on beam-off background events.
This provides an accurate description of the dominant cosmic backgrounds pertinent to surface detectors using real data.
Neutrino interactions are simulated using the $\texttt{GENIE v3.0.6}$ event generator~\cite{Andreopoulos:2009rq,Andreopoulos:2015wxa}.
The CC QE and CC meson exchange current (MEC) neutrino interaction models have been tuned to T2K $\nu_{\mu}$-$^{12}$C CC0$\pi$ data~\cite{PhysRevD.93.112012,GENIEKnobs}.
Predictions for more complex interactions, such as resonances, remain unaltered.
No additional MC constraints are applied.
We refer to the corresponding prediction as $\texttt{G18}$.
The latter configuration is used to simulate both CC1p0$\pi$ signal and non-CC1p0$\pi$ background events.
$\texttt{GENIE}$ generates all final-state particles associated with the primary neutrino interaction and propagates them through the nucleus, accounting for FSI.
The particle propagation outside the nucleus is simulated using $\texttt{GEANT4}$~\cite{Geant4}, with the \uB\ detector response modeled using the $\texttt{LArSoft}$ framework~\cite{Pordes:2016ycs,Snider:2017wjd}. 
Based on this simulation, we estimate that our efficiency for selecting fully contained \CCIpOpi\ events is $\approx$ 10\%, with a purity of $\approx$ 70\%.

The total covariance matrix $E$ = $E^{\mathrm{stat}}$ + $E^{\mathrm{syst}}$ used in the Wiener-SVD filter includes the statistical and systematic uncertainties associated with our measurement.
$E^{\mathrm{stat}}$ is a diagonal covariance matrix including the statistical uncertainties and $E^{\mathrm{syst}}$ is a covariance matrix incorporating the total systematic uncertainties.
More details on the sources of systematic uncertainty and the construction of these matrices can be found in~\cite{RefPRD}.
These matrices include uncertainties on the integrated cross section due to the neutrino flux prediction (7.3\%)~\cite{Aguilar-Arevalo:2013dva}, neutrino interaction cross section modeling (6\%)~\cite{Andreopoulos:2009rq,Andreopoulos:2015wxa,GENIEKnobs}, detector response modeling (4.9\%)~\cite{WireMod}, beam exposure (2.3\%), statistics (1.5\%), number-of-scattering-targets (1.15\%), reinteractions (1\%)~\cite{Calcutt_2021}, and out-of-cryostat interaction modeling (0.2\%).
The full fractional uncertainty on the integrated total cross section sums to 11\%.

Across the results reported in this Letter, statistical uncertainties are shown by the inner error bars on the final results.
The systematic uncertainties were decomposed into shape- and normalization-related sources following the procedure outlined in~\cite{MatrixDecomv}.
The cross-term uncertainties were incorporated in the normalization part.
The outer error bars on the reported cross sections correspond to statistical and shape uncertainties added in quadrature.
The normalization uncertainties are presented with the gray band at the bottom of our results.

\begin{figure*}[htb!]
\centering 
\includegraphics[width=0.32\linewidth]{\figures 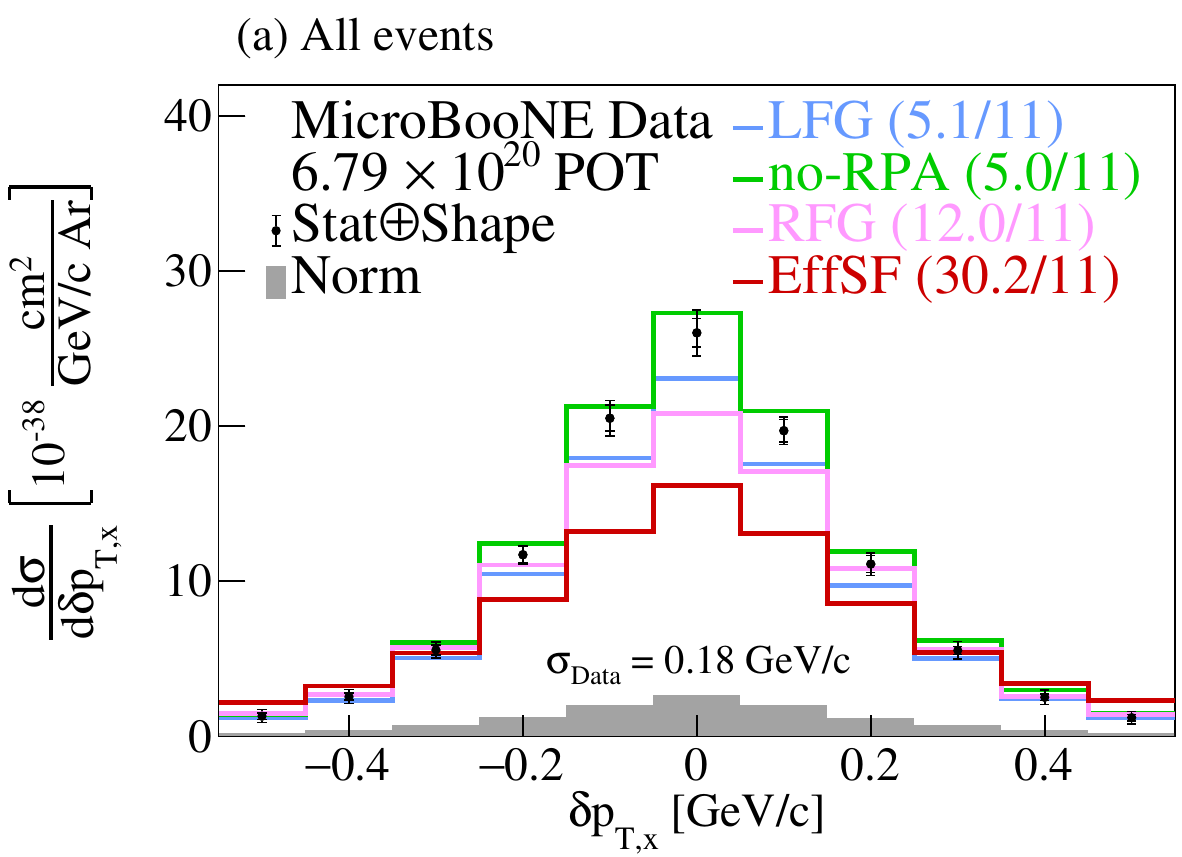}
\includegraphics[width=0.32\linewidth]{\figures 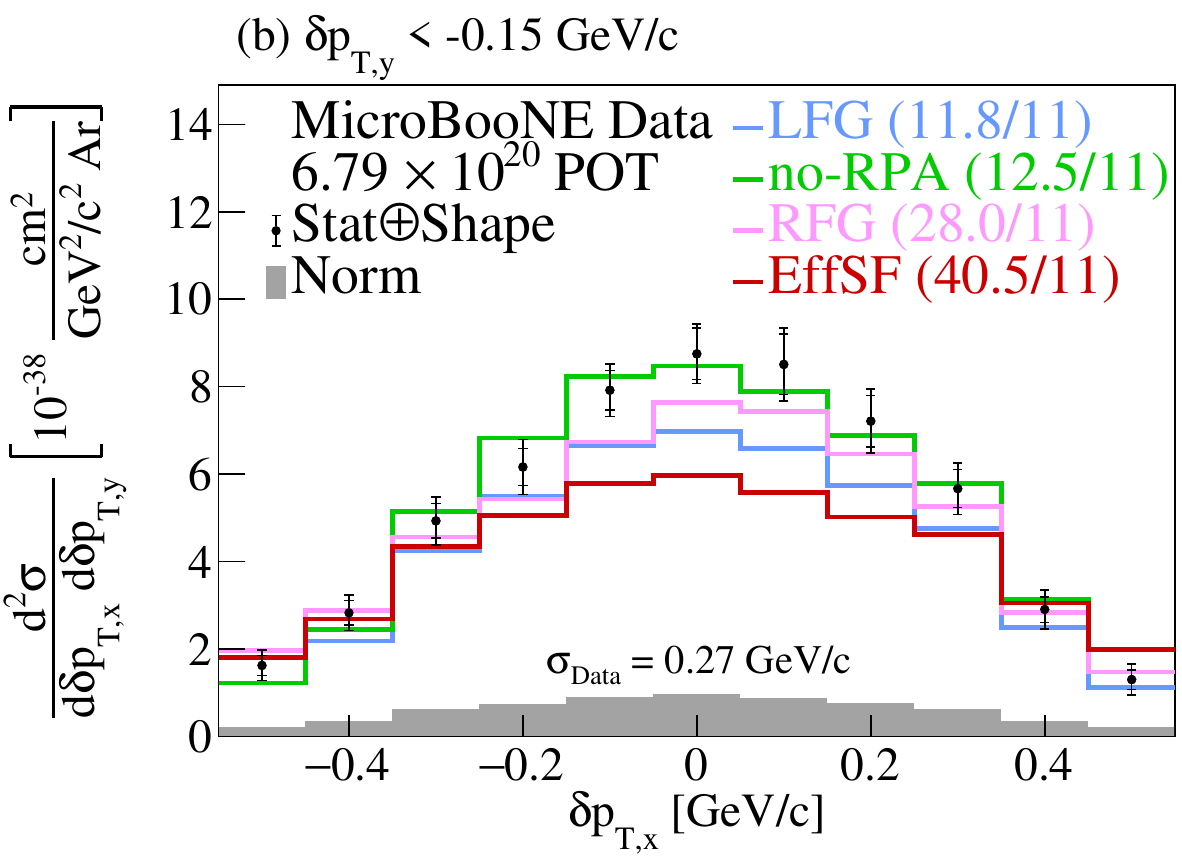}
\includegraphics[width=0.32\linewidth]{\figures 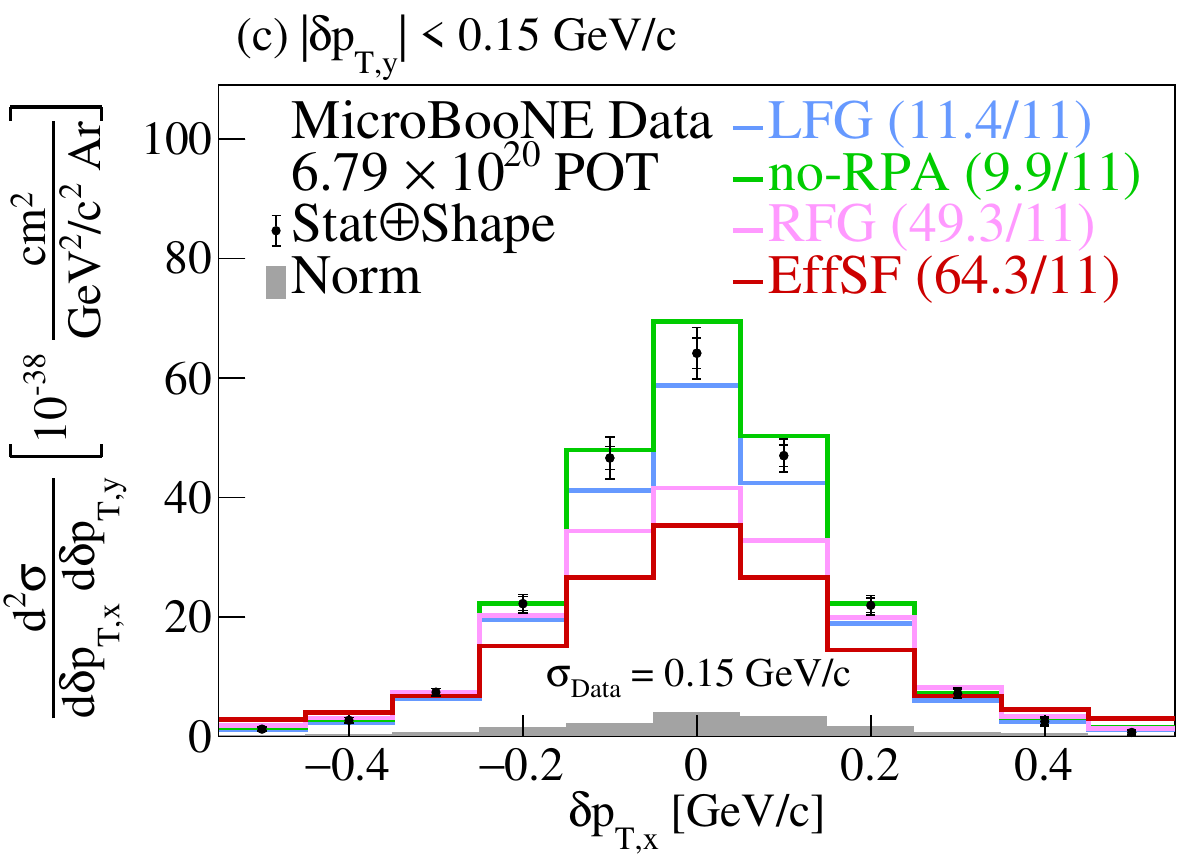}
\caption{
The flux-integrated (a) single- and (b-c) double- (in $\delta p_{T,y}$ bins) differential \CCIpOpi\ cross sections as a function of the transverse three-momentum transfer component, $\delta p_{T,x}$. 
Inner and outer error bars show the statistical and total (statistical and shape systematic) uncertainty at the 1$\sigma$, or 68\%, confidence level. 
The gray band shows the separate normalization systematic uncertainty.
Colored lines show the results of theoretical cross section calculations with a number of event generators.
The standard deviation ($\sigma_\mathrm{Data}$) of a Gaussian fit to the data is shown on each panel. 
}
\label{DeltaPtxInDeltaPty}
\end{figure*}


The single- and double-differential results as a function of $\delta p_{T}$ are presented in Fig.~\ref{DeltaPTInDeltaAlphaT}.
They are compared with $\texttt{G18}$ and the theory-driven $\texttt{GiBUU 2021 (GiB)}$ event generator.
Additional comparisons to the corresponding event generators when FSI are turned off are also included ($\texttt{G18 no-FSI}$ and $\texttt{GiBUU no-FSI}$).
$\texttt{G18}$ uses the local Fermi gas (LFG) model of the nuclear ground state~\cite{Carrasco:1989vq} and the Nieves CCQE scattering prescription~\cite{Nieves:2012yz} with Coulomb corrections for the outgoing muon~\cite{Engel:1997fy} and random phase approximation (RPA) corrections~\cite{RPA}.
It also uses the Nieves MEC model~\cite{Schwehr:2016pvn}, the KLN-BS resonance (RES)~\cite{Nowak:2009se,Kuzmin:2003ji,Berger:2007rq,Graczyk:2007bc} and Berger-Sehgal coherent (COH)~\cite{Berger:2008xs} scattering models.
Furthermore, the hA2018 FSI model~\cite{Ashery:1981tq} and the MicroBooNE-specific tuning of model parameters~\cite{GENIEKnobs} are utilized.
$\texttt{GiBUU}$ uses somewhat similar models, but, unlike $\texttt{GENIE}$, they are implemented in a coherent way by solving the Boltzmann-Uehling-Uhlenbeck transport equation~\cite{Mosel:2019vhx}. 
The simulation includes the LFG model~\cite{Carrasco:1989vq}, a standard CCQE expression~\cite{Leitner:2006ww}, an empirical MEC model and a dedicated spin-dependent resonances amplitude calculation following the $\texttt{MAID}$ analysis~\cite{Mosel:2019vhx}. 
The deep inelastic (DIS) model is from $\texttt{PYTHIA}$~\cite{Sjostrand:2006za}.
The FSI treatment is different as the hadrons propagate through the residual nucleus in a nuclear potential which is consistent with the initial state.

The single-differential results as a function of $\delta p_{T}$ using all the events that satisfy our selection are shown in Fig.~\ref{DeltaPTInDeltaAlphaT}a.
The $\chi^{2}$/bins data comparison for each generator shown on all the results takes into account the total covariance matrix, including the off-diagonal elements.
Theoretical uncertainties on the models themselves are not included.
The peak height of both generator predictions is $\approx$ 30\% higher when FSI effects are turned off.
Yet, all distributions illustrate a transverse missing momentum tail that extends beyond the Fermi momentum ($\approx$ 250\,MeV/$c$) whether FSI effects are incorporated or not.
The double-differential result using events with $\delta\alpha_{T} < 45^{\circ}$ shown in Fig.~\ref{DeltaPTInDeltaAlphaT}b is dominated by events that primarily occupy the region up to the Fermi momentum and do not exhibit a high-momentum tail.
The double-differential results using events with $135^{\circ} < \delta\alpha_{T} < 180^{\circ}$ are shown in Fig.~\ref{DeltaPTInDeltaAlphaT}c and illustrate high transverse missing momentum up to 1\,GeV/$c$. 
The prediction without FSI effects is strongly disfavored.
The region around 0.3\,GeV/c in Fig.~\ref{DeltaPTInDeltaAlphaT}c shows a noticeable difference between the $\texttt{G18}$ and $\texttt{GiBUU}$ predictions.
This behavior could be driven by the different approaches of simulating the MEC and FSI effects between the two event generators, as can be seen in the interaction breakdown of the relevant cross sections in the Supplemental Material.
Therefore, the high $\delta p_{T}$ region is an appealing candidate for neutrino experiments to benchmark and tune the FSI modeling in event generators. 
The same single- and double-differential cross section comparisons as a function of $\delta p_{T}$ using different FSI variations are included in the Supplemental Material.

Extracted cross sections as a function of $\delta\alpha_{T}$ are shown in Fig.~\ref{DeltaAlphaTInDeltaPT}.
Here we perform comparisons to the recently added theory driven $\texttt{GENIE v3.0.6 G21\_11b\_00\_000}$ configuration ($\texttt{G21 hN}$)~\cite{geniev3highlights}.
This configuration uses the SuSAv2 model for CCQE and CCMEC interactions~\cite{PhysRevD.101.033003}, and the hN2018 FSI model~\cite{hN2018}.
The modeling choices for RES, DIS, and COH interactions are the same as for $\texttt{G18}$.	
We investigated the effect of the FSI-modeling choice by comparing the $\texttt{G21 hN}$ results to the ones obtained with $\texttt{G21 hA}$, where the hA2018 FSI model was used instead, and to $\texttt{G21 G4}$ with the recently coupled $\texttt{GEANT4}$ FSI framework~\cite{Wright:2015xia}.
The prediction where the FSI effects have been turned off ($\texttt{G21 no-FSI}$) is also included for comparison.
The impact of different QE modeling options as a function of the same variables is investigated in the Supplemental Material.

The single-differential results as a function of $\delta\alpha_{T}$ using all the events that satisfy our selection are shown in Fig.~\ref{DeltaAlphaTInDeltaPT}a.
The prediction without FSI shows a uniform behavior as a function of $\delta\alpha_{T}$ and is disfavored by the data.
The addition of FSI effects leads to a $\approx$ 30\% asymmetry around $\delta\alpha_{T} = 90^{\circ}$.
The three FSI models used here for comparison yield a consistent behavior. 
The double-differential result shown in Fig.~\ref{DeltaAlphaTInDeltaPT}b using events with $\delta p_{T} <$ 0.2\,GeV/$c$ illustrates a uniform distribution indicative of the suppressed FSI impact in that part of the phase-space.
The $\texttt{G21 no-FSI}$ prediction is higher than the other FSI predictions.
The difference comes from the generation of multiple particles above detection threshold due to reinteraction effects in the FSI-rich samples.
Such events do not satisfy the signal definition and therefore introduce the difference in the absolute scale.
The double-differential results using events with $\delta p_{T} >$ 0.4\,GeV/$c$ are shown in Fig.~\ref{DeltaAlphaTInDeltaPT}c and illustrate the presence of strong FSI effects with a significantly enhanced asymmetry around $90^{\circ}$.
Thus, the high $\delta\alpha_{T}$ region is highly informative for the FSI-modeling performance in event generators.
See Supplemental Material for details on the interaction breakdown of the aforementioned results and~\cite{RefPRD} for further double-differential results.

Finally, Fig.~\ref{DeltaPtxInDeltaPty} shows the single- and double-differential results as a function of $\delta p_{T,x}$.
The result shows the comparison between the nominal $\texttt{G18}$ model using the local Fermi gas ($\texttt{LFG}$) and predictions using the same $\texttt{G18}$ interaction modeling but different nuclear ground-state model options available in the $\texttt{GENIE}$ event generator, namely the Bodek-Ritchie Fermi Gas ($\texttt{RFG}$)~\cite{PhysRevD.23.1070} and an effective spectral function ($\texttt{EffSF}$)~\cite{PhysRevC.74.054316}.
Furthermore, the prediction without RPA effects is shown for comparison ($\texttt{no-RPA}$)~\cite{RPA}.
The FSI impact on the same results in investigated in the Supplemental Material.

The single-differential result (Fig.~\ref{DeltaPtxInDeltaPty}a) illustrates a fairly broad symmetric distribution centered around 0\,GeV/$c$.
The double-differential result for events where $\delta p_{T,y} <$ -0.15\,GeV/$c$ (Fig.~\ref{DeltaPtxInDeltaPty}b) illustrates an even broader distribution, as can be seen in the widths ($\sigma_\mathrm{Data}$) of Gaussian fits on the data distributions.
Conversely, the double-differential result for events with $|\delta p_{T,y}| <$ 0.15\,GeV/$c$ (Fig.~\ref{DeltaPtxInDeltaPty}c) shows a much narrower peak which strongly depends on the choice of the underlying model and the inclusion or absence of nuclear effects such as RPA.
The $\texttt{LFG}$ and $\texttt{no-RPA}$ predictions are favored in both parts of the phase-space.
Both the RFG and EffSF predictions illustrate a poor performance in the double-differential measurements and particularly in the QE-dominated $|\delta p_{T,y}| <$ 0.15\,GeV/$c$ region.
The FSI-modeling impact on the same $\delta p_{T,x}$ cross sections is presented in the Supplemental Material.
The latter further contains details on the interaction breakdown of various generator predictions for the results reported here, and further single- and double-differential results can be found in~\cite{RefPRD}.


In summary, we report the first measurement of muon neutrino double-differential cross sections on argon as a function of kinematic imbalance variables for event topologies with a single muon and a single proton detected in the final state.
We identify parts of the phase space where the Fermi motion can be largely disentangled from FSI and multi-nucleon effects. 
This disentanglement provides leverage to improve separate parts of the complicated neutrino interaction models that affect single-differential distributions in similar ways.
Therefore, the reported results pave the path to substantially reducing cross section systematic uncertainties which will enable precision measurements of fundamental neutrino properties.

\begin{acknowledgments}
This document was prepared by the MicroBooNE collaboration using the
resources of the Fermi National Accelerator Laboratory (Fermilab), a
U.S. Department of Energy, Office of Science, HEP User Facility.
Fermilab is managed by Fermi Research Alliance, LLC (FRA), acting
under Contract No. DE-AC02-07CH11359.  MicroBooNE is supported by the
following: the U.S. Department of Energy, Office of Science, Offices
of High Energy Physics and Nuclear Physics; the U.S. National Science
Foundation; the Swiss National Science Foundation; the Science and
Technology Facilities Council (STFC), part of the United Kingdom Research 
and Innovation; the Royal Society (United Kingdom); the UK Research 
and Innovation (UKRI) Future Leaders Fellowship; and The European 
Union’s Horizon 2020 Marie Sklodowska-Curie Actions. Additional support 
for the laser calibration system and cosmic ray tagger was provided by 
the Albert Einstein Center for Fundamental Physics, Bern, Switzerland. 
We also acknowledge the contributions of technical and scientific staff 
to the design, construction, and operation of the MicroBooNE detector 
as well as the contributions of past collaborators to the development 
of MicroBooNE analyses, without whom this work would not have been 
possible. For the purpose of open access, the authors have applied a 
Creative Commons Attribution (CC BY) license to any Author Accepted 
Manuscript version arising from this submission.
\end{acknowledgments}


\bibliography{main}


\end{document}


\centering
\large{\textbf{First Double-Differential Measurement of Kinematic Imbalance\texorpdfstring{\\}{}in Neutrino Interactions with the MicroBooNE Detector}}

(Dated: \today)


\justify
\section{Data Release}

Overflow (underflow) values are included in the last (first) bin.
The additional smearing matrix $A_{C}$ should first be applied to the event distribution of an independent theoretical prediction when a comparison is performed to the data reported herein, and then divided by the bin width.
The double-differential cross sections include correlations between the phase-space slices.
The data release with the data results, the covariance matrices, and the additional smearing matrices is included in the DataRelease.root file.
Instructions on how to use the data release and the description of the binning scheme are included in the README file.
More single- and double-differential cross section results can be found in~\cite{RefPRD}.

\raggedbottom
\begin{table}[H]
\raggedright
\begin{adjustbox}{width=\textwidth}
\small
\begin{tabular}{ |c|c|c|c|c| }
\hline
\multicolumn{5}{|c|}{Cross Section $\delta p_{T}$, $All\,events$} \\
\hline
\hline
Bin \# & Low edge [GeV/$\textit{c}$] & High edge [GeV/$\textit{c}$] & Cross Section [$10^{-38}\frac{cm^{2}}{(GeV/\textit{c})\,^{40}Ar}$] & Uncertainty [$10^{-38}\frac{cm^{2}}{(GeV/\textit{c})\,^{40}Ar}$] \\
\hline
\hline
1 & 0 & 0.05 & 17.62 & 2.5914\\
2 & 0.05 & 0.1 & 31.268 & 3.4669\\
3 & 0.1 & 0.15 & 38.993 & 3.8931\\
4 & 0.15 & 0.2 & 36.591 & 3.6679\\
5 & 0.2 & 0.25 & 26.615 & 2.9078\\
6 & 0.25 & 0.3 & 15.445 & 2.285\\
7 & 0.3 & 0.35 & 11.569 & 2.2447\\
8 & 0.35 & 0.4 & 10.864 & 2.1028\\
9 & 0.4 & 0.47 & 9.5846 & 1.6583\\
10 & 0.47 & 0.55 & 7.7026 & 1.3719\\
11 & 0.55 & 0.65 & 4.696 & 0.97287\\
12 & 0.65 & 0.75 & 2.5541 & 0.77974\\
13 & 0.75 & 0.9 & 1.117 & 0.50573\\
\hline
\end{tabular}
\end{adjustbox}
\end{table}

\begin{table}[H]
\raggedright
\begin{adjustbox}{width=\textwidth}
\small
\begin{tabular}{ |c|c|c|c|c| }
\hline
\multicolumn{5}{|c|}{Cross Section $\delta p_{T}$, $\delta\alpha_{T}\,<\,45^{o}$} \\
\hline
\hline
Bin \# & Low edge [GeV/$\textit{c}$] & High edge [GeV/$\textit{c}$] & Cross Section [$10^{-38}\frac{cm^{2}}{deg\,(GeV/\textit{c})\,^{40}Ar}$] & Uncertainty [$10^{-38}\frac{cm^{2}}{deg\,(GeV/\textit{c})\,^{40}Ar}$] \\
\hline
\hline
1 & 0 & 0.05 & 0.092432 & 0.017847\\
2 & 0.05 & 0.1 & 0.1565 & 0.020584\\
3 & 0.1 & 0.15 & 0.18956 & 0.02176\\
4 & 0.15 & 0.2 & 0.17479 & 0.022882\\
5 & 0.2 & 0.25 & 0.12334 & 0.018392\\
6 & 0.25 & 0.3 & 0.05485 & 0.011418\\
7 & 0.3 & 0.35 & 0.028104 & 0.010931\\
8 & 0.35 & 0.4 & 0.013506 & 0.008218\\
9 & 0.4 & 0.47 & 0.0076037 & 0.0046992\\
10 & 0.47 & 0.55 & 0.0070471 & 0.0029157\\
11 & 0.55 & 0.9 & 0.0013794 & 0.00037879\\
\hline
\end{tabular}
\end{adjustbox}
\end{table}

\begin{table}[H]
\raggedright
\begin{adjustbox}{width=\textwidth}
\small
\begin{tabular}{ |c|c|c|c|c| }
\hline
\multicolumn{5}{|c|}{Cross Section $\delta p_{T}$, $135^{o}\,<\,\delta\alpha_{T}\,<\,180^{o}$} \\
\hline
\hline
Bin \# & Low edge [GeV/$\textit{c}$] & High edge [GeV/$\textit{c}$] & Cross Section [$10^{-38}\frac{cm^{2}}{deg\,(GeV/\textit{c})\,^{40}Ar}$] & Uncertainty [$10^{-38}\frac{cm^{2}}{deg\,(GeV/\textit{c})\,^{40}Ar}$] \\
\hline
\hline
1 & 0 & 0.05 & 0.066626 & 0.014819\\
2 & 0.05 & 0.1 & 0.16071 & 0.021217\\
3 & 0.1 & 0.15 & 0.22967 & 0.025077\\
4 & 0.15 & 0.2 & 0.23265 & 0.024401\\
5 & 0.2 & 0.25 & 0.17848 & 0.022261\\
6 & 0.25 & 0.3 & 0.11622 & 0.019362\\
7 & 0.3 & 0.35 & 0.10121 & 0.019008\\
8 & 0.35 & 0.4 & 0.10497 & 0.018705\\
9 & 0.4 & 0.47 & 0.09877 & 0.016175\\
10 & 0.47 & 0.55 & 0.091046 & 0.013986\\
11 & 0.55 & 0.65 & 0.067952 & 0.010618\\
12 & 0.65 & 0.75 & 0.049075 & 0.0090352\\
13 & 0.75 & 0.9 & 0.027264 & 0.0062172\\
\hline
\end{tabular}
\end{adjustbox}
\end{table}

\begin{table}[H]
\raggedright
\begin{adjustbox}{width=\textwidth}
\small
\begin{tabular}{ |c|c|c|c|c| }
\hline
\multicolumn{5}{|c|}{Cross Section $\delta\alpha_{T}$, $All\,events$} \\
\hline
\hline
Bin \# & Low edge [deg] & High edge [deg] & Cross Section [$10^{-38}\frac{cm^{2}}{deg\,^{40}Ar}$] & Uncertainty [$10^{-38}\frac{cm^{2}}{deg\,^{40}Ar}$] \\
\hline
\hline
1 & 0 & 22 & 0.047296 & 0.0069867\\
2 & 22 & 44 & 0.044963 & 0.00586\\
3 & 44 & 66 & 0.044549 & 0.0056673\\
4 & 66 & 88 & 0.050441 & 0.00704\\
5 & 88 & 110 & 0.066812 & 0.0084041\\
6 & 110 & 145 & 0.078273 & 0.0087228\\
7 & 145 & 180 & 0.090828 & 0.010114\\
\hline
\end{tabular}
\end{adjustbox}
\end{table}

\begin{table}[H]
\raggedright
\begin{adjustbox}{width=\textwidth}
\small
\begin{tabular}{ |c|c|c|c|c| }
\hline
\multicolumn{5}{|c|}{Cross Section $\delta\alpha_{T}$, $\delta p_{T}\,<\,0.2\,GeV/\textit{c}$} \\
\hline
\hline
Bin \# & Low edge [deg] & High edge [deg] & Cross Section [$10^{-38}\frac{cm^{2}}{deg\,(GeV/\textit{c})\,^{40}Ar}$] & Uncertainty [$10^{-38}\frac{cm^{2}}{deg\,(GeV/\textit{c})\,^{40}Ar}$] \\
\hline
\hline
1 & 0 & 22 & 0.12696 & 0.022679\\
2 & 22 & 44 & 0.14525 & 0.018476\\
3 & 44 & 66 & 0.14735 & 0.01792\\
4 & 66 & 88 & 0.14898 & 0.01917\\
5 & 88 & 110 & 0.18662 & 0.020387\\
6 & 110 & 145 & 0.17278 & 0.018093\\
7 & 145 & 180 & 0.13438 & 0.020788\\
\hline
\end{tabular}
\end{adjustbox}
\end{table}

\begin{table}[H]
\raggedright
\begin{adjustbox}{width=\textwidth}
\small
\begin{tabular}{ |c|c|c|c|c| }
\hline
\multicolumn{5}{|c|}{Cross Section $\delta\alpha_{T}$, $\delta p_{T}\,>\,0.4\,GeV/\textit{c}$} \\
\hline
\hline
Bin \# & Low edge [deg] & High edge [deg] & Cross Section [$10^{-38}\frac{cm^{2}}{deg\,(GeV/\textit{c})\,^{40}Ar}$] & Uncertainty [$10^{-38}\frac{cm^{2}}{deg\,(GeV/\textit{c})\,^{40}Ar}$] \\
\hline
\hline
1 & 0 & 22 & 0.0022705 & 0.0014401\\
2 & 22 & 44 & 0.0009548 & 0.0016242\\
3 & 44 & 66 & 0.0026414 & 0.0021136\\
4 & 66 & 88 & 0.0068414 & 0.0027825\\
5 & 88 & 110 & 0.0068027 & 0.0037867\\
6 & 110 & 145 & 0.022076 & 0.0052629\\
7 & 145 & 180 & 0.045358 & 0.0087483\\
\hline
\end{tabular}
\end{adjustbox}
\end{table}

\begin{table}[H]
\raggedright
\begin{adjustbox}{width=\textwidth}
\small
\begin{tabular}{ |c|c|c|c|c| }
\hline
\multicolumn{5}{|c|}{Cross Section $\delta p_{T,x}$, $All\,events$} \\
\hline
\hline
Bin \# & Low edge [GeV/$\textit{c}$] & High edge [GeV/$\textit{c}$] & Cross Section [$10^{-38}\frac{cm^{2}}{(GeV/\textit{c})\,^{40}Ar}$] & Uncertainty [$10^{-38}\frac{cm^{2}}{(GeV/\textit{c})\,^{40}Ar}$] \\
\hline
\hline
1 & -0.55 & -0.45 & 1.3017107 & 0.46826726\\
2 & -0.45 & -0.35 & 2.563368 & 0.59039691\\
3 & -0.35 & -0.25 & 5.5731445 & 0.89649636\\
4 & -0.25 & -0.15 & 11.707776 & 1.3957739\\
5 & -0.15 & -0.05 & 20.511688 & 2.3000876\\
6 & -0.05 & 0.05 & 26.020455 & 3.0236695\\
7 & 0.05 & 0.15 & 19.708245 & 2.2296454\\
8 & 0.15 & 0.25 & 11.103336 & 1.4077286\\
9 & 0.25 & 0.35 & 5.5393912 & 0.9180937\\
10 & 0.35 & 0.45 & 2.5182752 & 0.62433322\\
11 & 0.45 & 0.55 & 1.2015199 & 0.45891559\\
\hline
\end{tabular}
\end{adjustbox}
\end{table}

\begin{table}[H]
\raggedright
\begin{adjustbox}{width=\textwidth}
\small
\begin{tabular}{ |c|c|c|c|c| }
\hline
\multicolumn{5}{|c|}{Cross Section $\delta p_{T,x}$, $\delta p_{T,y}\,<\,-0.15\,GeV/\textit{c}$} \\
\hline
\hline
Bin \# & Low edge [GeV/$\textit{c}$] & High edge [GeV/$\textit{c}$] & Cross Section [$10^{-38}\frac{cm^{2}}{(GeV/\textit{c})^2\,^{40}Ar}$] & Uncertainty [$10^{-38}\frac{cm^{2}}{(GeV/\textit{c})^2\,^{40}Ar}$] \\
\hline
\hline
1 & -0.55 & -0.45 & 1.6208121 & 0.40811656\\
2 & -0.45 & -0.35 & 2.8270055 & 0.53216682\\
3 & -0.35 & -0.25 & 4.9291254 & 0.82899734\\
4 & -0.25 & -0.15 & 6.1606674 & 0.9663476\\
5 & -0.15 & -0.05 & 7.917514 & 1.0822931\\
6 & -0.05 & 0.05 & 8.7498076 & 1.1891885\\
7 & 0.05 & 0.15 & 8.5082736 & 1.2186225\\
8 & 0.15 & 0.25 & 7.2098121 & 1.0521901\\
9 & 0.25 & 0.35 & 5.6680756 & 0.864424\\
10 & 0.35 & 0.45 & 2.9003432 & 0.56377234\\
11 & 0.45 & 0.55 & 1.2978947 & 0.41077731\\
\hline
\end{tabular}
\end{adjustbox}
\end{table}

\begin{table}[H]
\raggedright
\begin{adjustbox}{width=\textwidth}
\small
\begin{tabular}{ |c|c|c|c|c| }
\hline
\multicolumn{5}{|c|}{Cross Section $\delta p_{T,x}$, $|\delta p_{T,y}|\,<\,0.15\,GeV/\textit{c}$} \\
\hline
\hline
Bin \# & Low edge [GeV/$\textit{c}$] & High edge [GeV/$\textit{c}$] & Cross Section [$10^{-38}\frac{cm^{2}}{(GeV/\textit{c})^2\,^{40}Ar}$] & Uncertainty [$10^{-38}\frac{cm^{2}}{(GeV/\textit{c})^2\,^{40}Ar}$] \\
\hline
\hline
1 & -0.55 & -0.45 & 1.2119704 & 0.61176744\\
2 & -0.45 & -0.35 & 2.6763573 & 0.82181016\\
3 & -0.35 & -0.25 & 7.3739087 & 1.055886\\
4 & -0.25 & -0.15 & 22.201232 & 2.2046742\\
5 & -0.15 & -0.05 & 46.579385 & 4.1454453\\
6 & -0.05 & 0.05 & 64.1127 & 5.923266\\
7 & 0.05 & 0.15 & 46.971297 & 4.3486902\\
8 & 0.15 & 0.25 & 21.927763 & 2.3914571\\
9 & 0.25 & 0.35 & 7.1648705 & 1.1876442\\
10 & 0.35 & 0.45 & 2.4895325 & 0.95614843\\
11 & 0.45 & 0.55 & 0.65303698 & 0.72564438\\
\hline
\end{tabular}
\end{adjustbox}
\end{table}


\section{Fake Data Studies}\label{FakeData}

For the purposes of our fake data studies, we treated the signal reconstructed events from an alternative model as if they were real data. 
We investigated three samples: (a) using NuWro events, (b) by removing the weights corresponding to the MicroBooNE tune (NoTune) and, (c) by multiplying the weight for the MEC events by 2 (TwiceMEC).
We then extracted the cross section from these fake data using our nominal MC response and covariance matrices and the Wiener-SVD filter both for the single- and double-differential cross sections. 
Given that we have the true cross section information from the fake data, we compared the fake data extracted cross section to the corresponding truth-level cross section predictions.  
The outcome agreed within an  expected  uncertainty.   
In  the  case  of  all  these  variations,  uncertainties  due  to  beam exposure, number-of-targets, detector, fluxes, and re-interactions are irrelevant and were not included in the fake-data uncertainty for comparison.
Only the uncorrelated uncertainties (cross section, statistical, MC statistical, alternative generator) were used for the minimization procedure and are shown on the plots to account for the unfolding process and the change in the event generator.
The combination of these uncertainties covers the differences between the unfolded fake data points (black) and the alternative-generator theory results (orange).
The relevant $A_{C}$ matrices have been applied to the generator predictions.

\begin{figure*}[htb!]
\centering 
\includegraphics[width=0.32\linewidth]{\figures 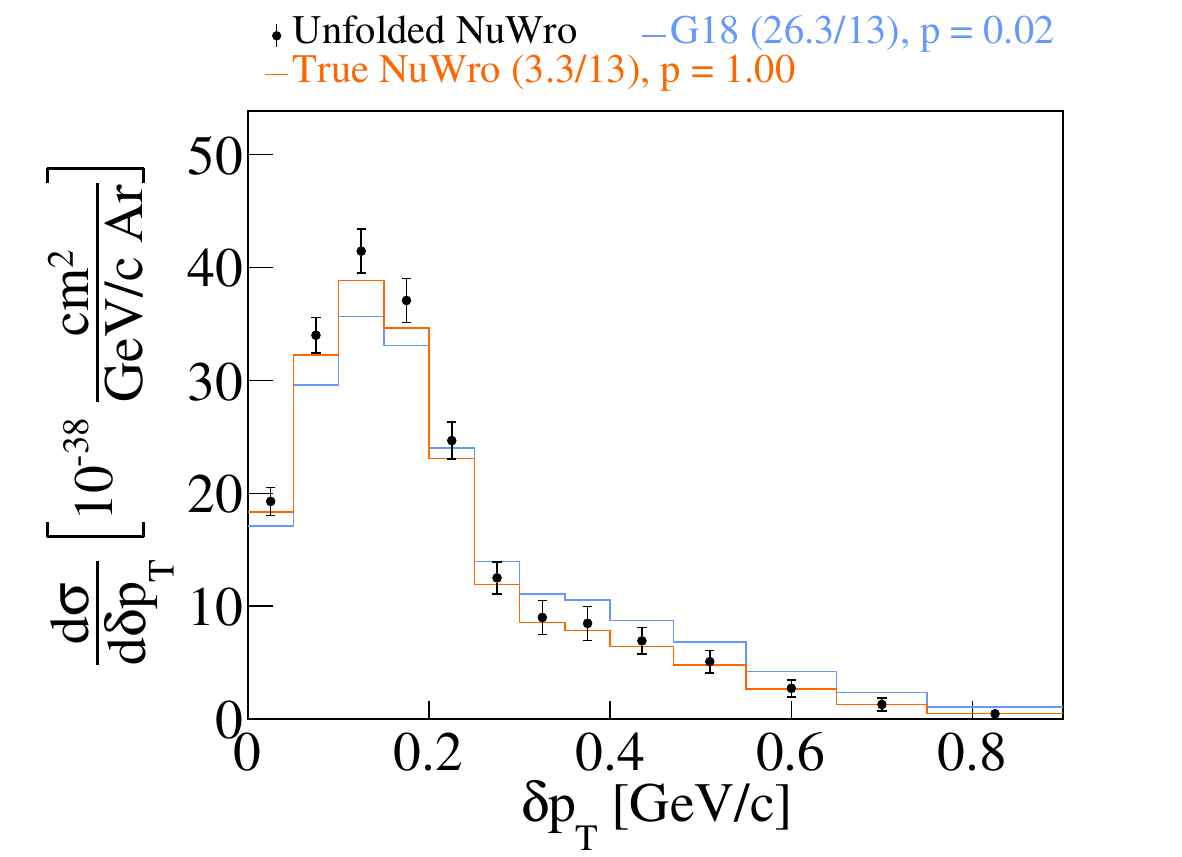}
\includegraphics[width=0.32\linewidth]{\figures 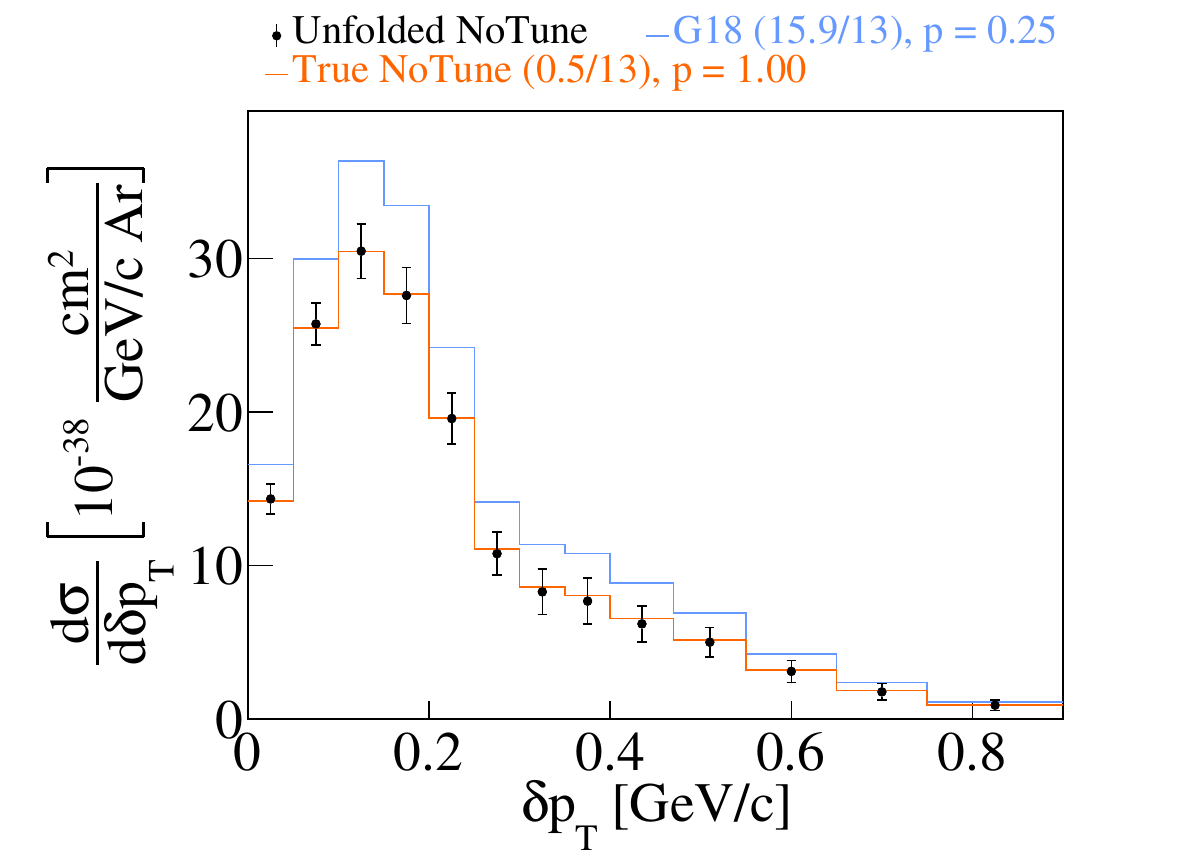}
\includegraphics[width=0.32\linewidth]{\figures 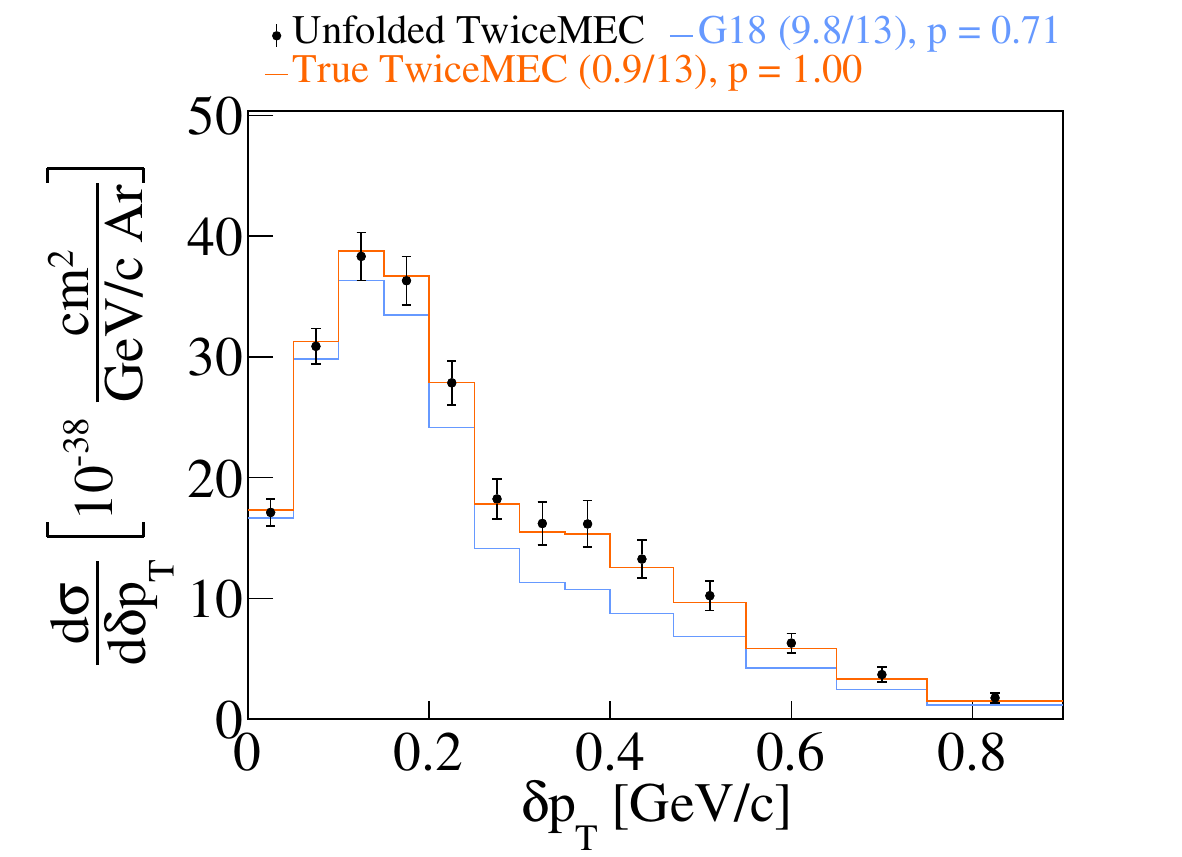}
\caption{
Fake data studies for $\delta p_{T}$ using (left) NuWro, (center) GENIE without the MicroBooNE tune (NoTune), and (right) twice the weights for MEC events (TwiceMEC) as fake data samples.
}
\label{DeltaPTFakeData}
\end{figure*}

\begin{figure*}[htb!]
\centering 
\includegraphics[width=0.32\linewidth]{\figures 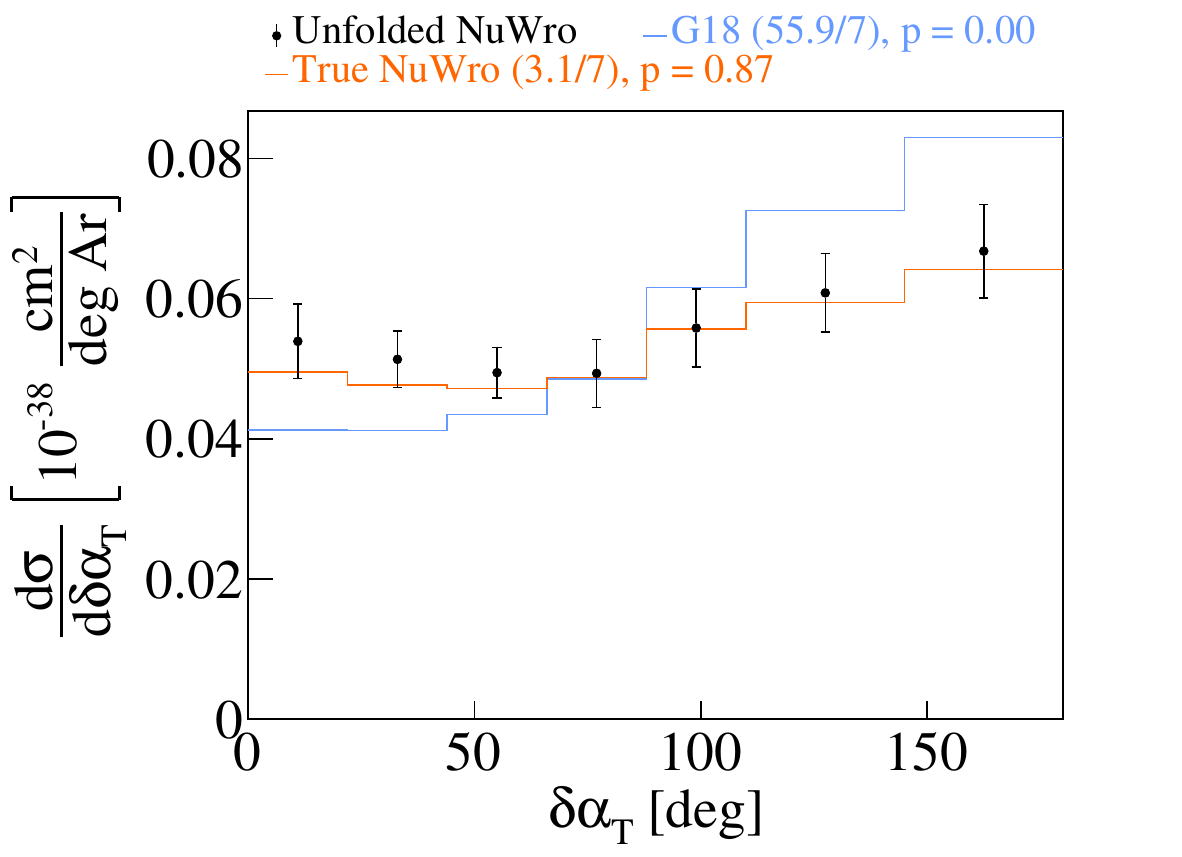}
\includegraphics[width=0.32\linewidth]{\figures 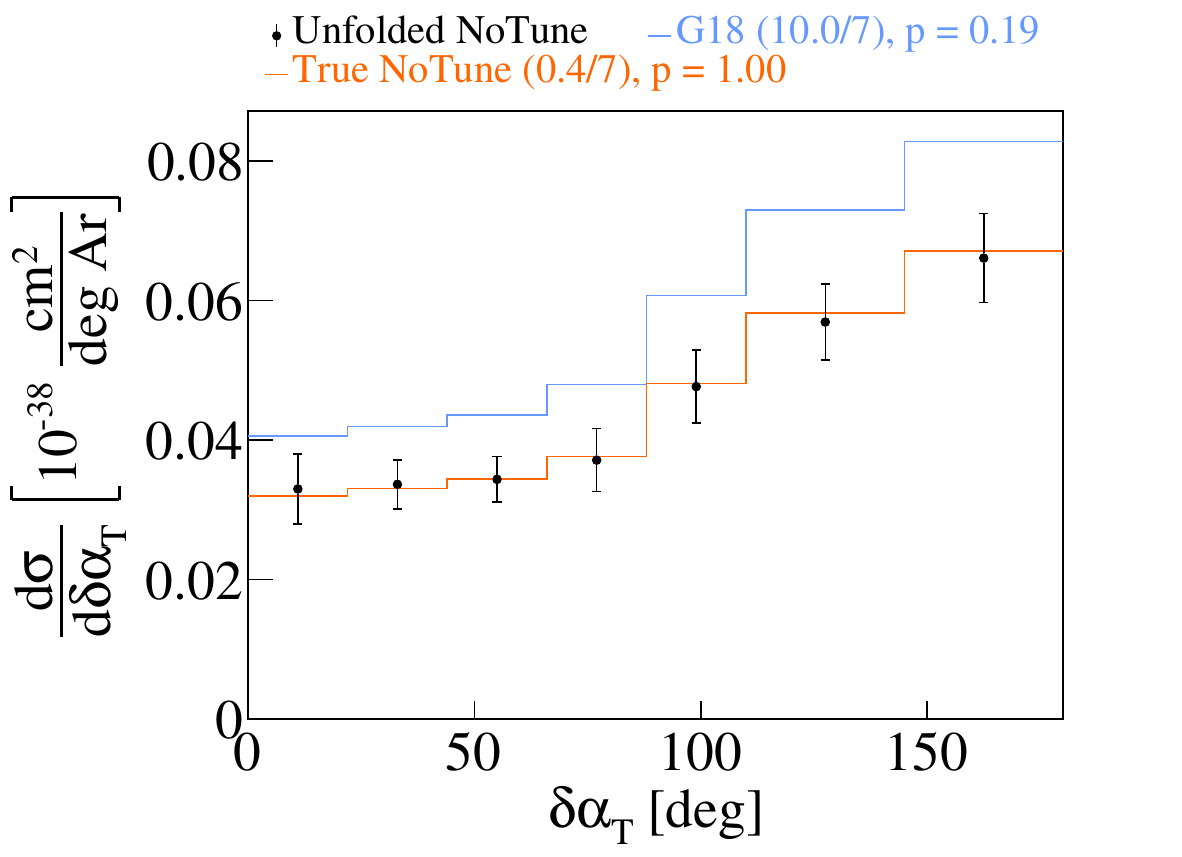}
\includegraphics[width=0.32\linewidth]{\figures 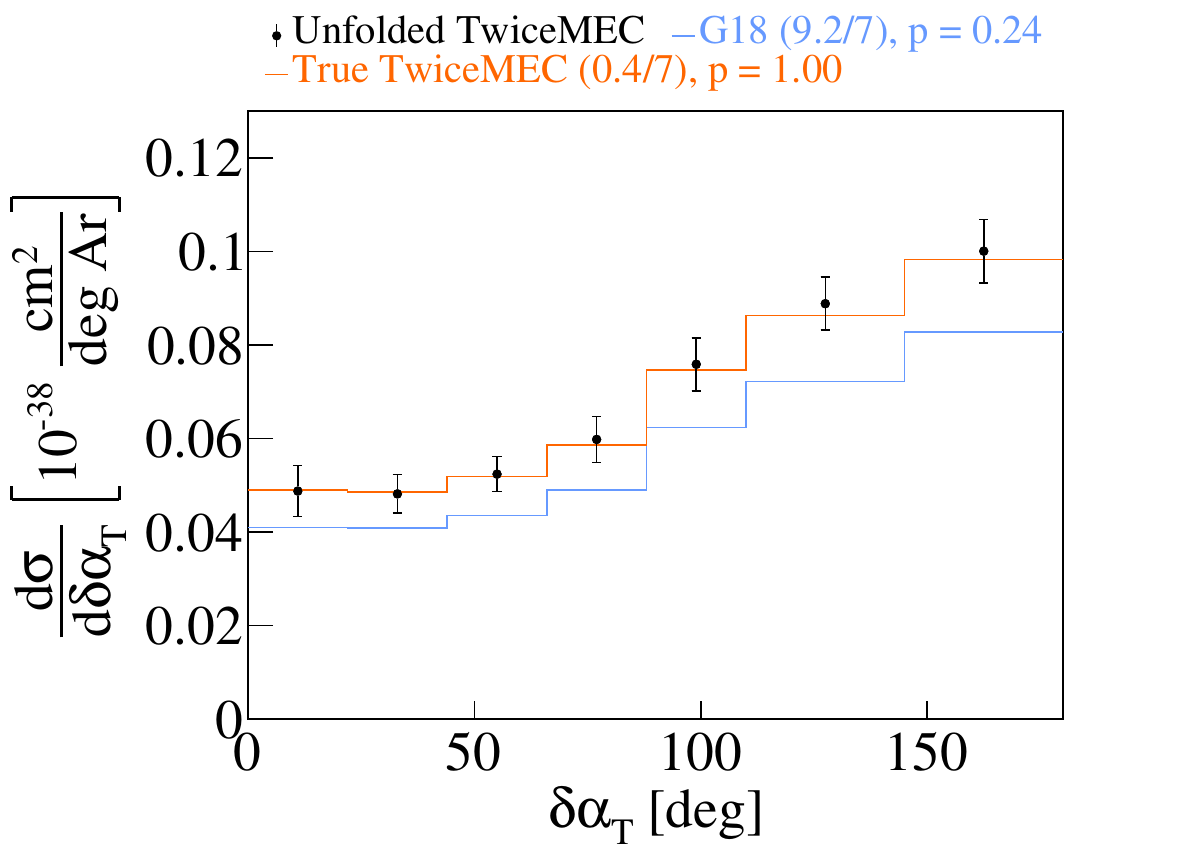}
\caption{
Fake data studies for $\delta\alpha_{T}$ using (left) NuWro, (center) GENIE without the MicroBooNE tune (NoTune), and (right) twice the weights for MEC events (TwiceMEC) as fake data samples.
}
\label{DeltaAlphaTFakeData}
\end{figure*}

\begin{figure*}[htb!]
\centering 
\includegraphics[width=0.32\linewidth]{\figures 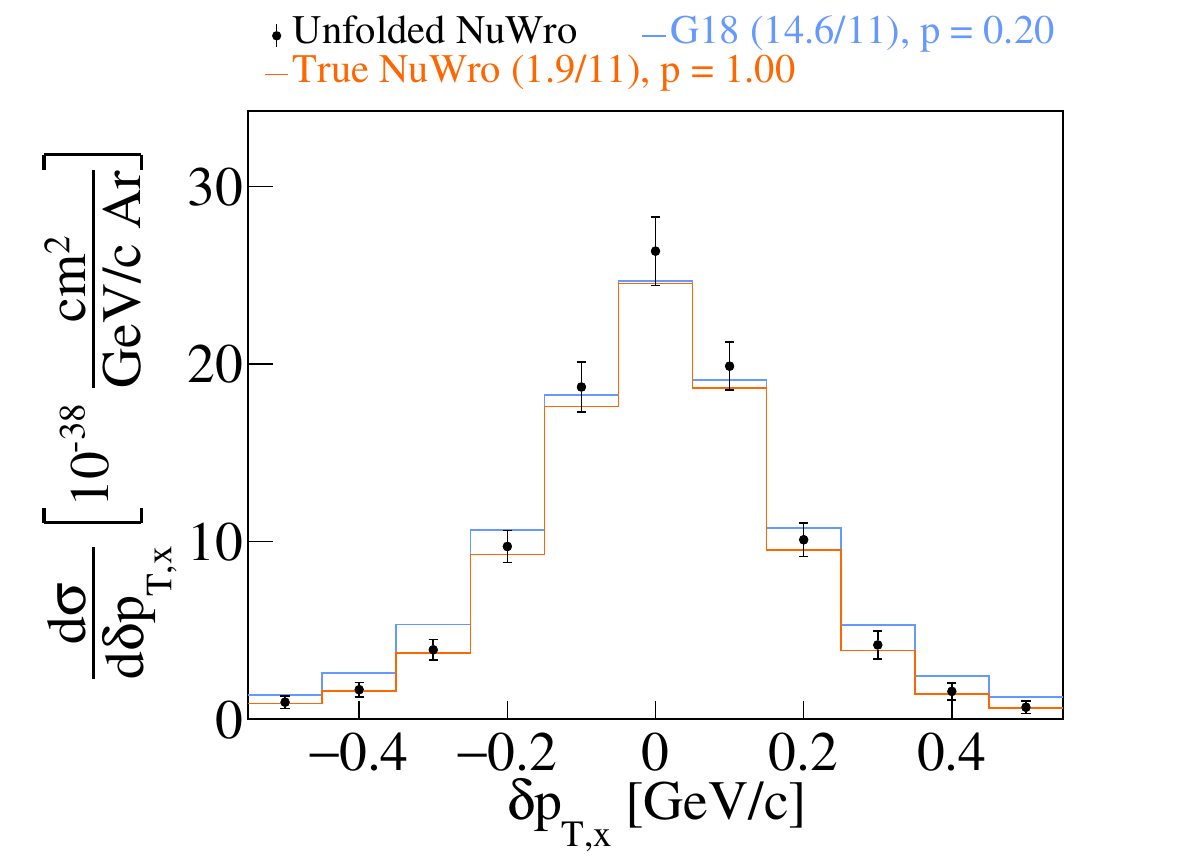}
\includegraphics[width=0.32\linewidth]{\figures 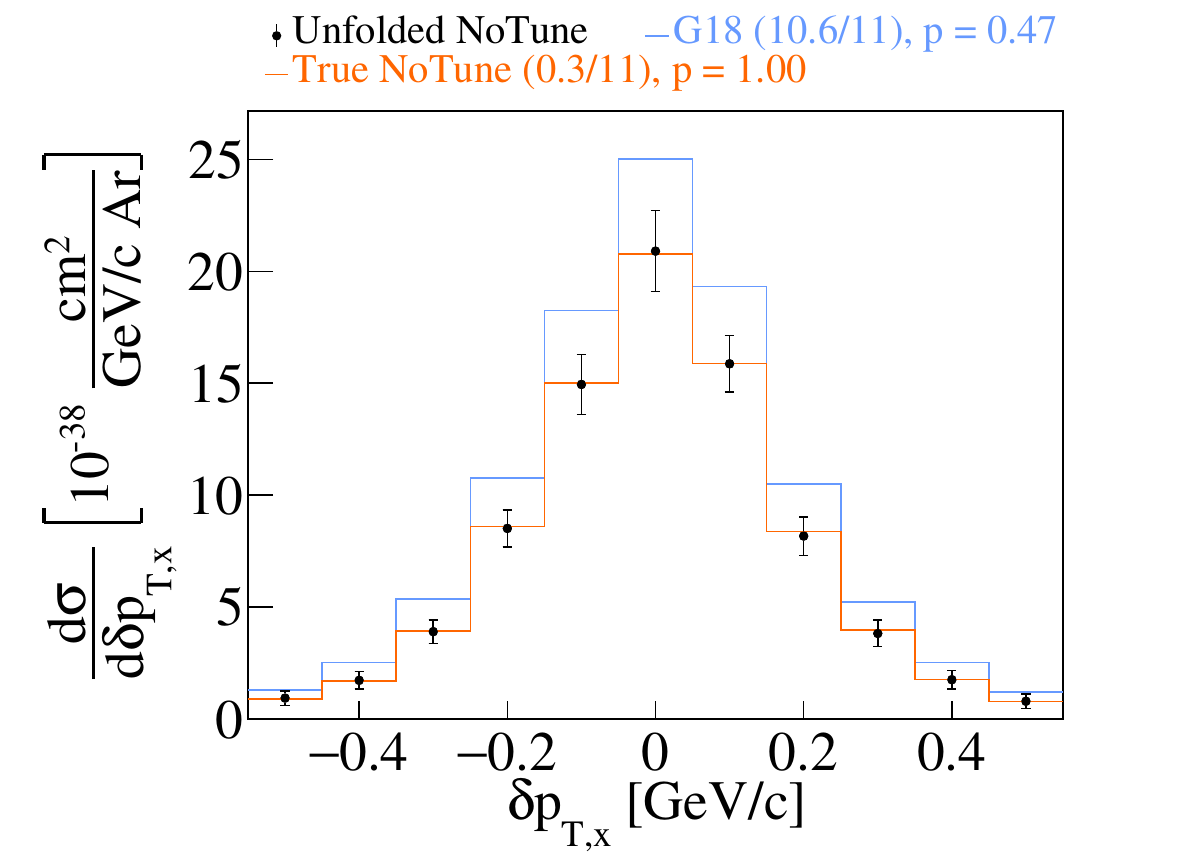}
\includegraphics[width=0.32\linewidth]{\figures 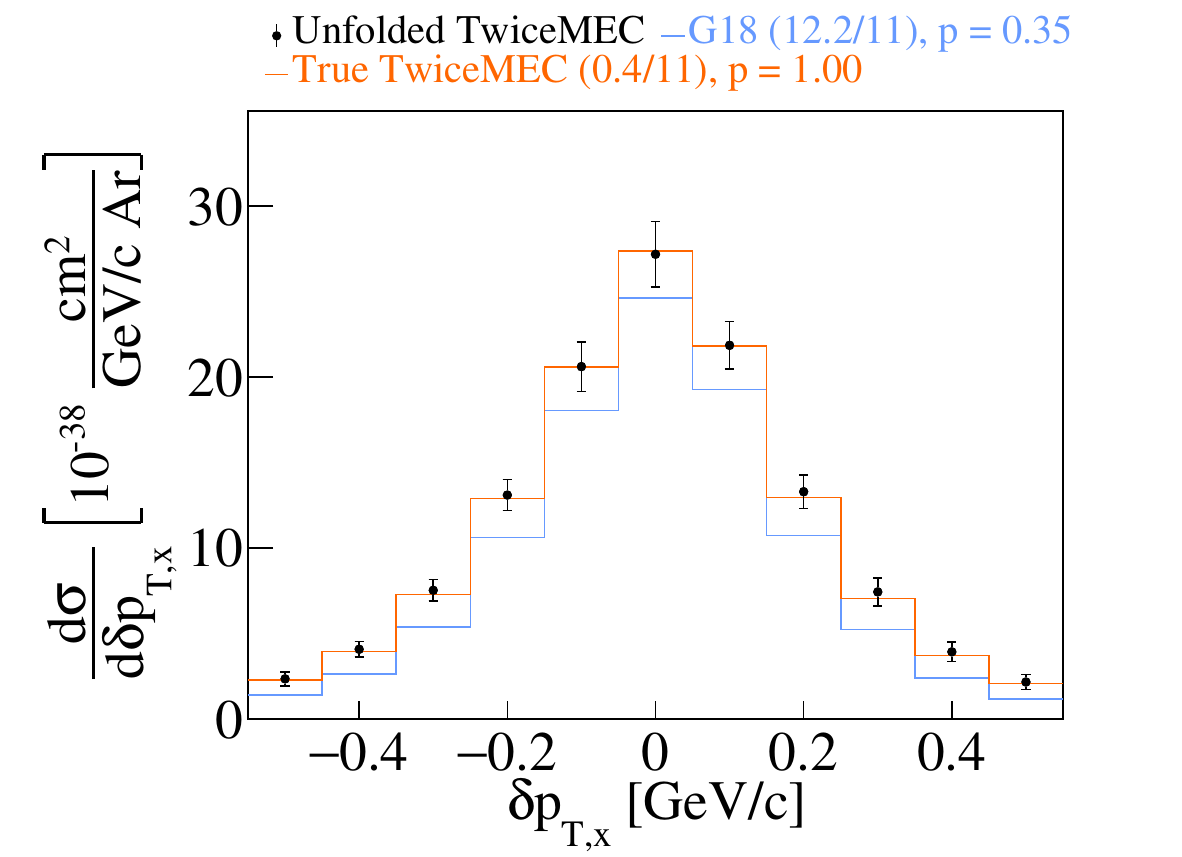}
\caption{
Fake data studies for $\delta p_{T,x}$ using (left) NuWro, (center) GENIE without the MicroBooNE tune (NoTune), and (right) twice the weights for MEC events (TwiceMEC) as fake data samples.
}
\label{DeltaPtxFakeData}
\end{figure*}

\begin{figure*}[htb!]
\centering 
\includegraphics[width=0.32\linewidth]{\figures 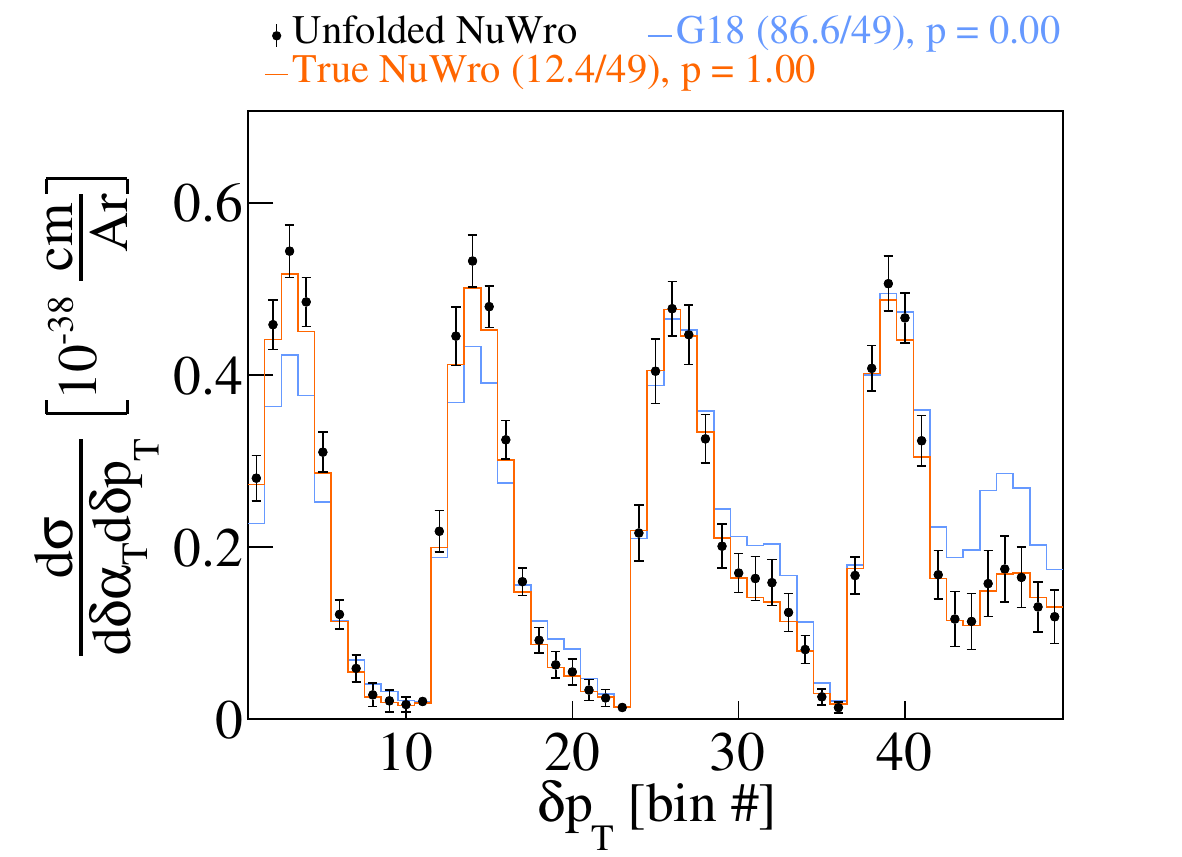}
\includegraphics[width=0.32\linewidth]{\figures 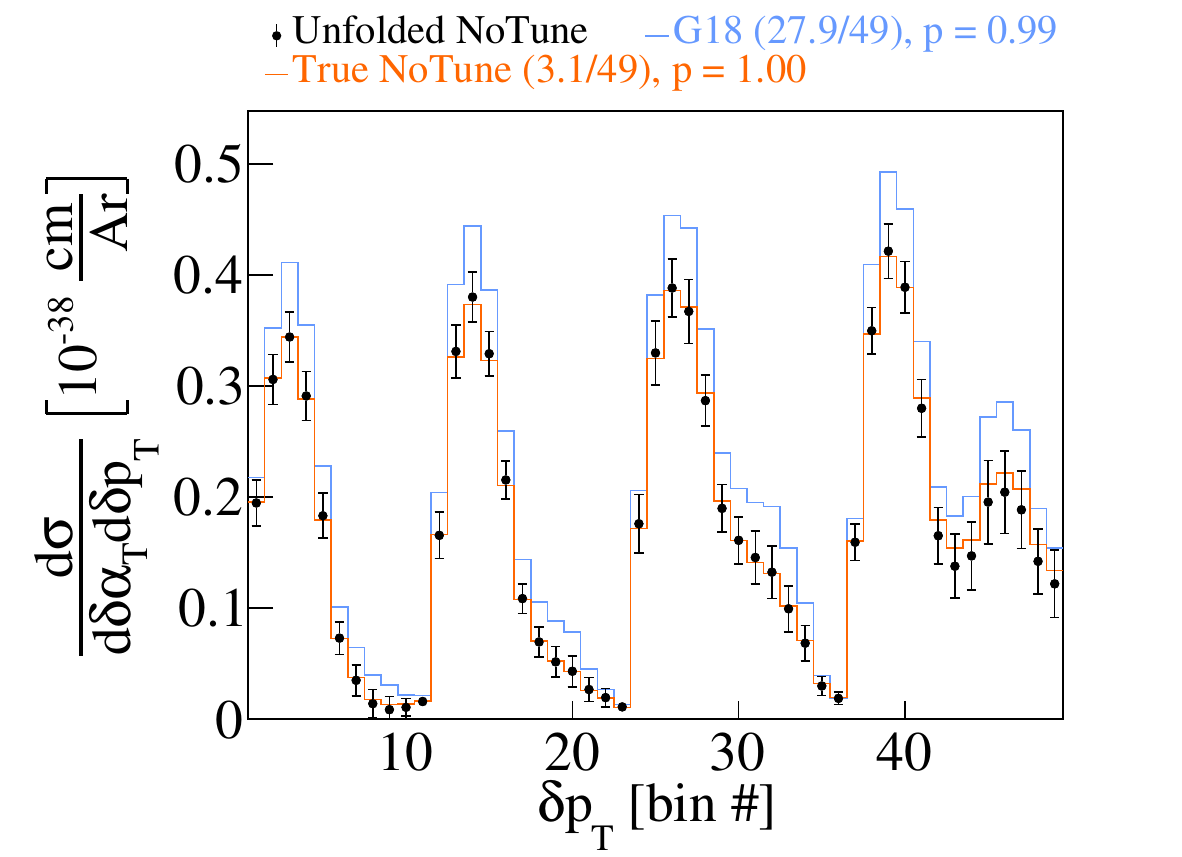}
\includegraphics[width=0.32\linewidth]{\figures 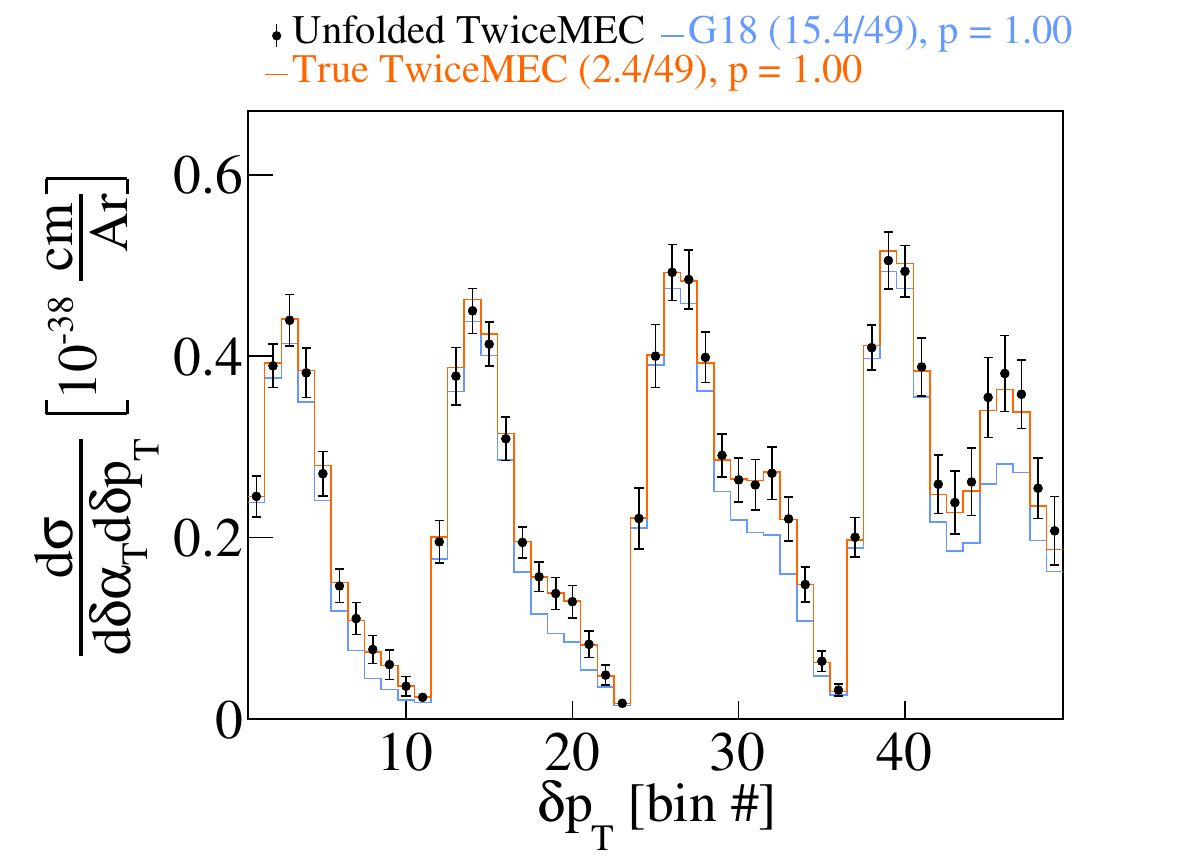}
\caption{
Fake data studies for $\delta p_{T}$ in $\delta\alpha_{T}$ bins using (left) NuWro, (center) GENIE without the MicroBooNE tune (NoTune), and (right) twice the weights for MEC events (TwiceMEC) as fake data samples.
}
\label{SerialDeltaPT_DeltaAlphaTFakeData}
\end{figure*}

\begin{figure*}[htb!]
\centering 
\includegraphics[width=0.32\linewidth]{\figures 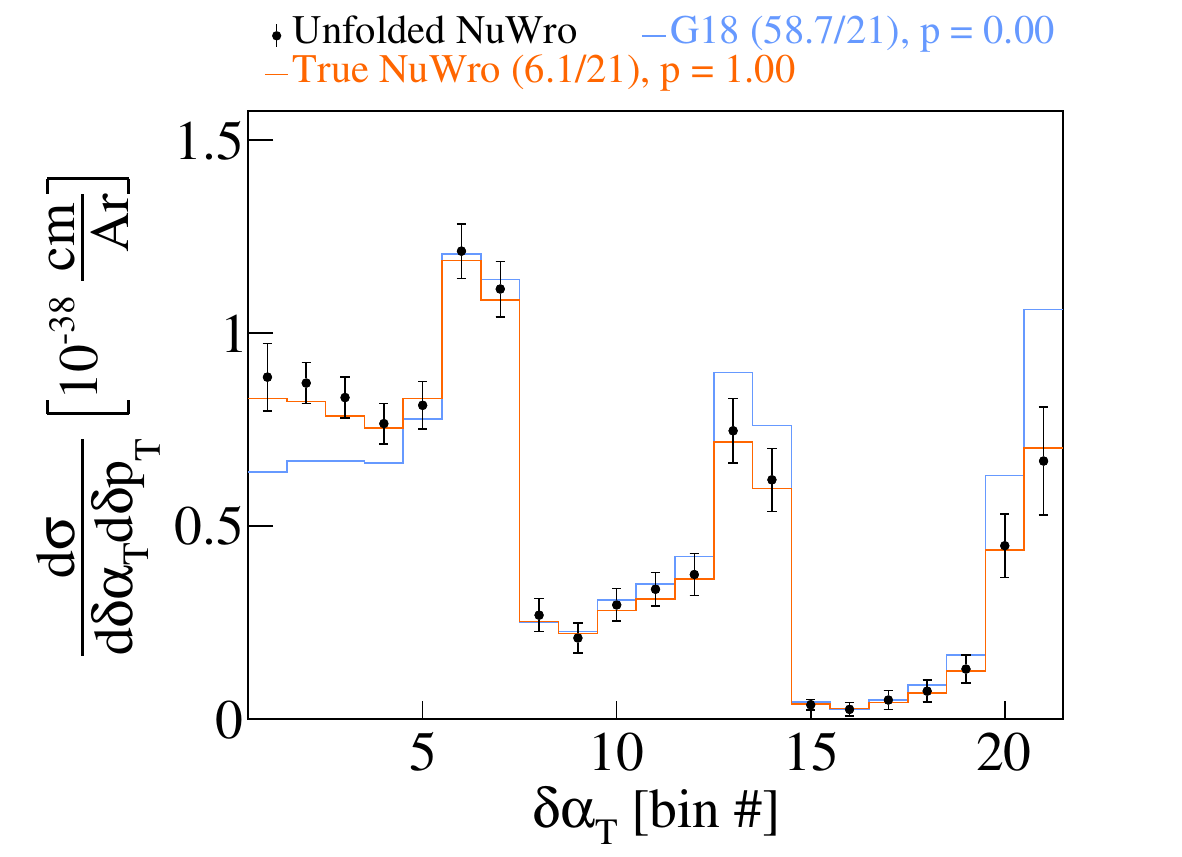}
\includegraphics[width=0.32\linewidth]{\figures 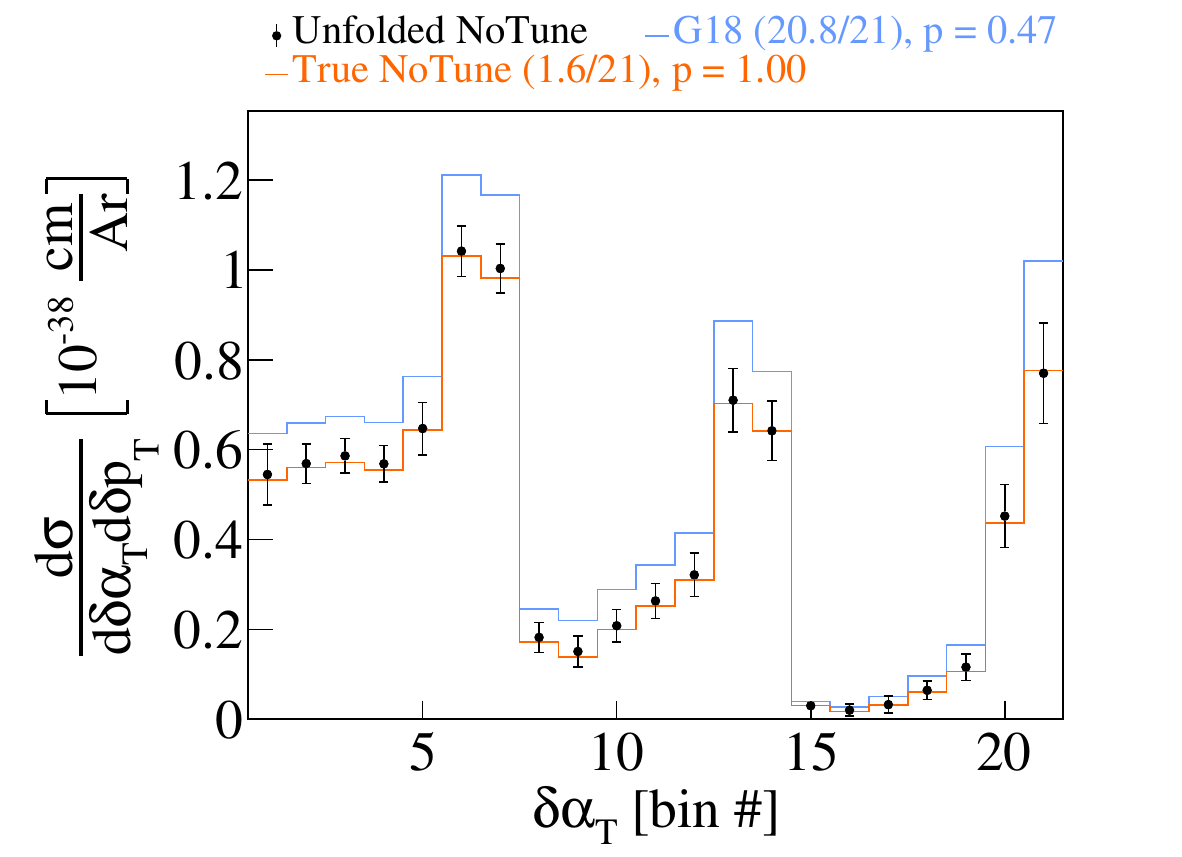}
\includegraphics[width=0.32\linewidth]{\figures 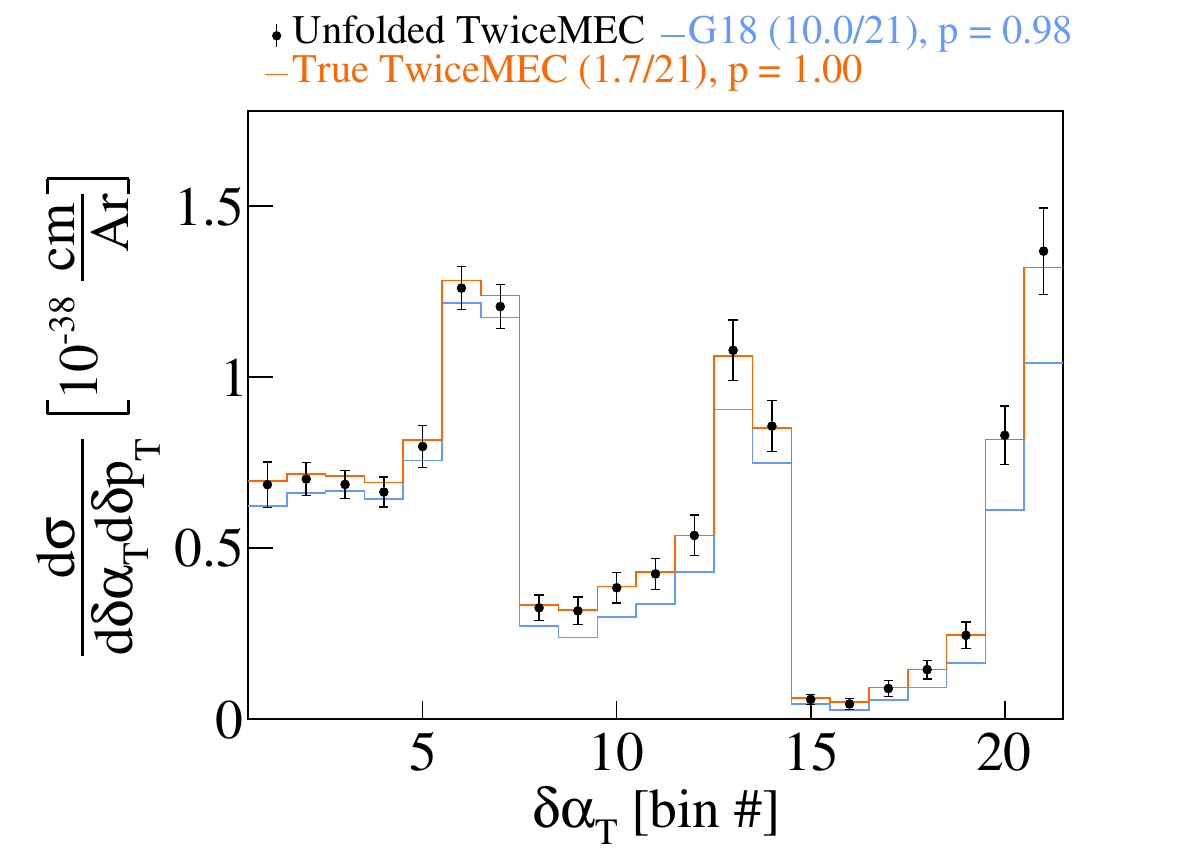}
\caption{
Fake data studies for $\delta\alpha_{T}$ in $\delta p_{T}$ bins using (left) NuWro, (center) GENIE without the MicroBooNE tune (NoTune), and (right) twice the weights for MEC events (TwiceMEC) as fake data samples.
}
\label{SerialDeltaAlphaT_DeltaPTFakeData}
\end{figure*}

\begin{figure*}[htb!]
\centering 
\includegraphics[width=0.32\linewidth]{\figures 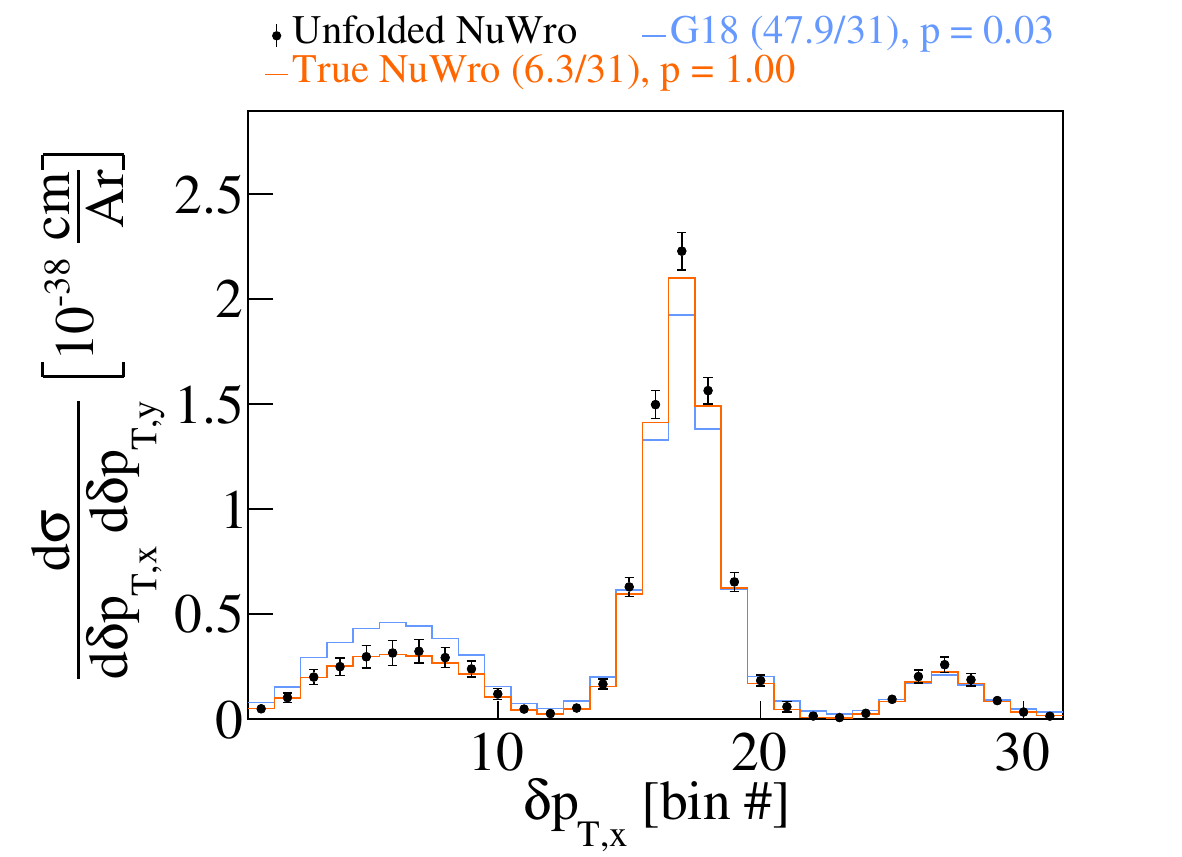}
\includegraphics[width=0.32\linewidth]{\figures 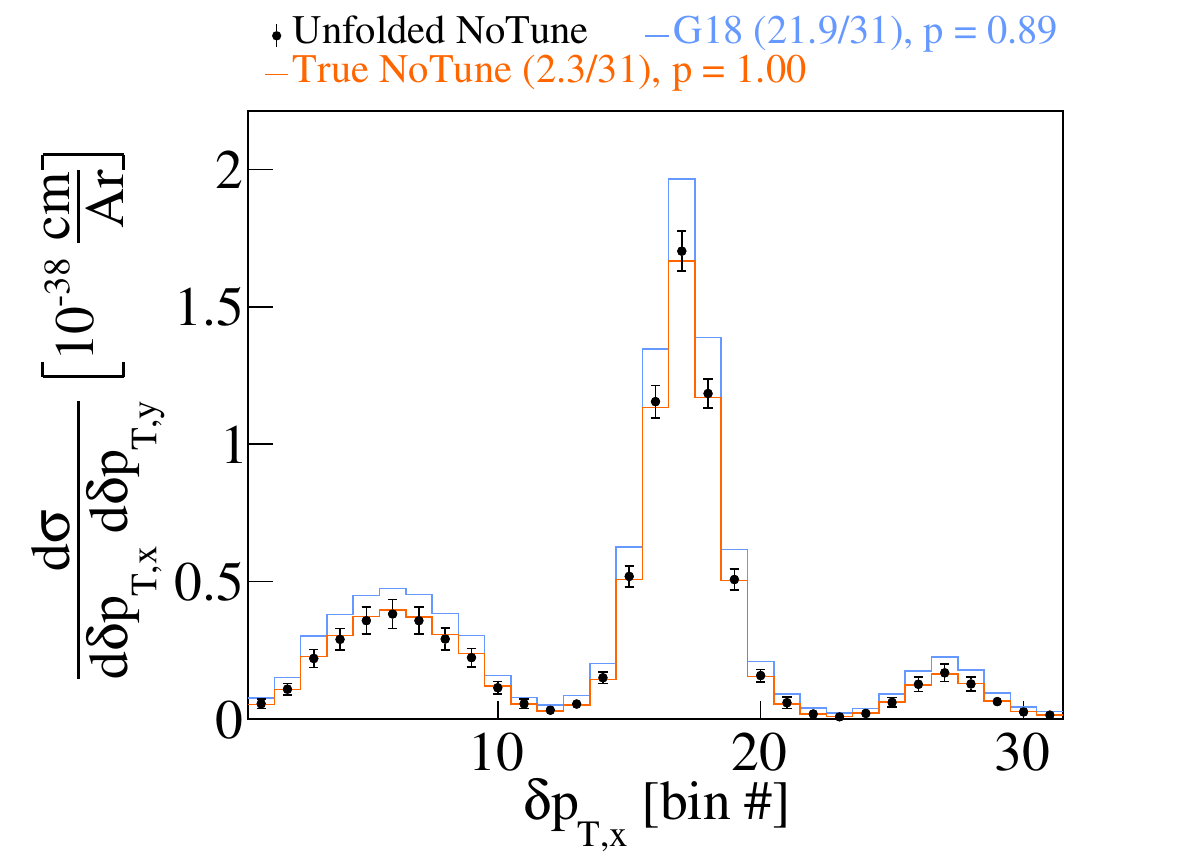}
\includegraphics[width=0.32\linewidth]{\figures 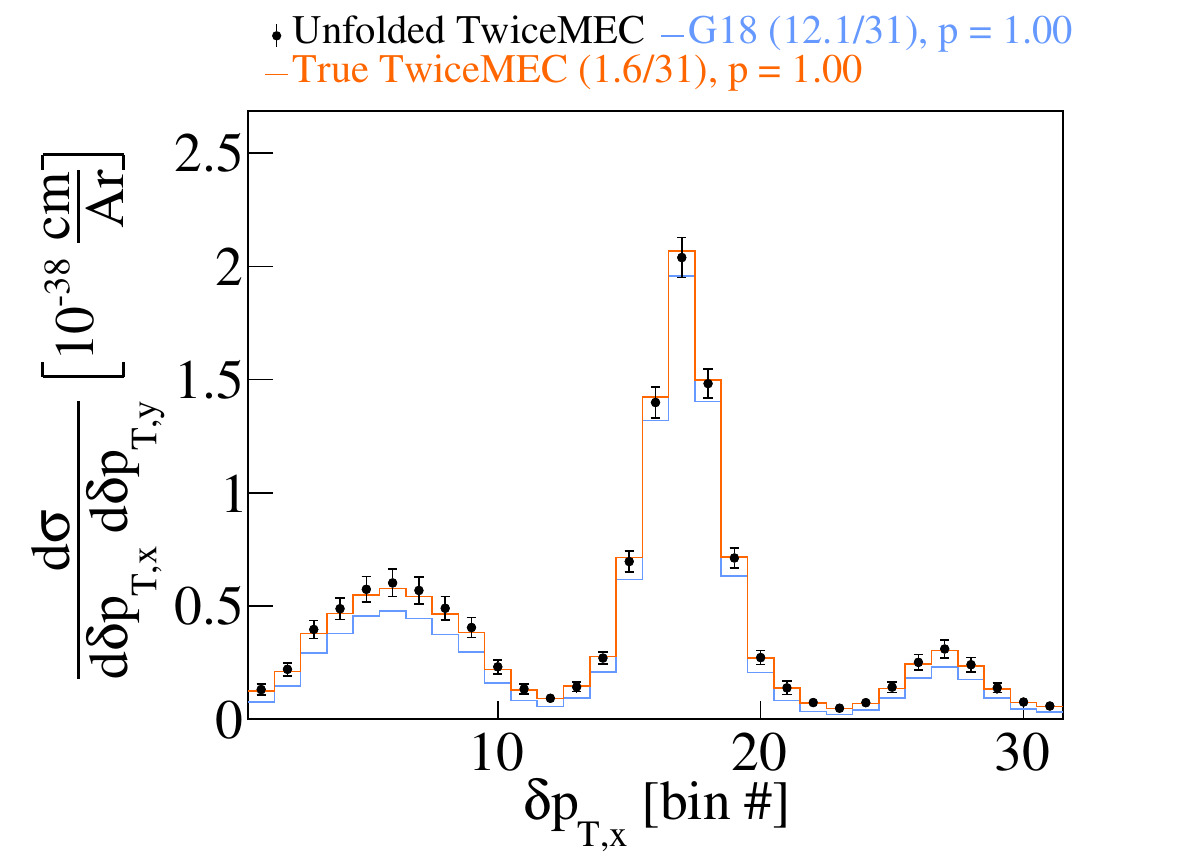}
\caption{
Fake data studies for $\delta p_{T,x}$ in $\delta p_{T,y}$ bins using (left) NuWro, (center) GENIE without the MicroBooNE tune (NoTune), and (right) twice the weights for MEC events (TwiceMEC) as fake data samples.
}
\label{SerialDeltaPtx_DeltaPTtyFakeData}
\end{figure*}


\section{Interaction Vertex Z Cross Section And Efficiency}\label{VertexZ}

\begin{figure*}[htb!]
\centering 
\includegraphics[width=0.49\linewidth]{\figures 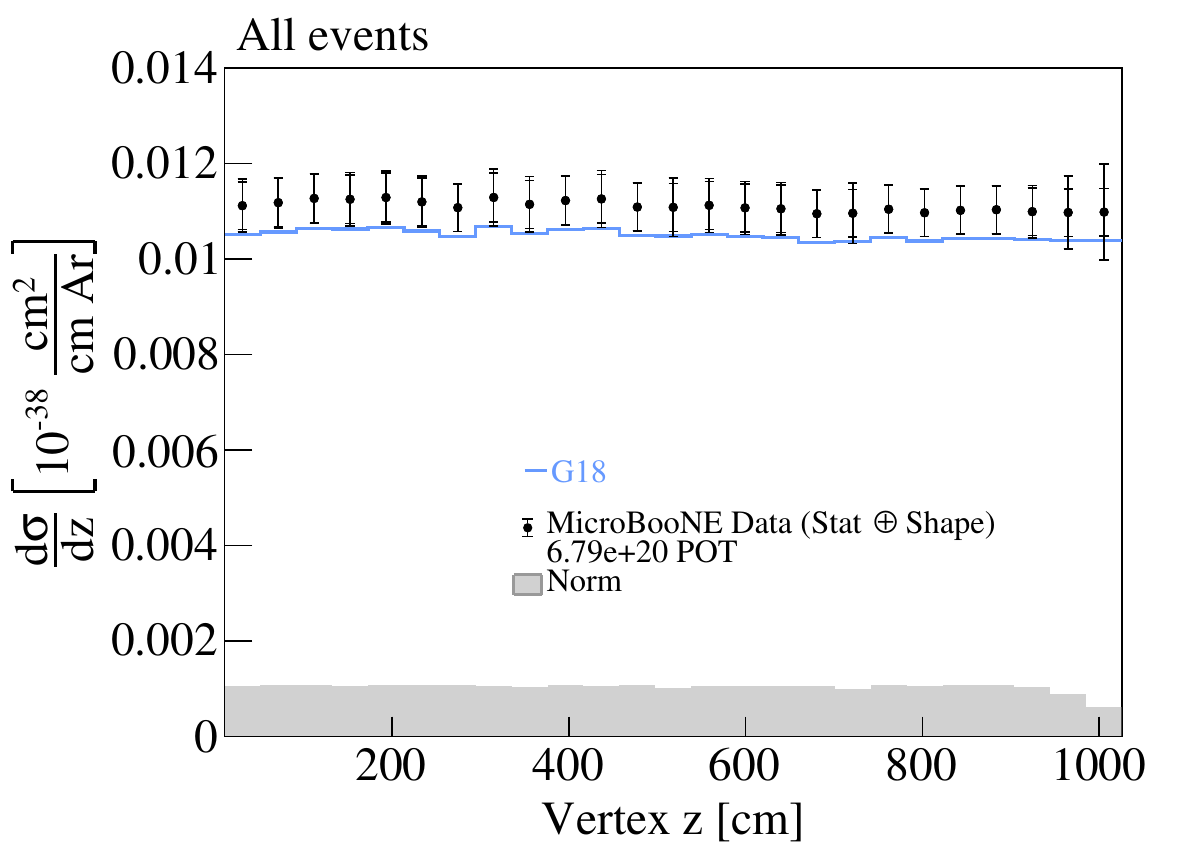}
\includegraphics[width=0.49\linewidth]{\figures 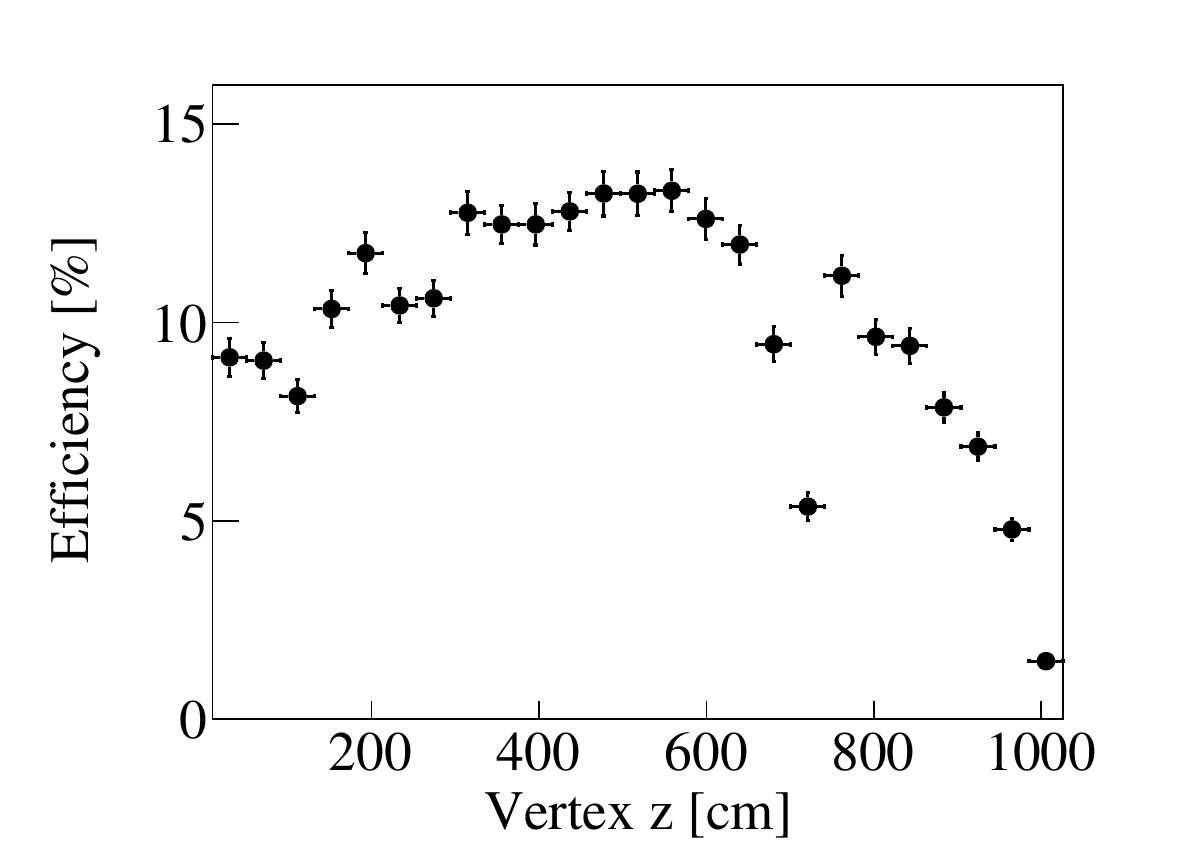}
\caption{
(Left) extracted cross section as a function of the vertex z distribution.
(Right) vertex z efficiency function.
}
\label{NominalVertexZ}
\end{figure*}

In order to ensure that the selection is unbiased against isolating specific parts of the detector and that the unfolding procedure yields reliable results, we verified that the z-vertex cross sections for the candidate signal events follow a uniform distribution.
Figure~\ref{NominalVertexZ} (left) shows the extracted cross section as a function of the z-vertex distribution of the signal events, illustrating the uniform behavior. 
Figure~\ref{NominalVertexZ} (right) shows the corresponding efficiency.
The deficit at z = 700\,cm is due to dead wires in our detector and its effect has been incorporated in our simulation.

\section{Cross Section Interaction Breakdown}\label{IntBreakDown}

\begin{figure}[htb!]
\centering 
\includegraphics[width=0.24\linewidth]{\figures 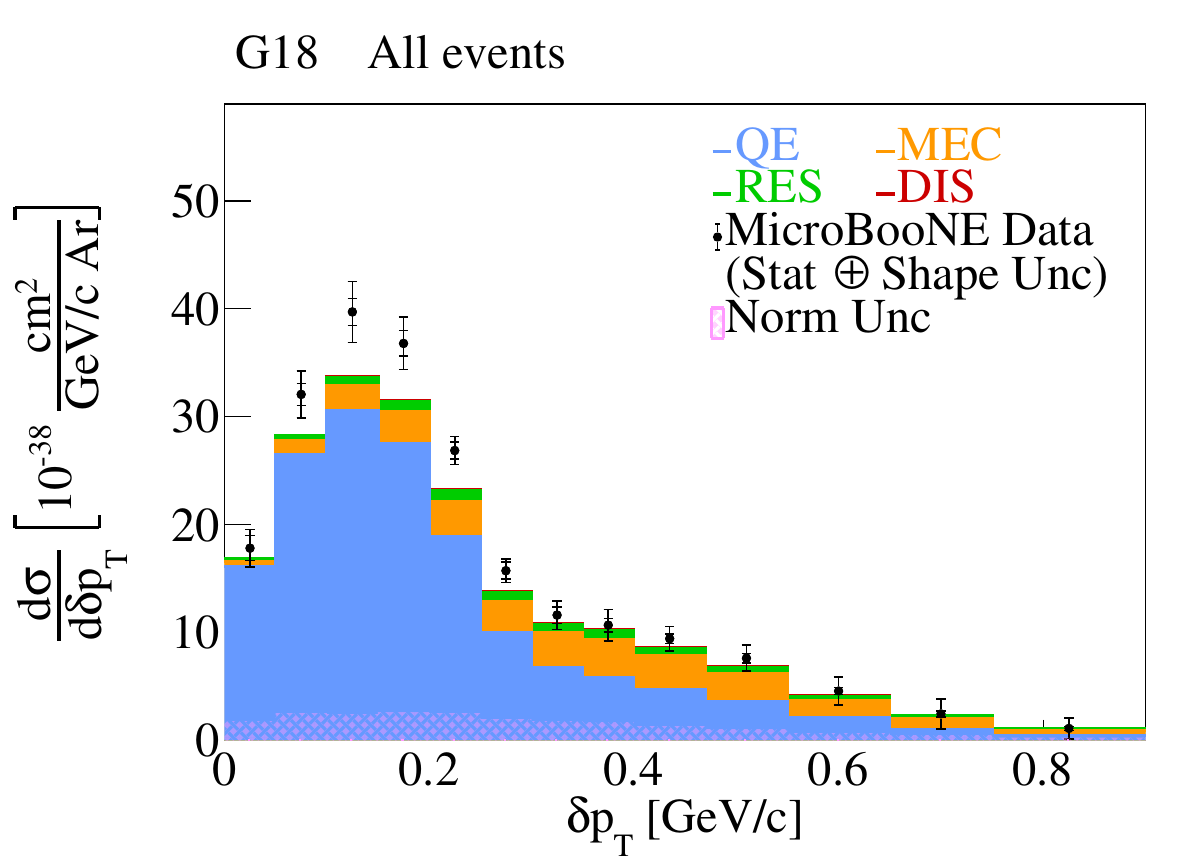}
\includegraphics[width=0.24\linewidth]{\figures 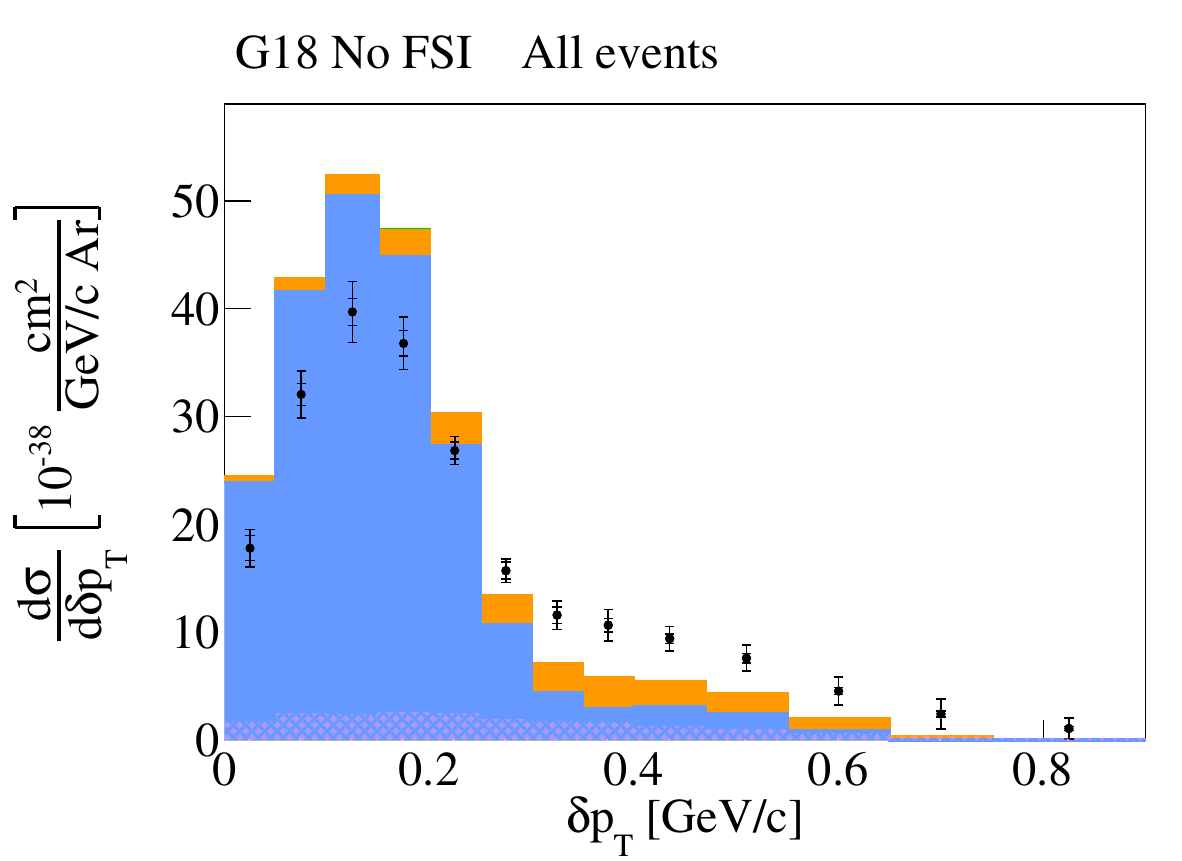}
\includegraphics[width=0.24\linewidth]{\figures 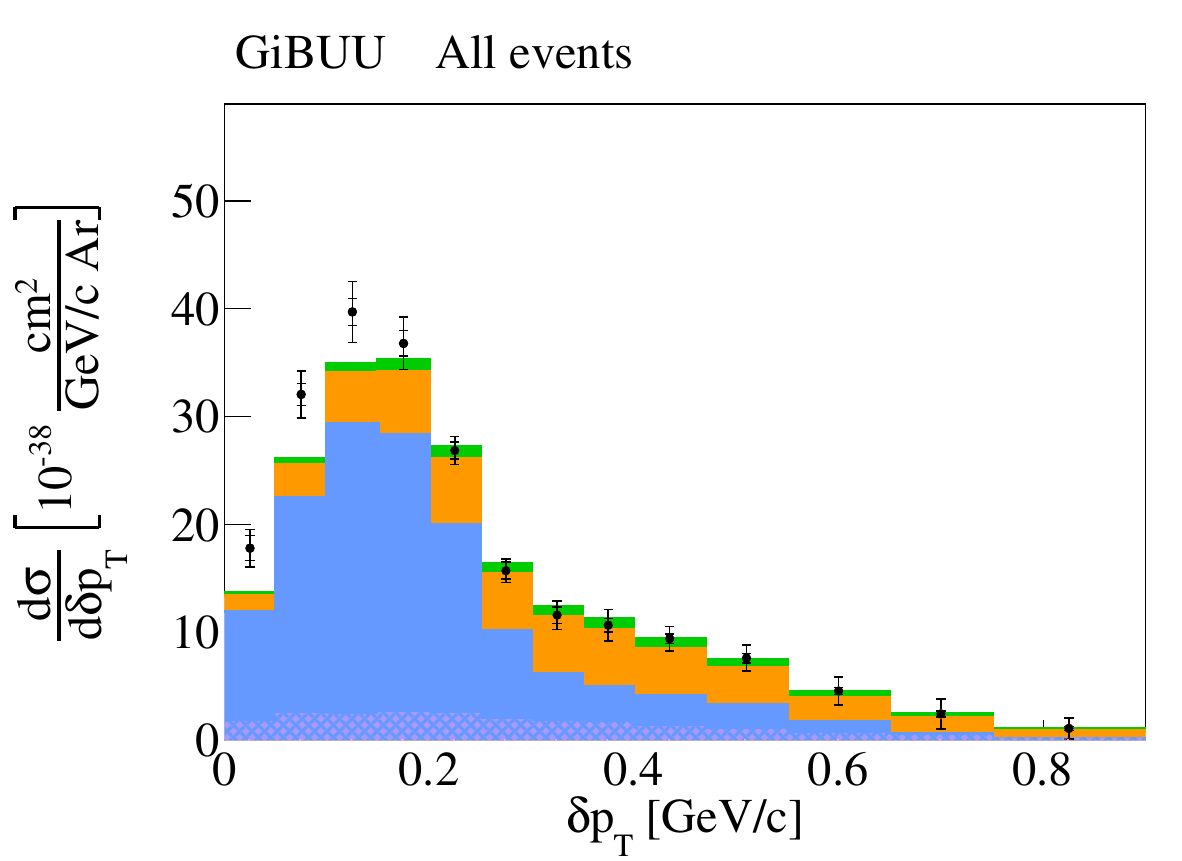}
\includegraphics[width=0.24\linewidth]{\figures 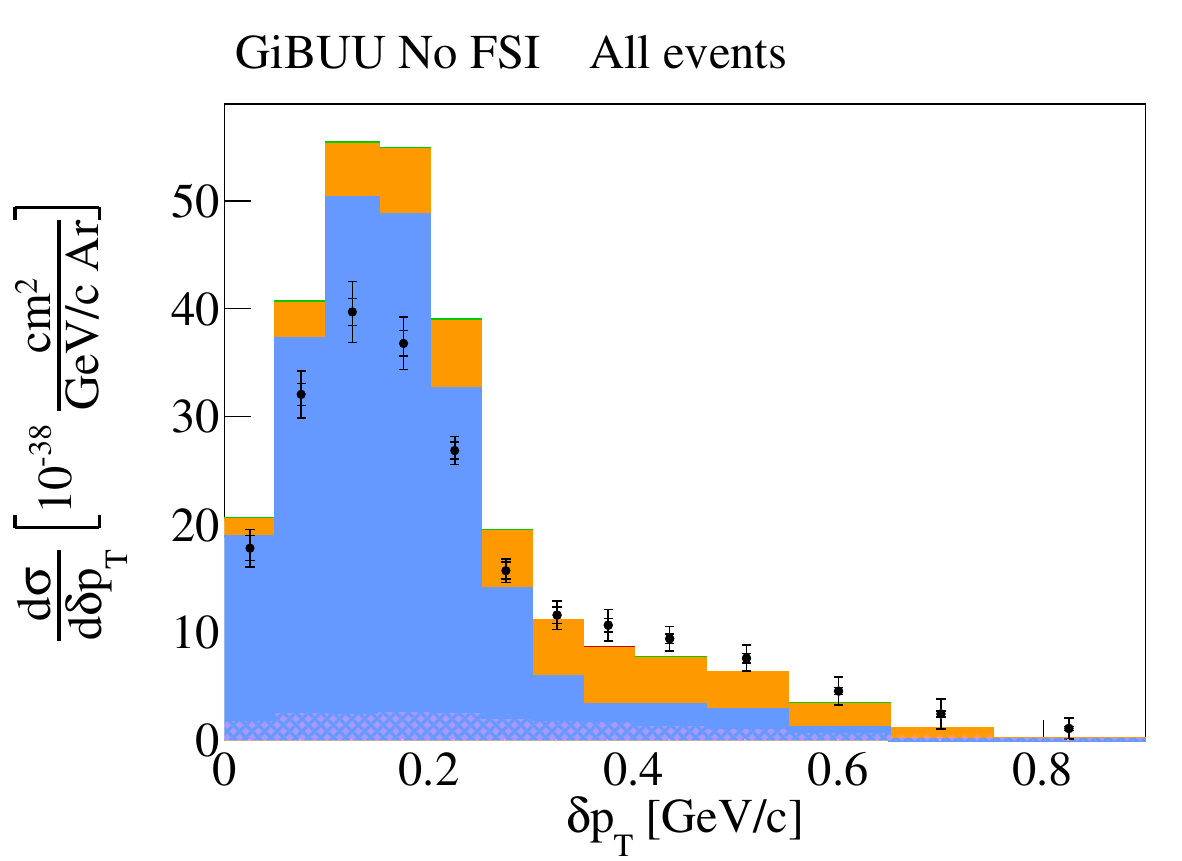}

\includegraphics[width=0.24\linewidth]{\figures 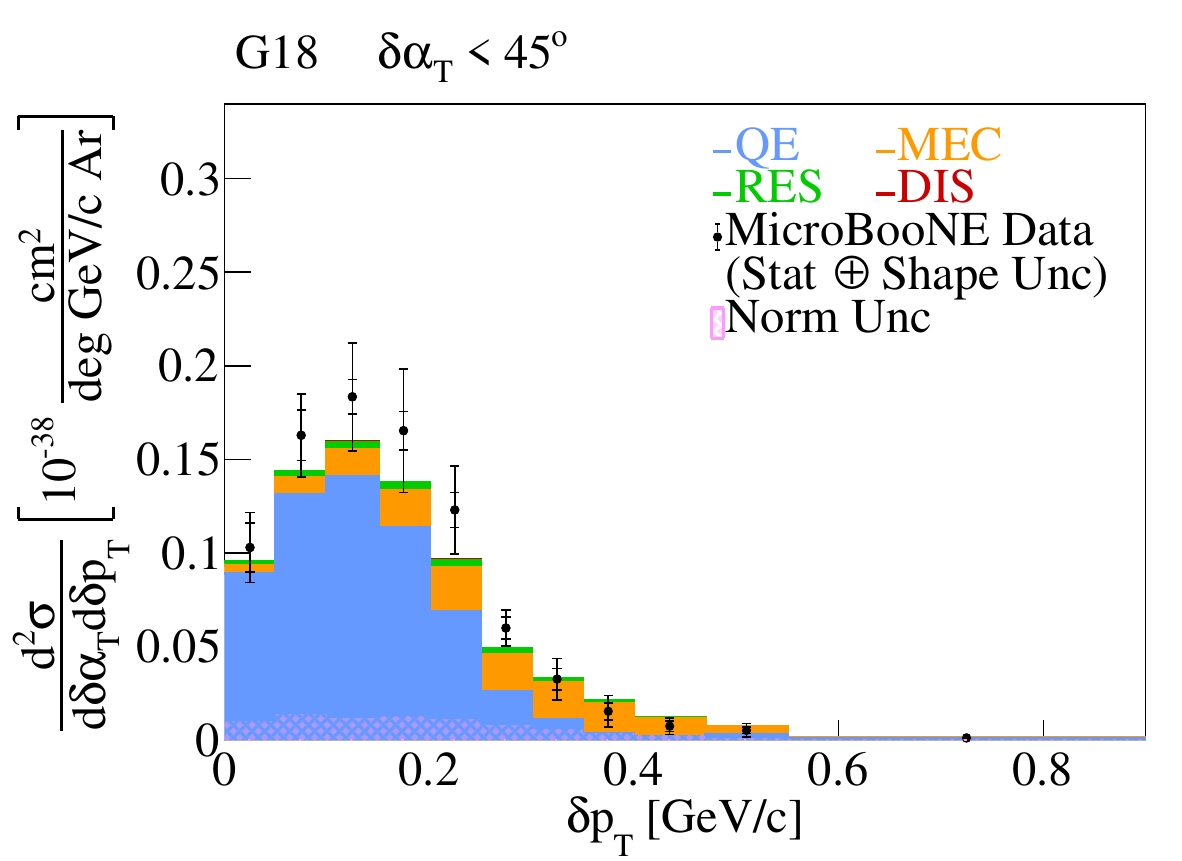}
\includegraphics[width=0.24\linewidth]{\figures 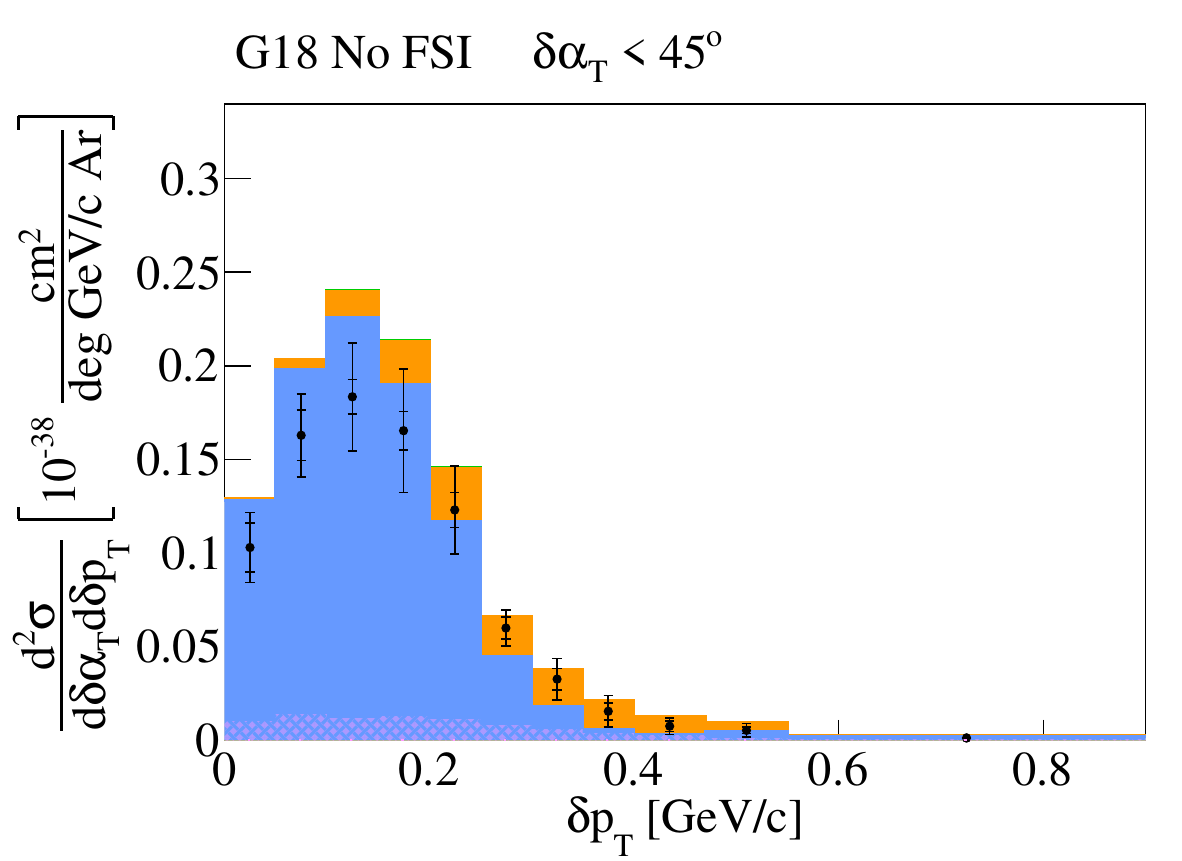}
\includegraphics[width=0.24\linewidth]{\figures 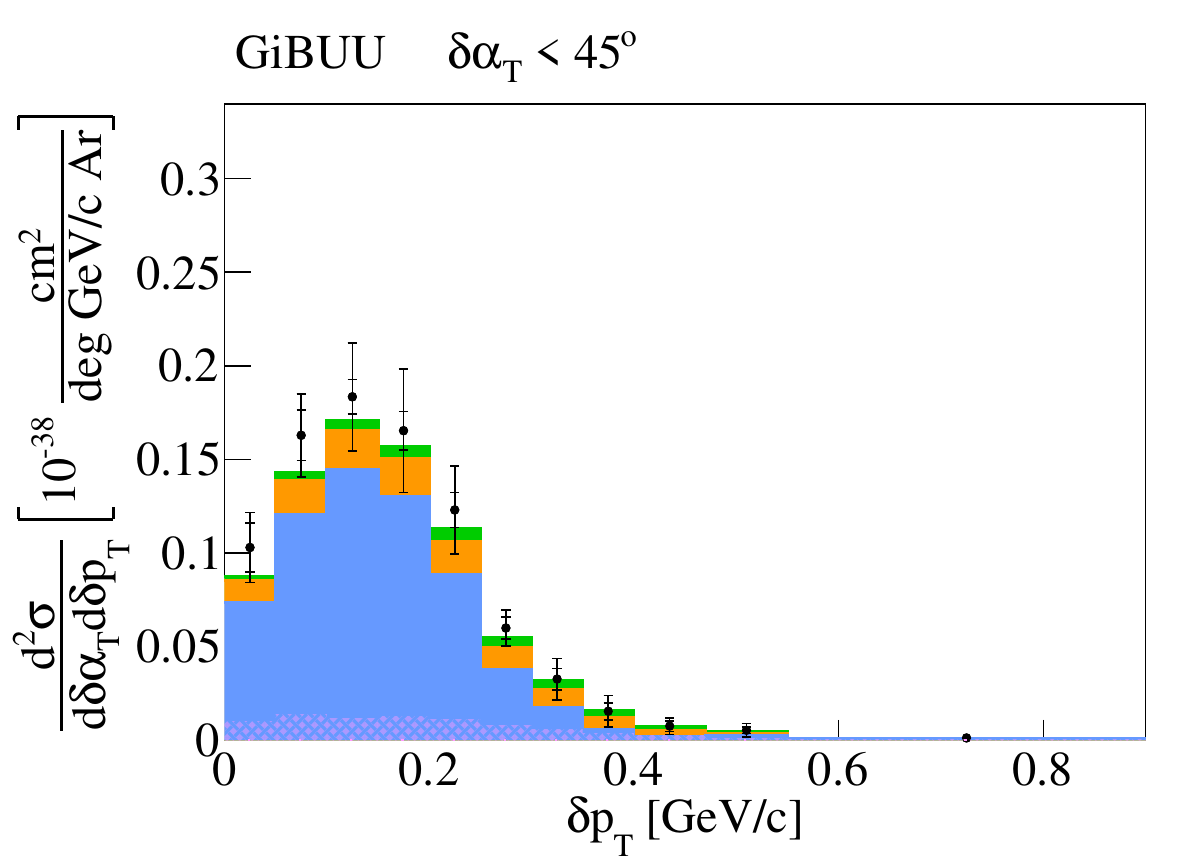}
\includegraphics[width=0.24\linewidth]{\figures 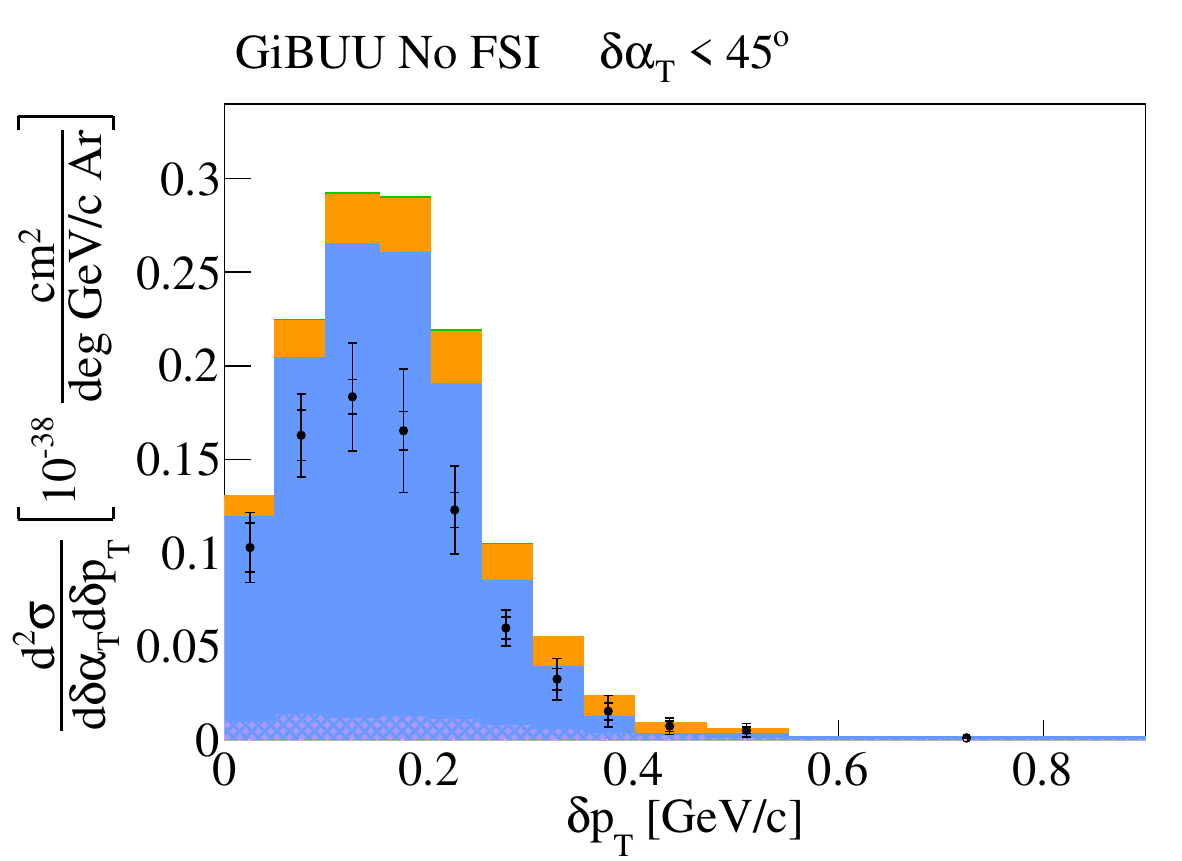}

\includegraphics[width=0.24\linewidth]{\figures 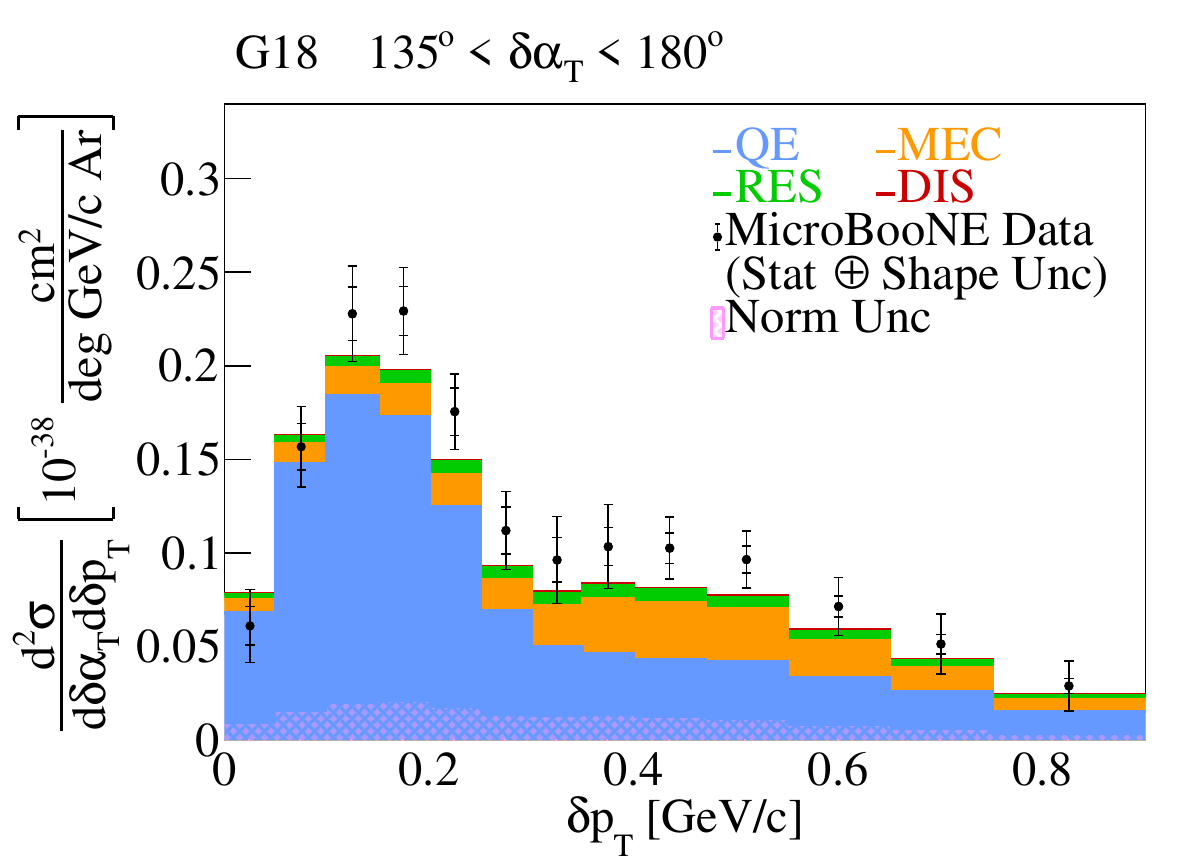}
\includegraphics[width=0.24\linewidth]{\figures 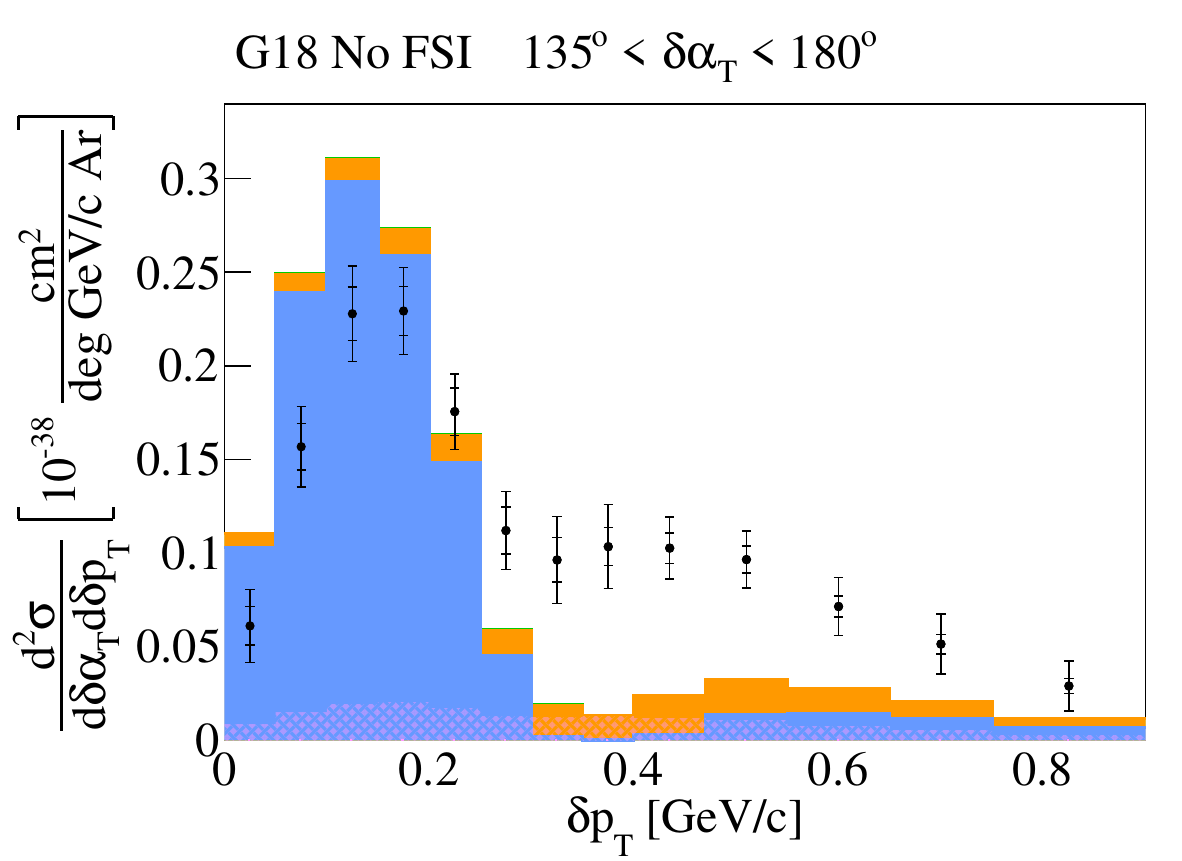}
\includegraphics[width=0.24\linewidth]{\figures 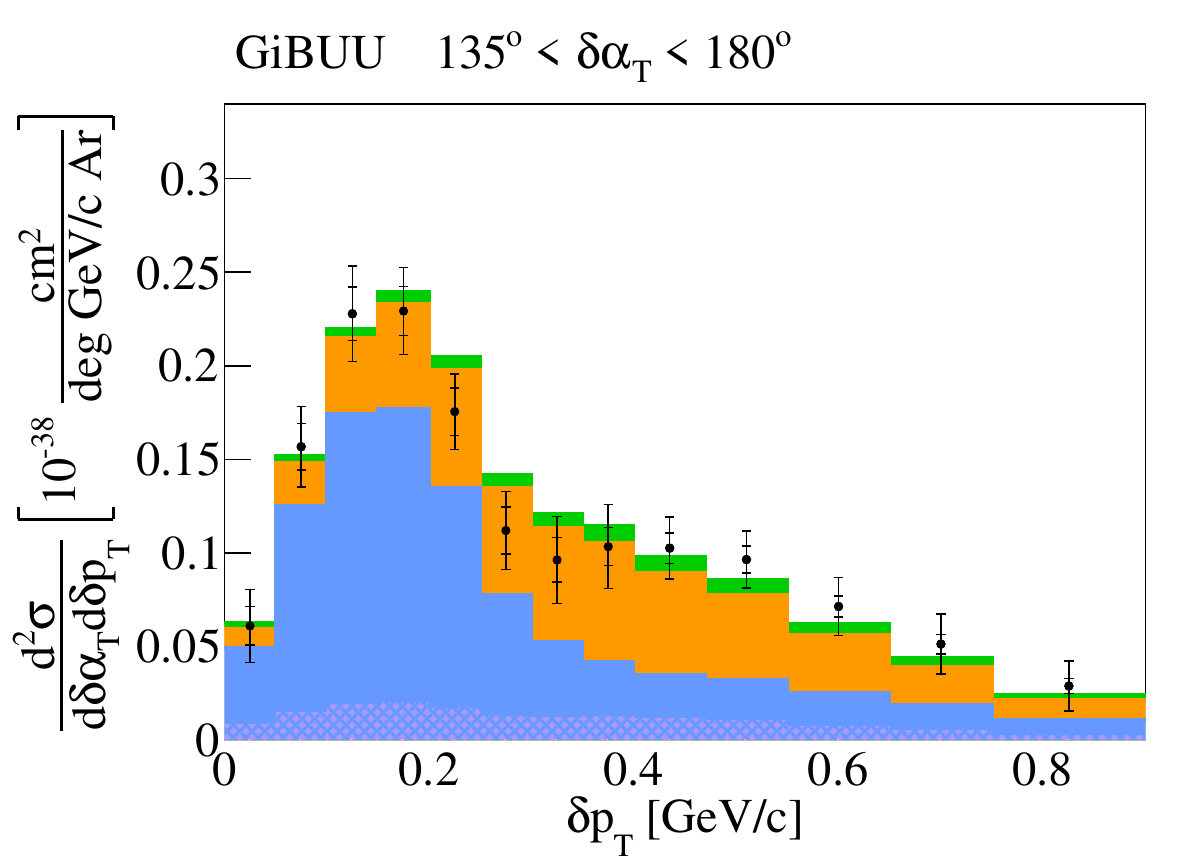}
\includegraphics[width=0.24\linewidth]{\figures 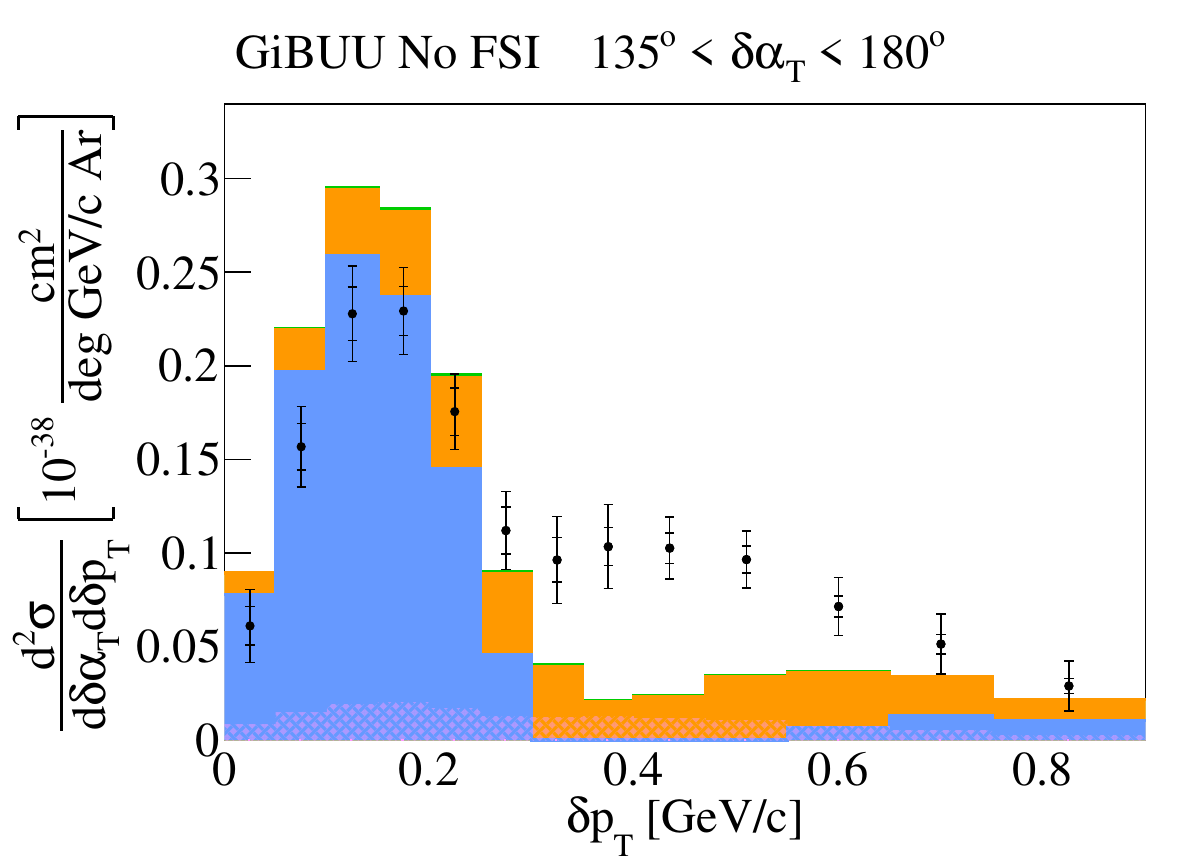}

\caption{Cross section interaction breakdown for (top) all the selected events, (middle) events with $\delta\alpha_{T} < 45^{o}$, and (bottom) events with $135^{o} < \delta\alpha_{T} < 180^{o}$. 
The breakdown is shown for (first column) the G18 configuration with FSI effects, (second column) the G18 configuration without FSI effects, (third column) GiB with FSI effects, and (forth column) GiB without FSI effects.
}

\label{DeltaPTBreakdown}
\end{figure}

\begin{figure}[htb!]
\centering 
\includegraphics[width=0.24\linewidth]{\figures 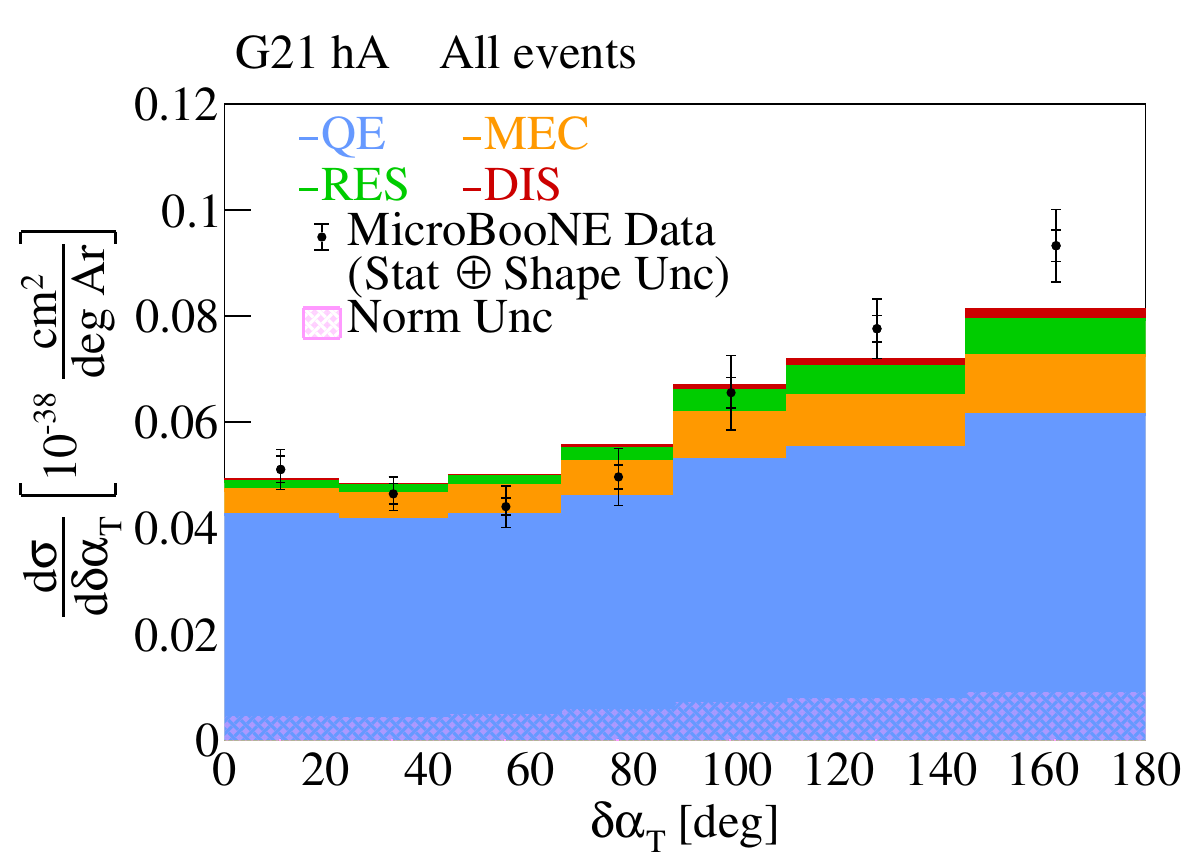}
\includegraphics[width=0.24\linewidth]{\figures 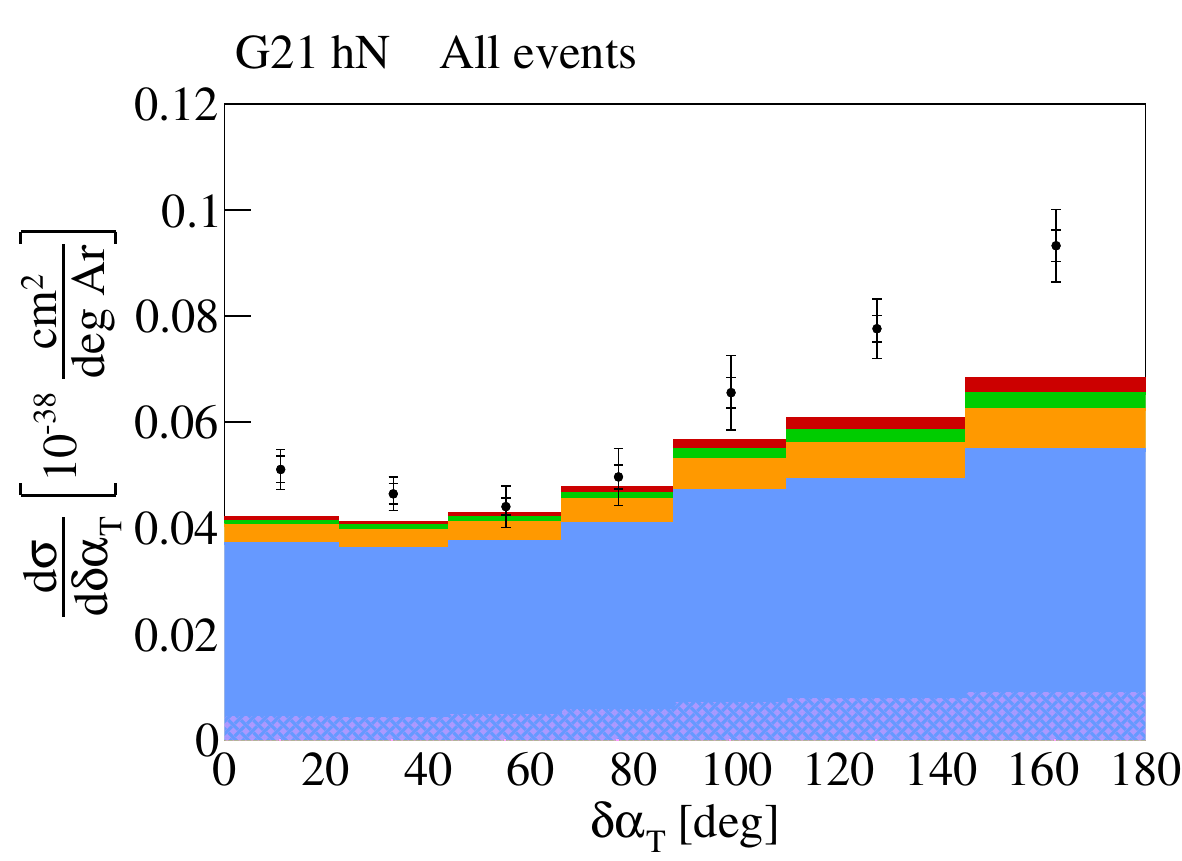}
\includegraphics[width=0.24\linewidth]{\figures 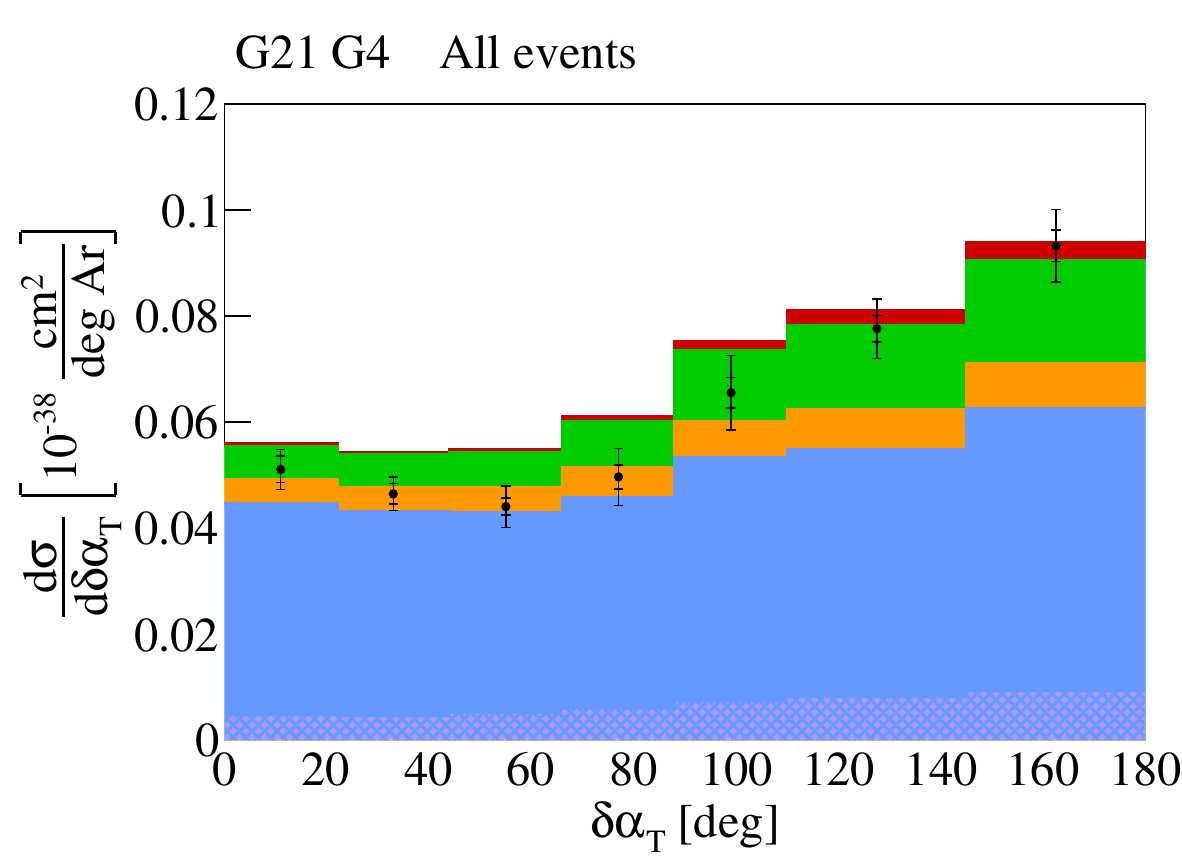}
\includegraphics[width=0.24\linewidth]{\figures 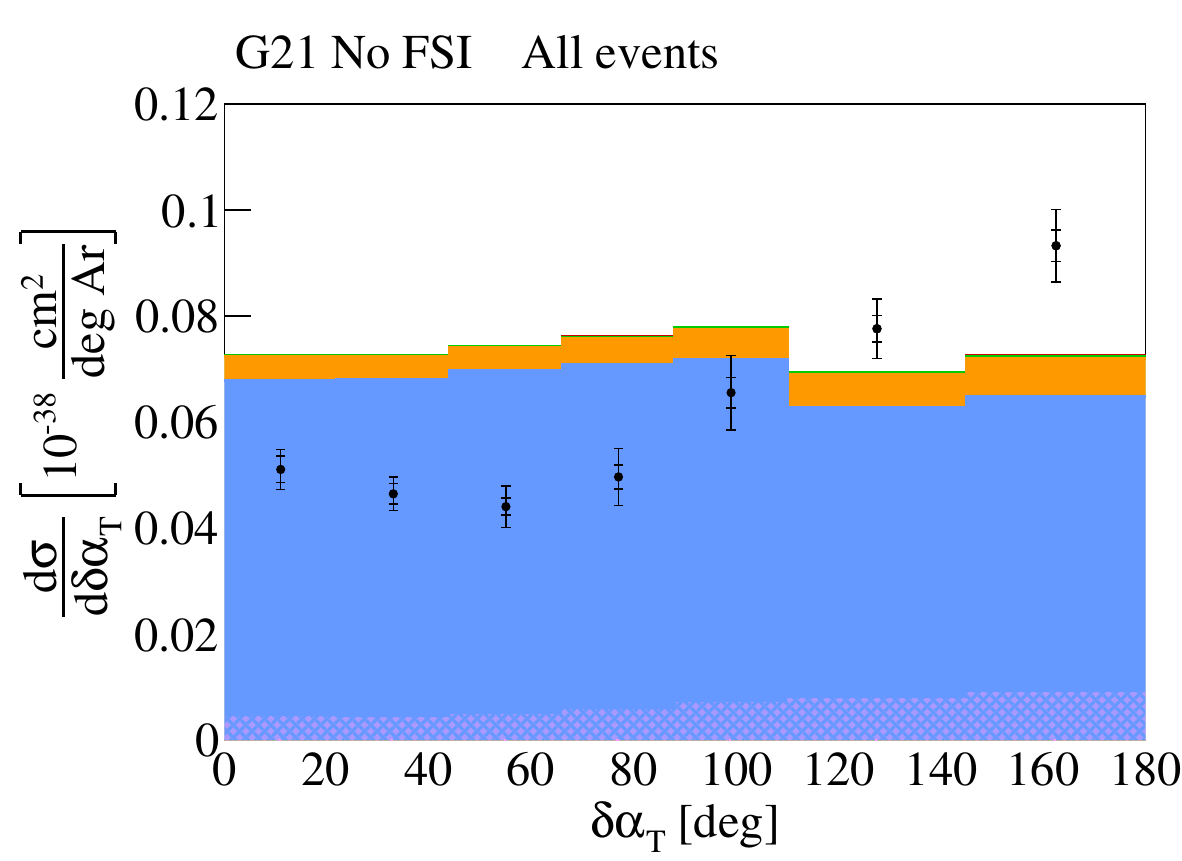}

\includegraphics[width=0.24\linewidth]{\figures 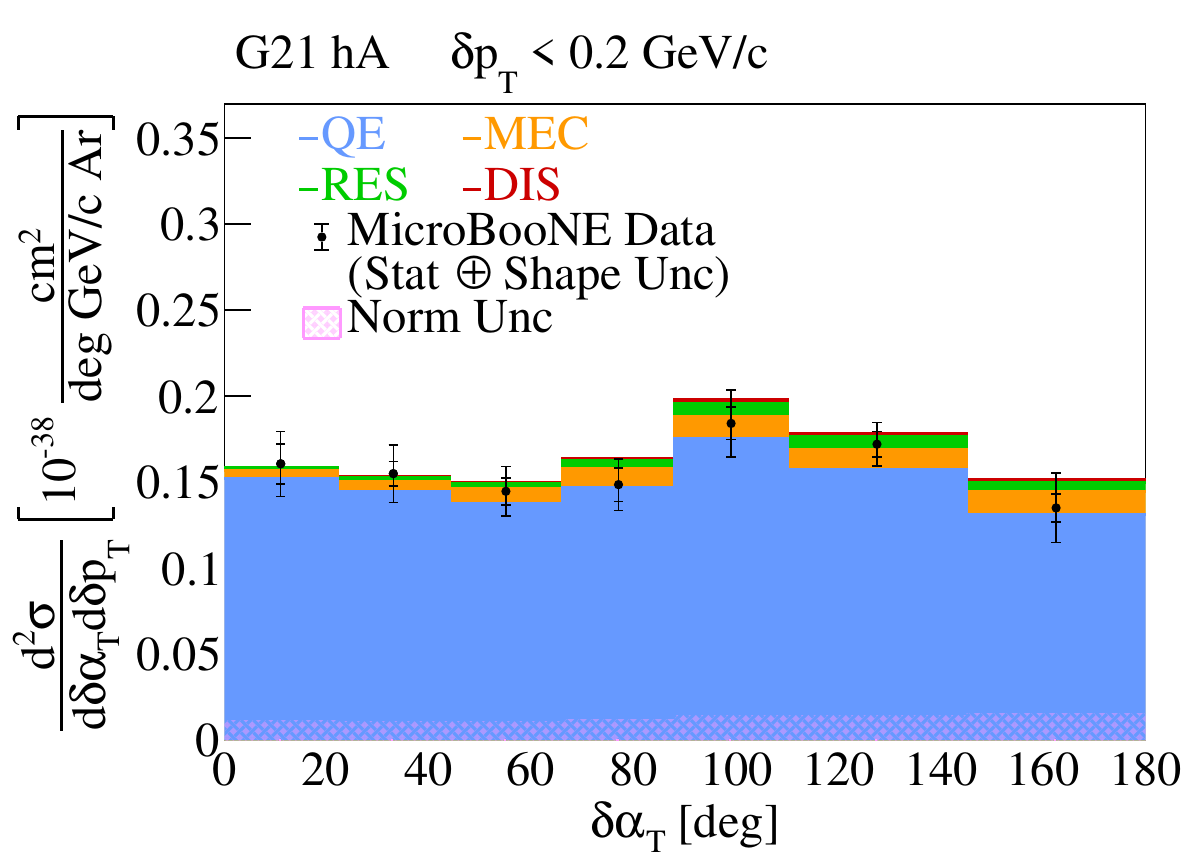}
\includegraphics[width=0.24\linewidth]{\figures 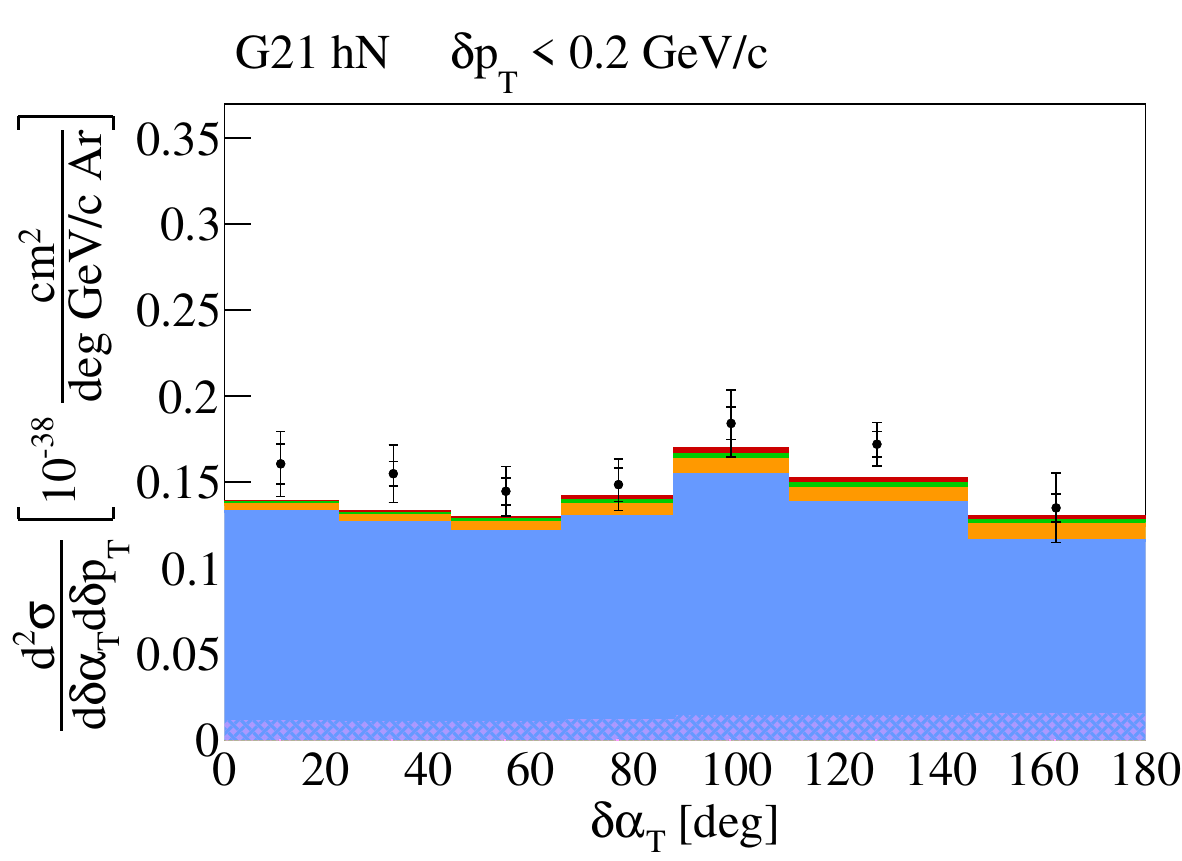}
\includegraphics[width=0.24\linewidth]{\figures 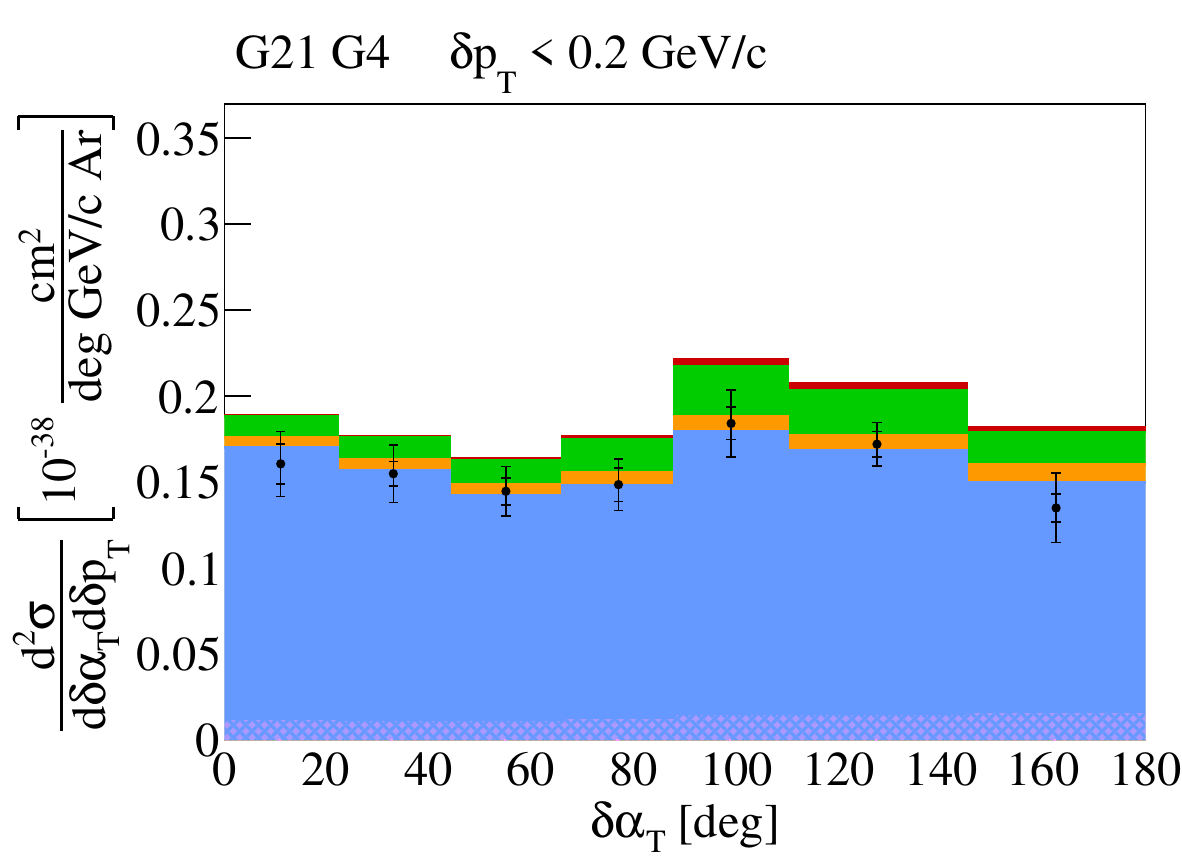}
\includegraphics[width=0.24\linewidth]{\figures 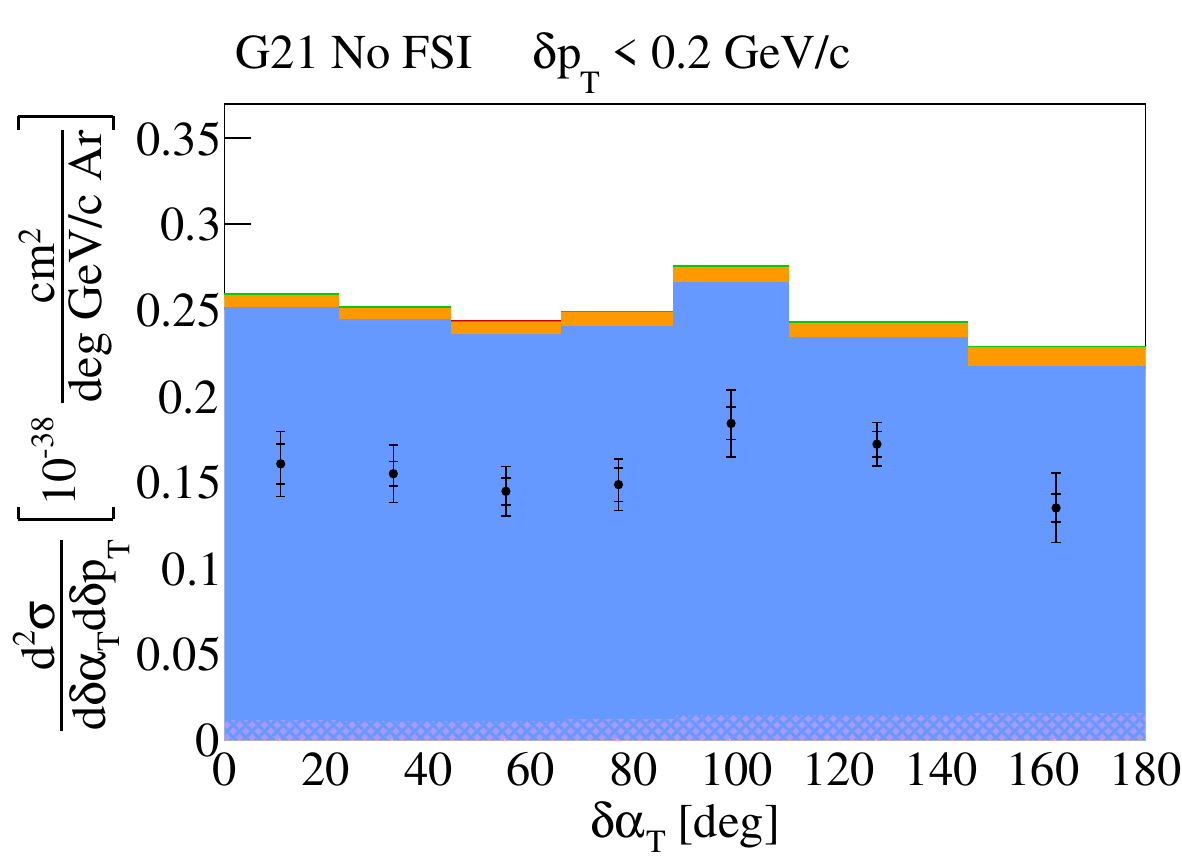}

\includegraphics[width=0.24\linewidth]{\figures 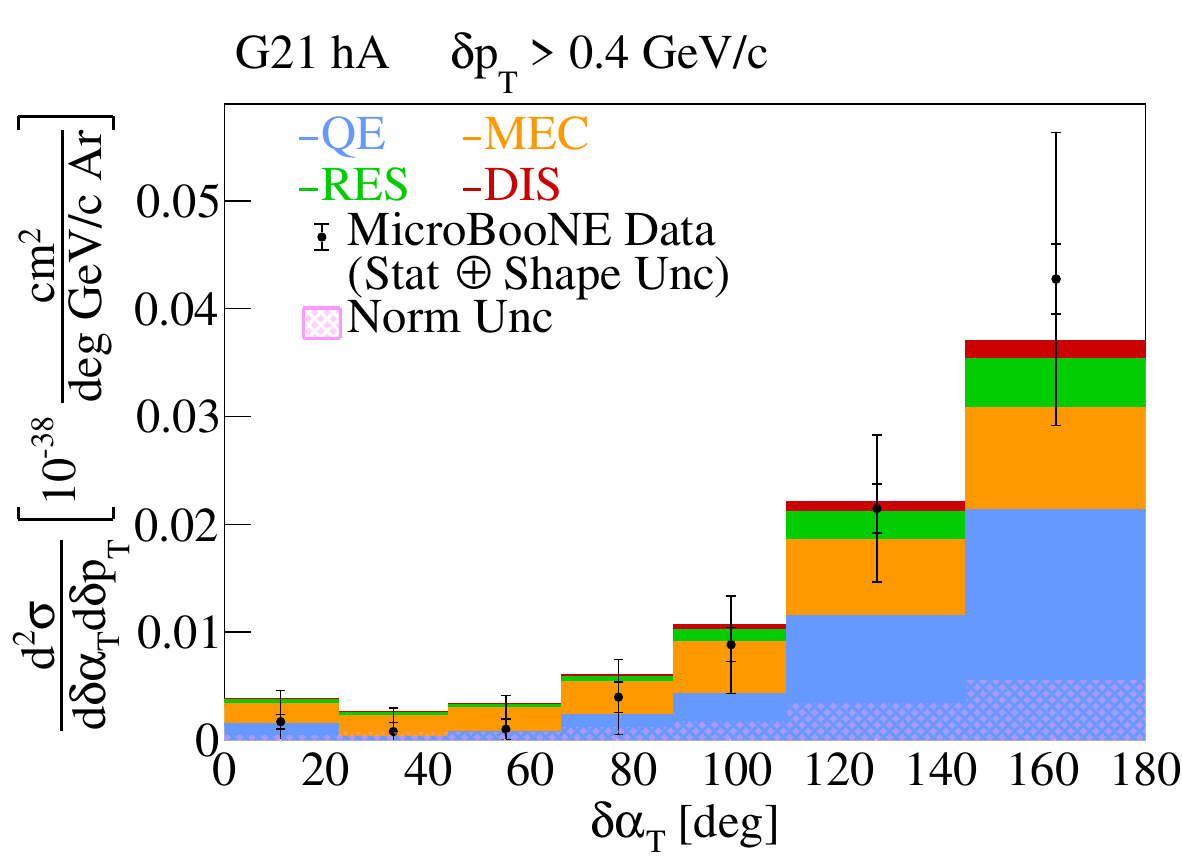}
\includegraphics[width=0.24\linewidth]{\figures 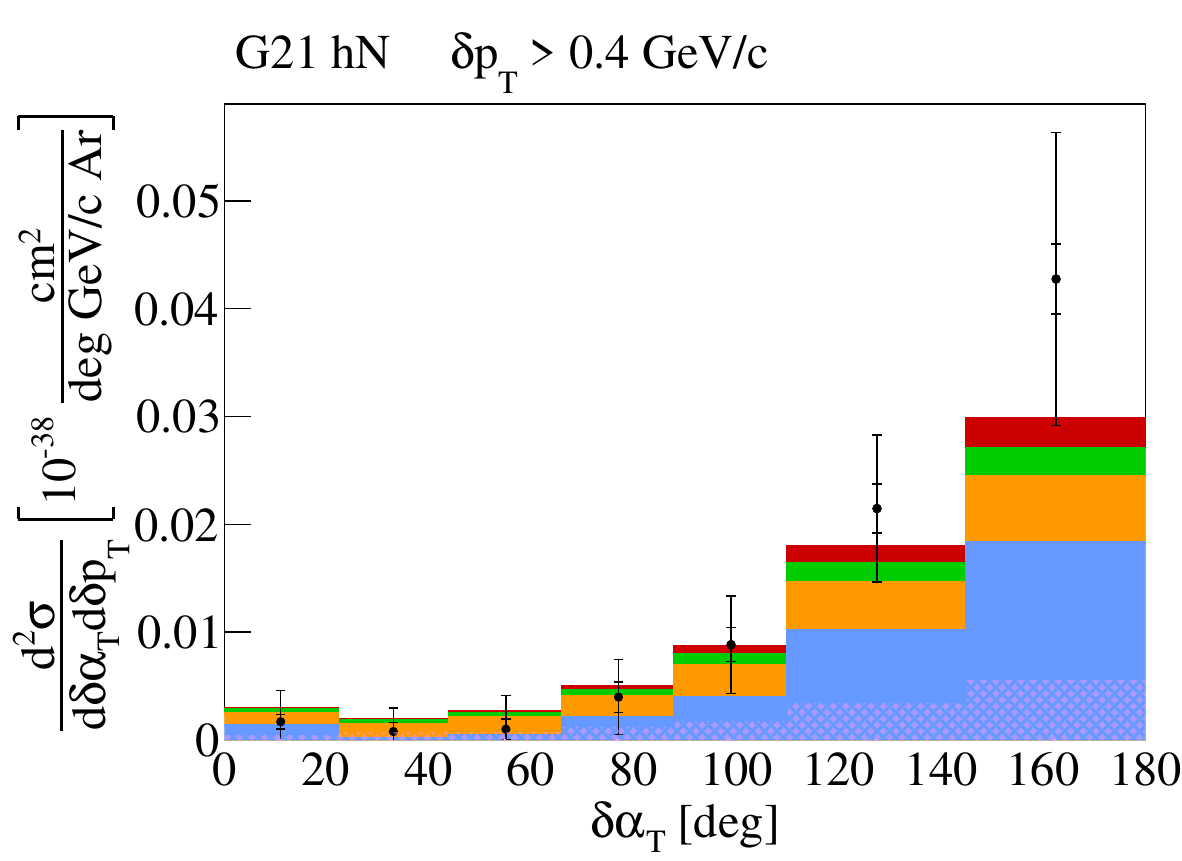}
\includegraphics[width=0.24\linewidth]{\figures 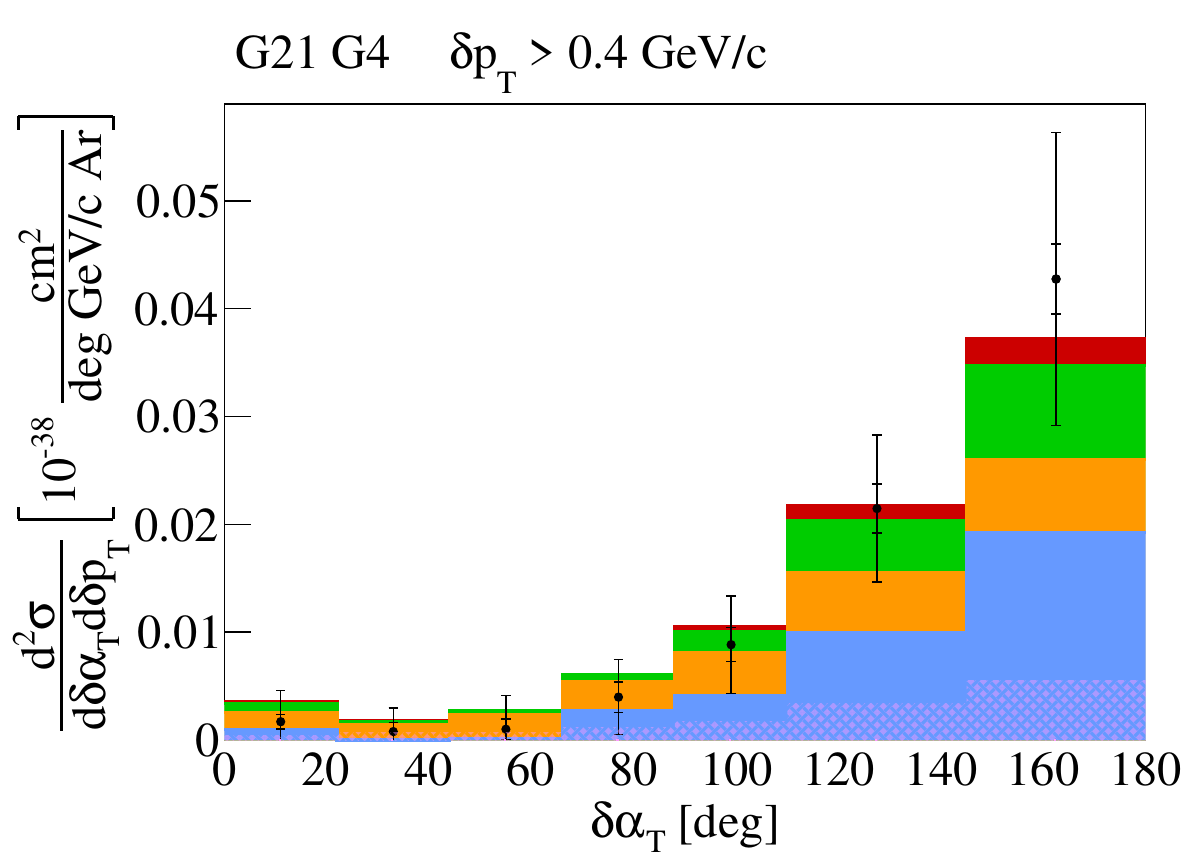}
\includegraphics[width=0.24\linewidth]{\figures 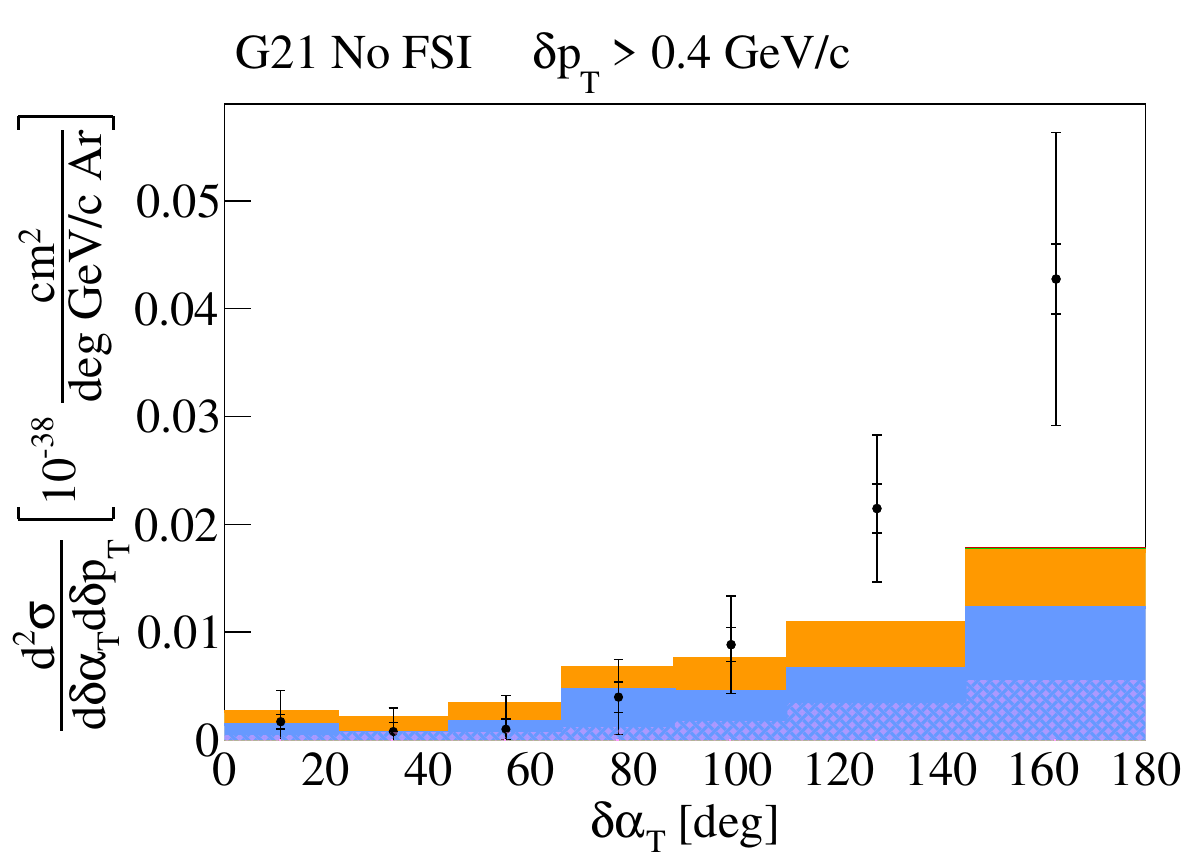}

\caption{Cross section interaction breakdown for (top) all the selected events, (middle) events with $\delta p_{T} <$ 0.2\,GeV/c, and (bottom) events with $\delta p_{T} >$ 0.4\,GeV/c. 
The breakdown is shown for (first column) the G21 hA configuration with the hA2018 FSI model, (second column) the G21 hN configuration with the hN FSI model, (third column) the G21 G4 configuration with the G4 FSI model, and (forth column) the G21 No FSI configuration without FSI effects.
}

\label{DeltaAlphaTBreakdown}
\end{figure}

\begin{figure}[htb!]
\centering 
\includegraphics[width=0.24\linewidth]{\figures 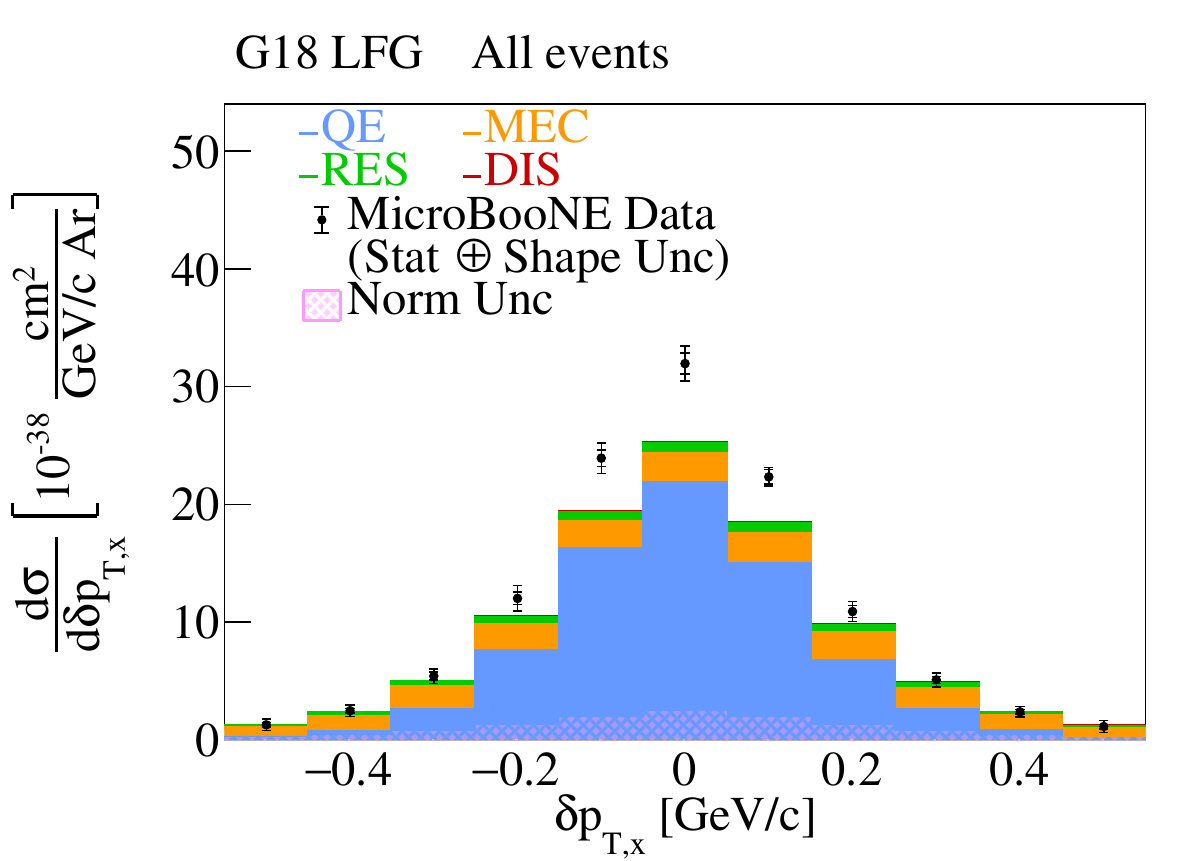}
\includegraphics[width=0.24\linewidth]{\figures 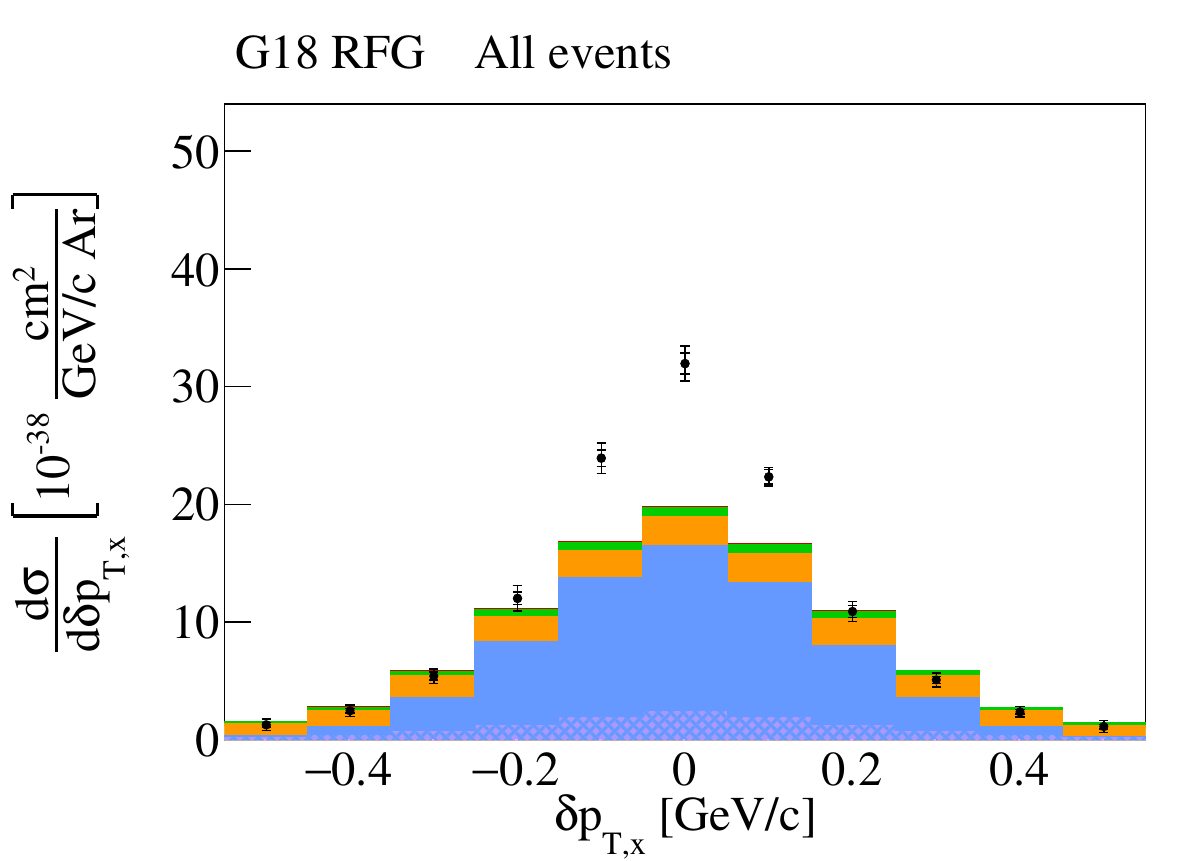}
\includegraphics[width=0.24\linewidth]{\figures 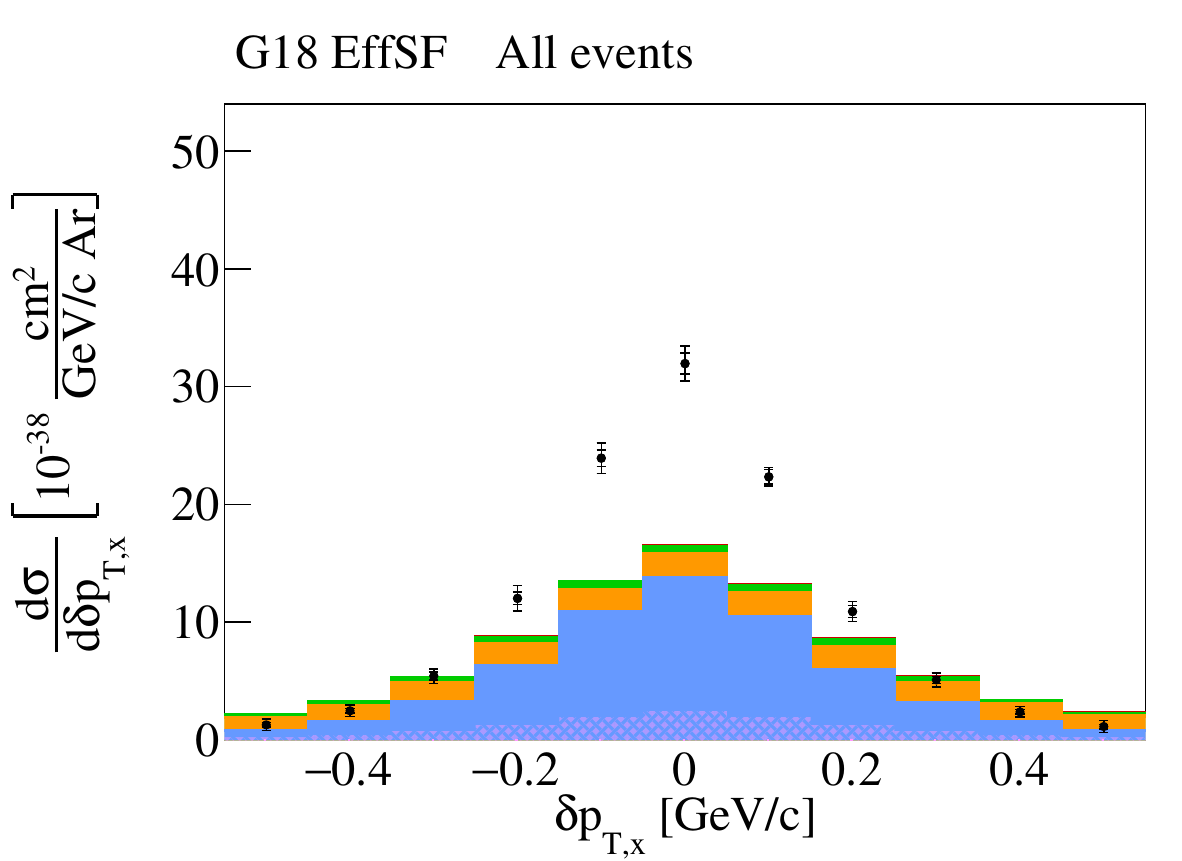}
\includegraphics[width=0.24\linewidth]{\figures 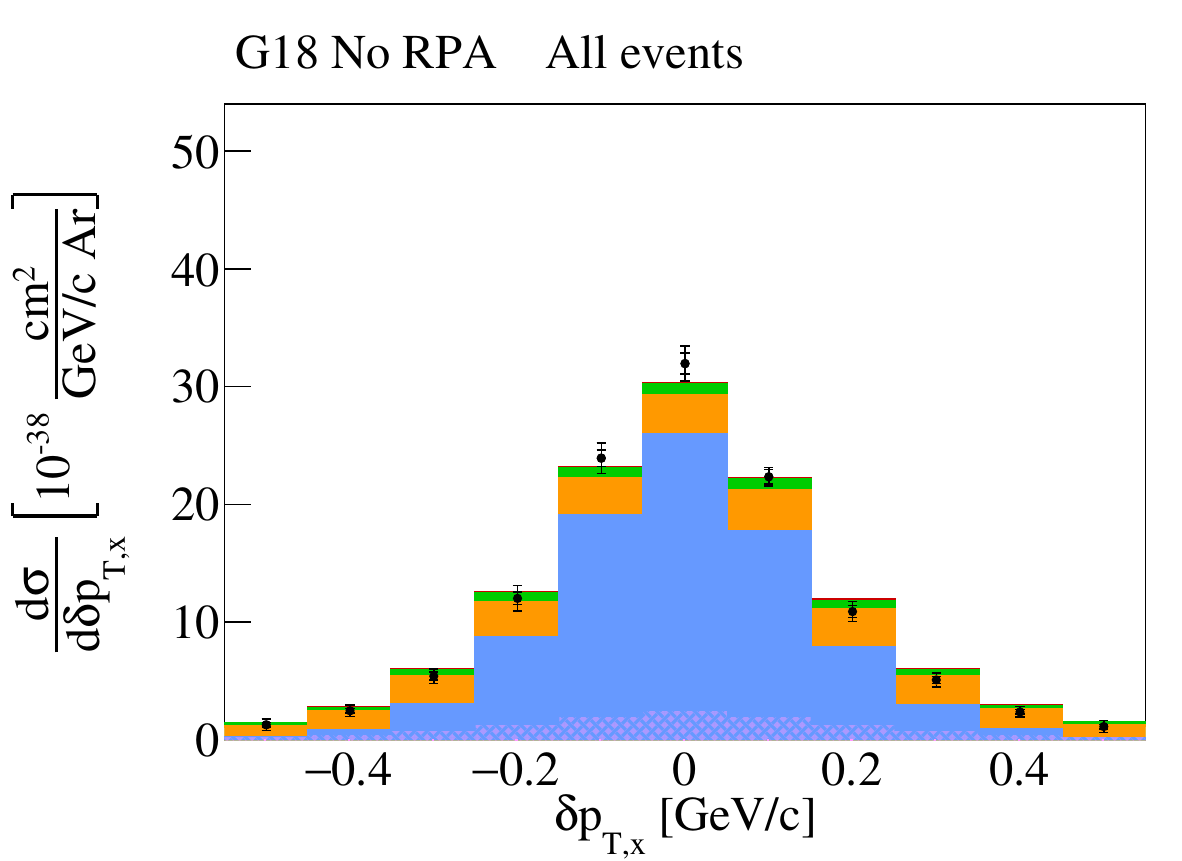}

\includegraphics[width=0.24\linewidth]{\figures 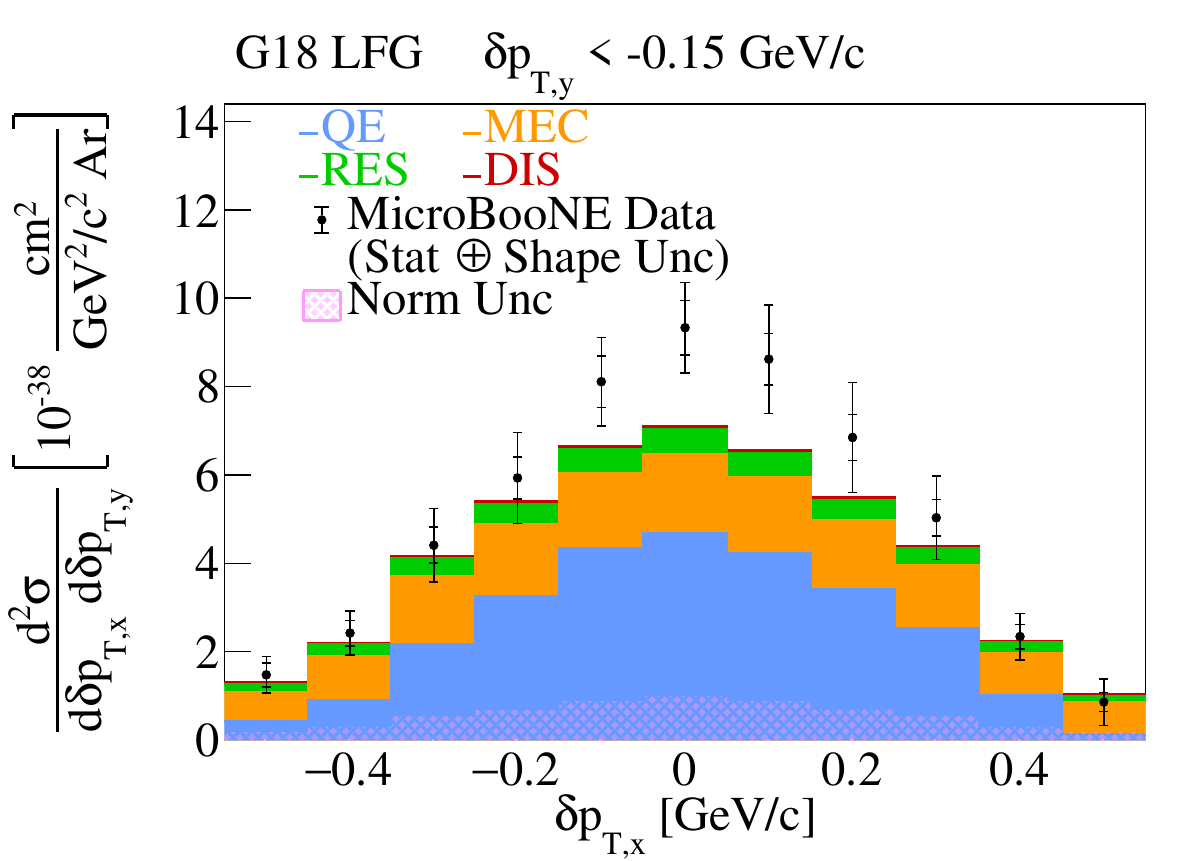}
\includegraphics[width=0.24\linewidth]{\figures 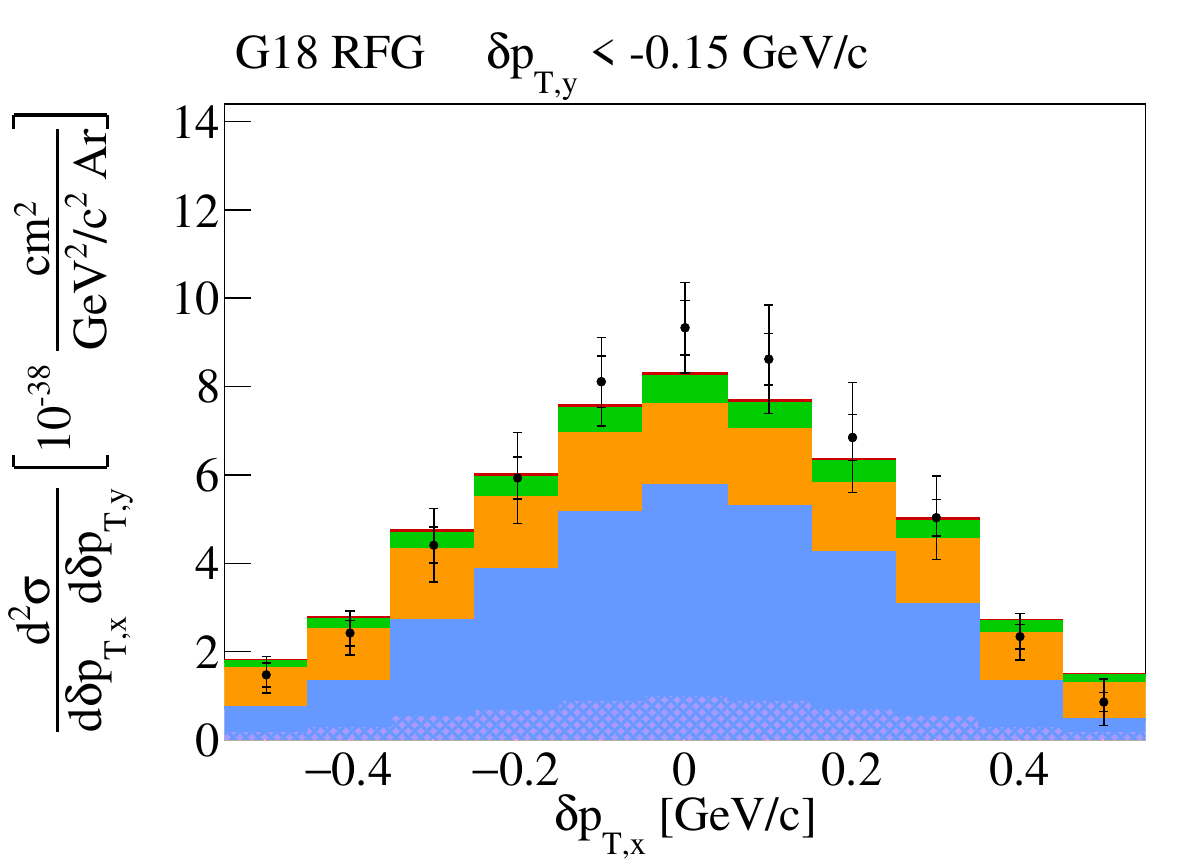}
\includegraphics[width=0.24\linewidth]{\figures 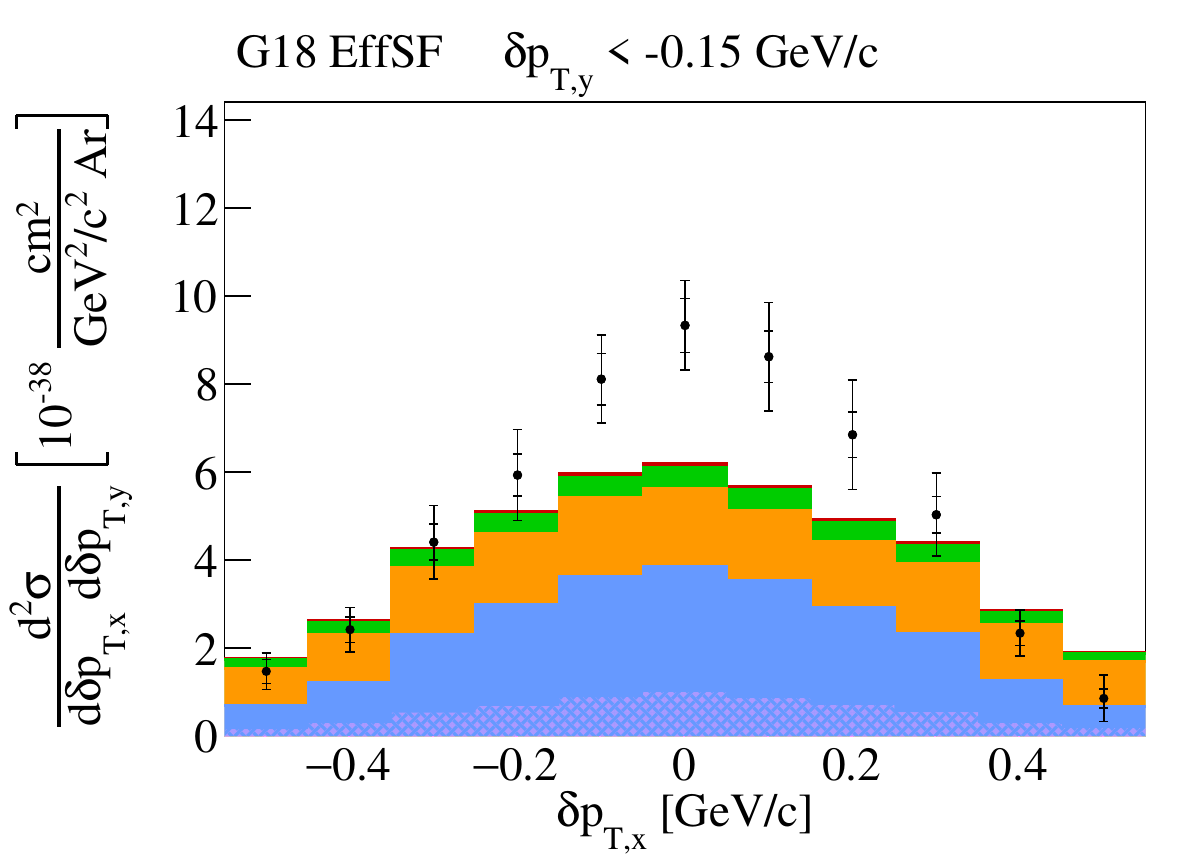}
\includegraphics[width=0.24\linewidth]{\figures 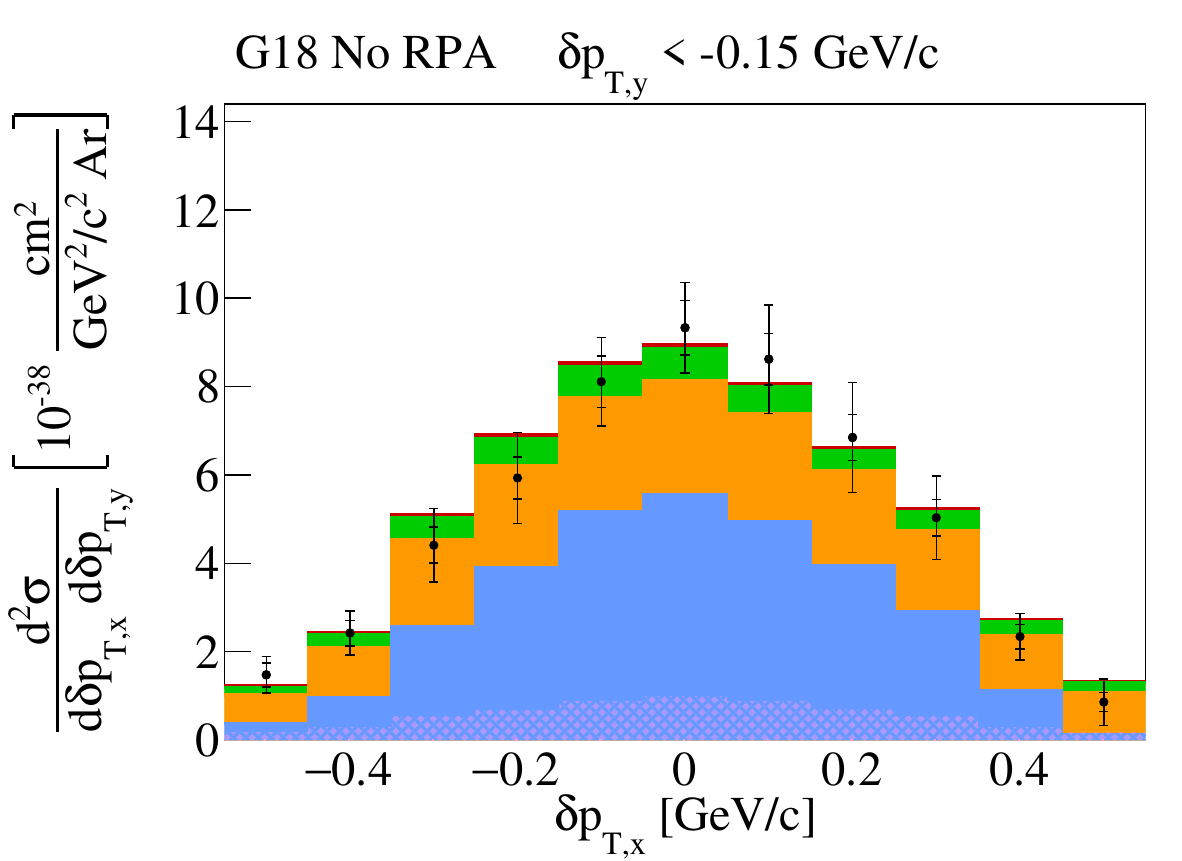}

\includegraphics[width=0.24\linewidth]{\figures 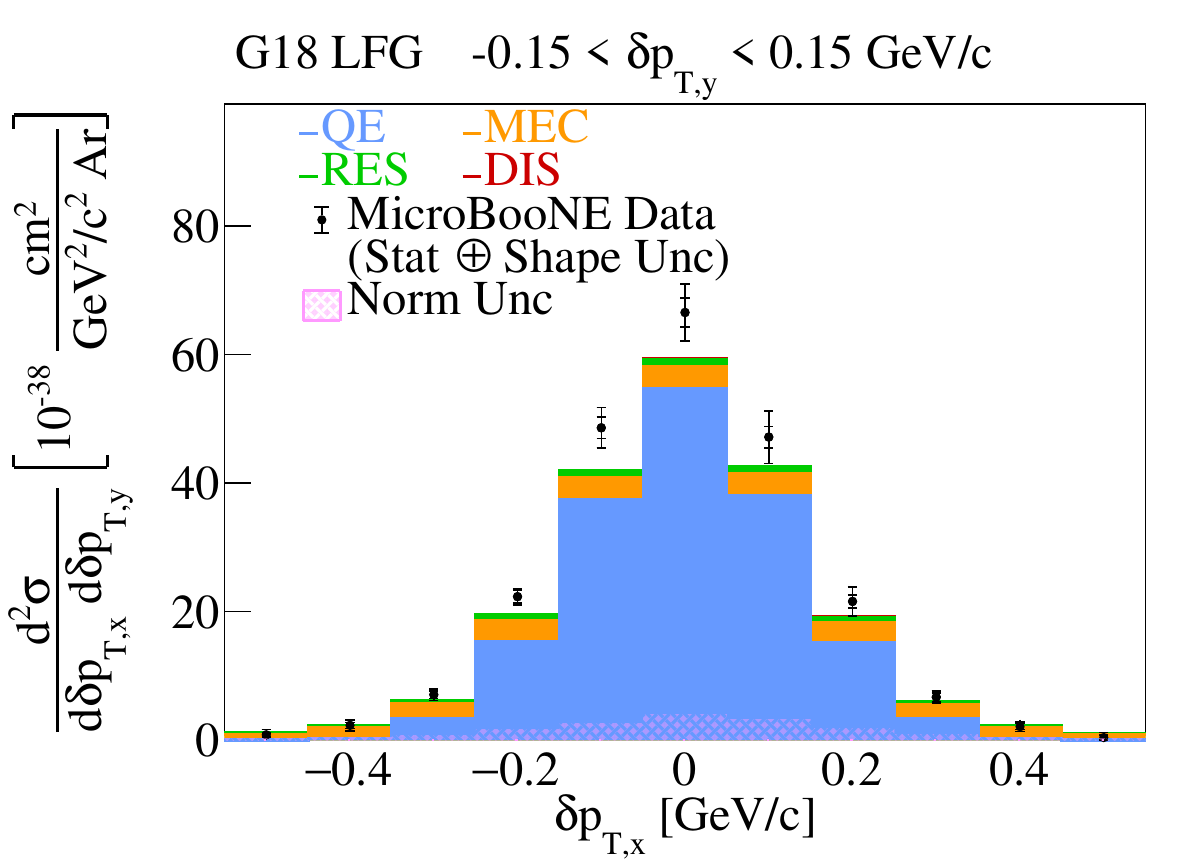}
\includegraphics[width=0.24\linewidth]{\figures 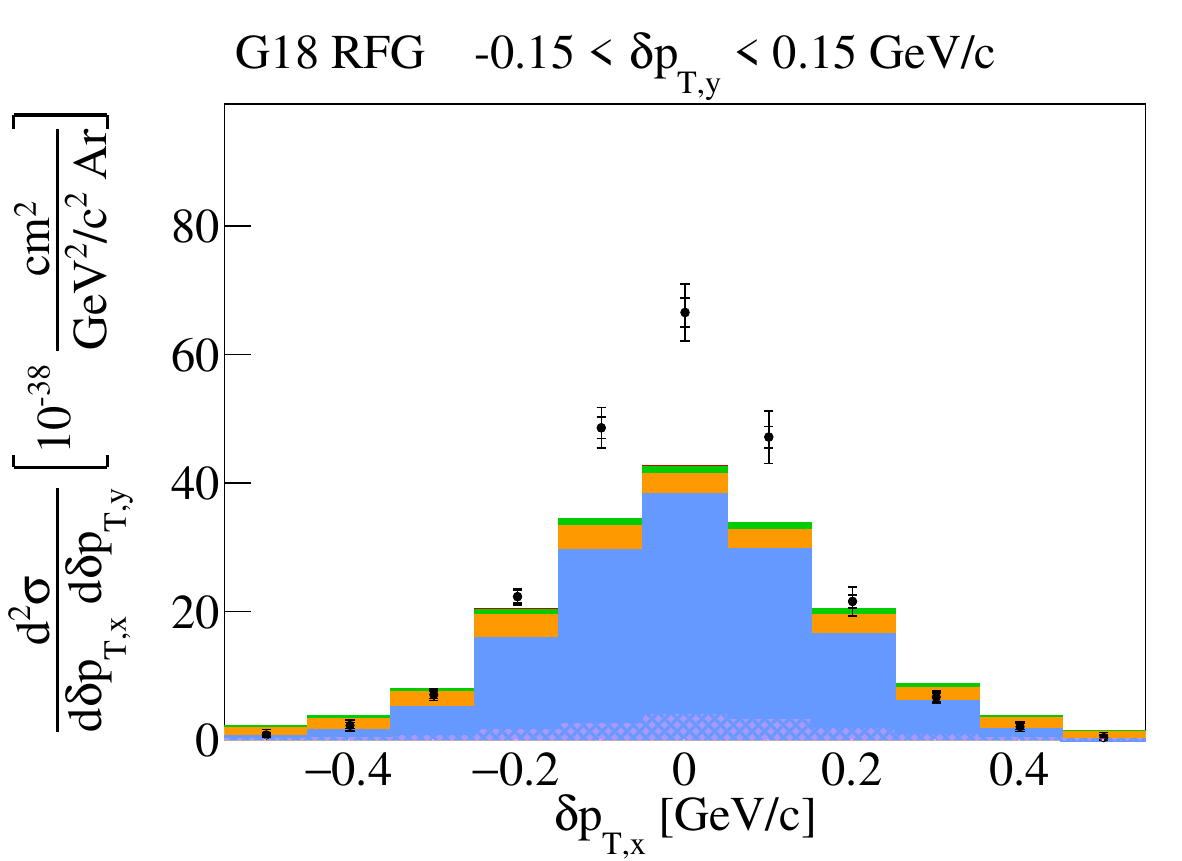}
\includegraphics[width=0.24\linewidth]{\figures 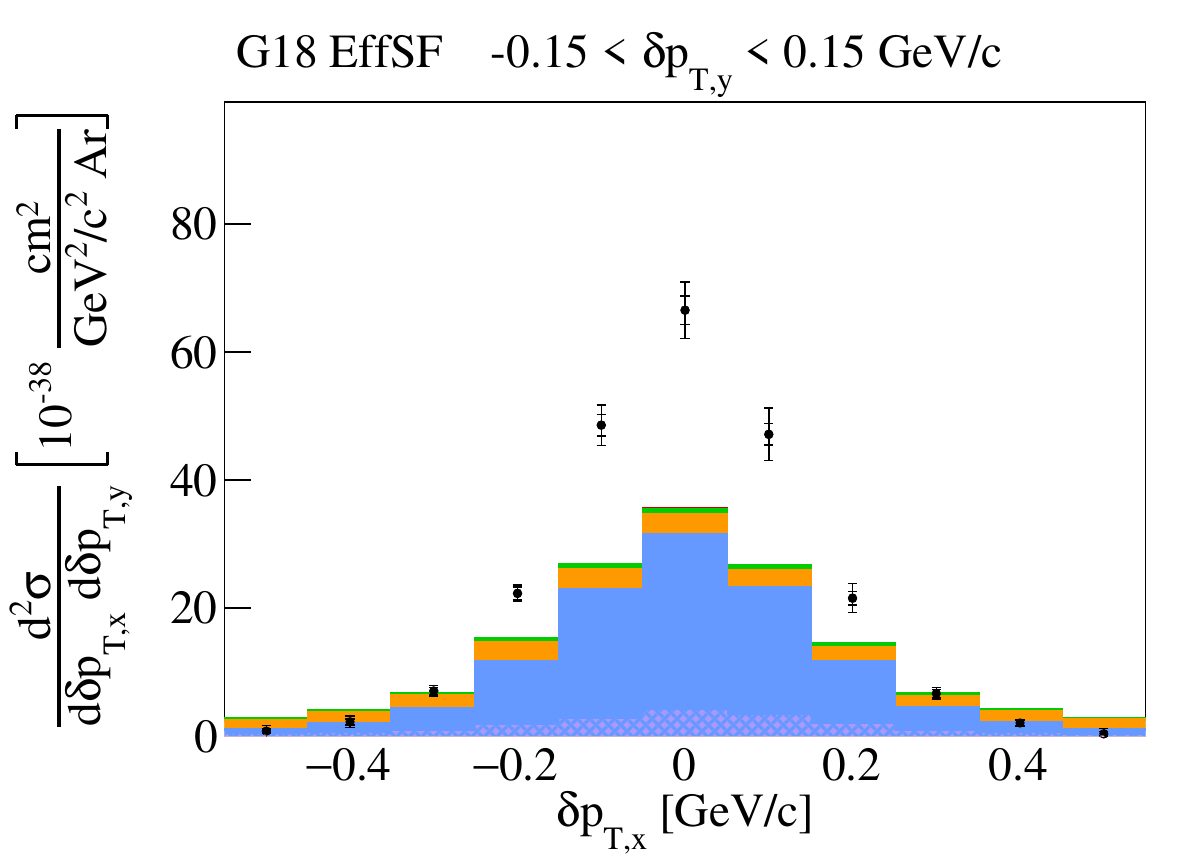}
\includegraphics[width=0.24\linewidth]{\figures 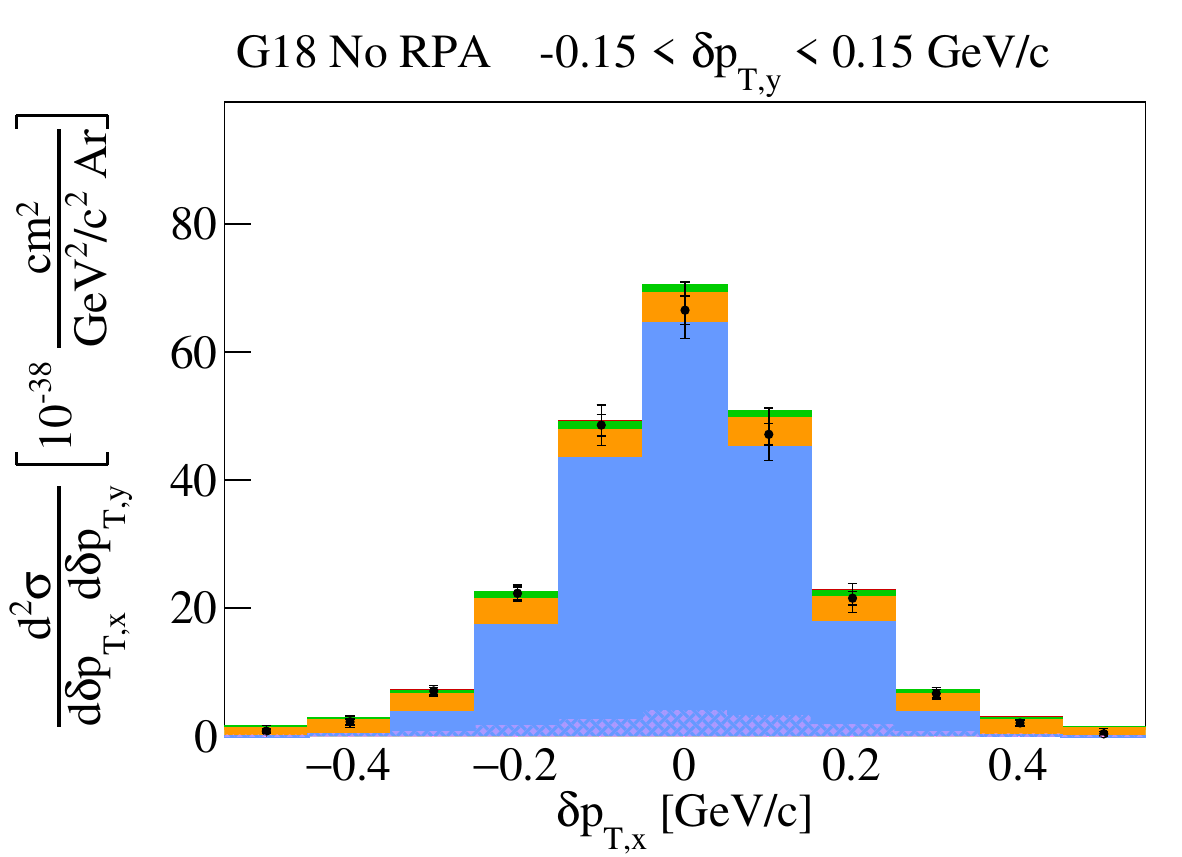}

\caption{Cross section interaction breakdown for (top) all the selected events, (middle) events with $\delta p_{T,y} <$ -0.15\,GeV/c, and (bottom) events with -0.15 $< \delta p_{T,y} <$ 0.15\,GeV/c. 
The breakdown is shown for (first column) the G18 LFG configuration, (second column) the G18 RFG configuration, (third column) the G18 EffSF configuration, and (forth column) the G18 No RPA configuration.
}

\label{DeltaPtxBreakdown}
\end{figure}

Figures~\ref{DeltaPTBreakdown}-\ref{DeltaPtxBreakdown} show the interaction breakdown into QE, MEC, RES, and DIS of the results presented in the main text, as labeled by the relevant event generators.

\clearpage
\section{FSI-modeling Impact on \texorpdfstring{$\delta p_{T}$}{deltapt}}\label{Fig1FSI}

\begin{figure*}[htb!]
\centering 
\includegraphics[width=0.32\linewidth]{\figures 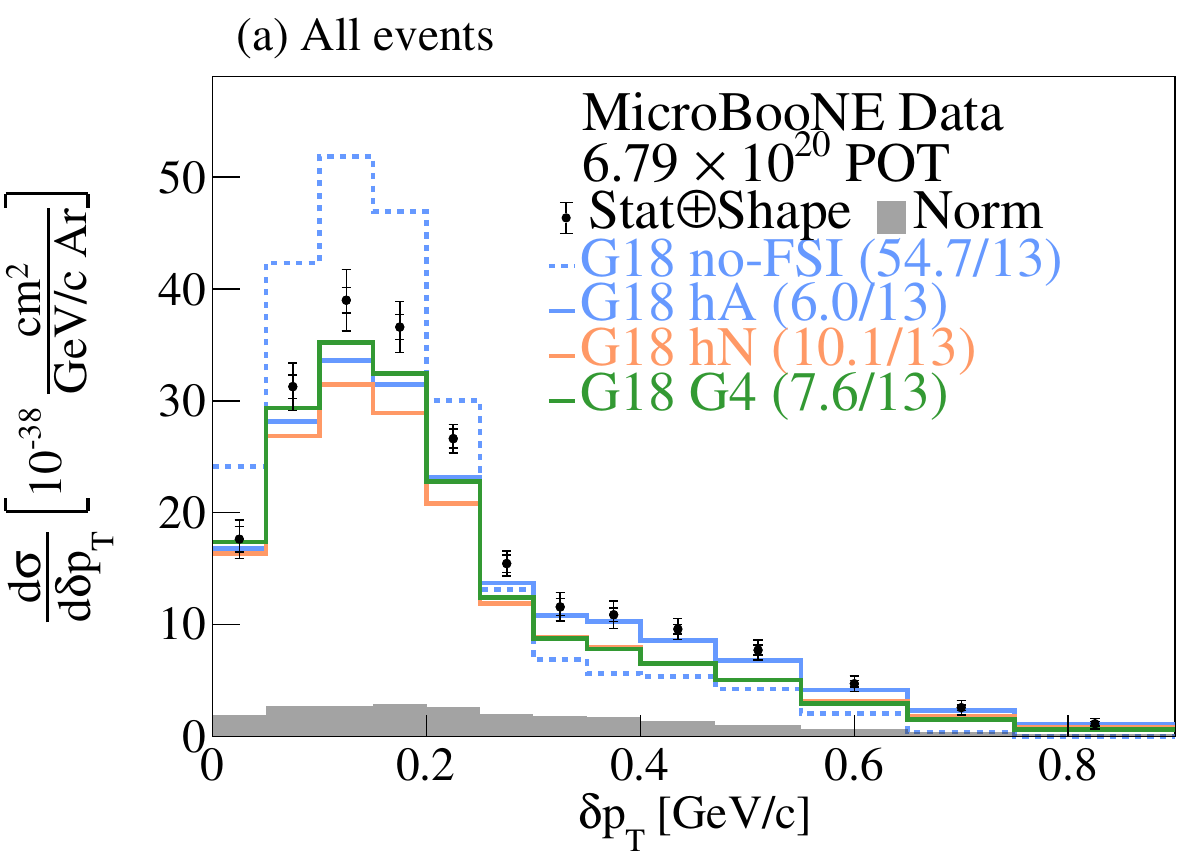}
\includegraphics[width=0.32\linewidth]{\figures 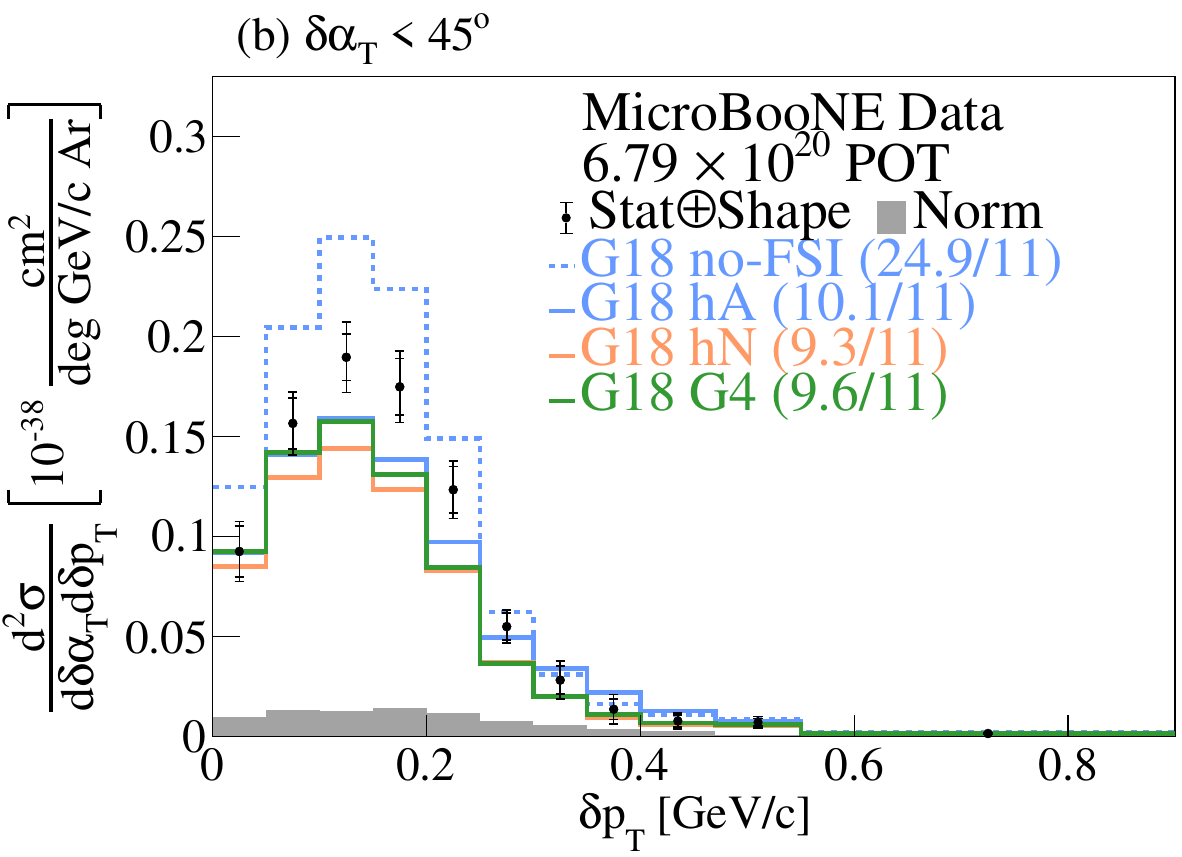}
\includegraphics[width=0.32\linewidth]{\figures 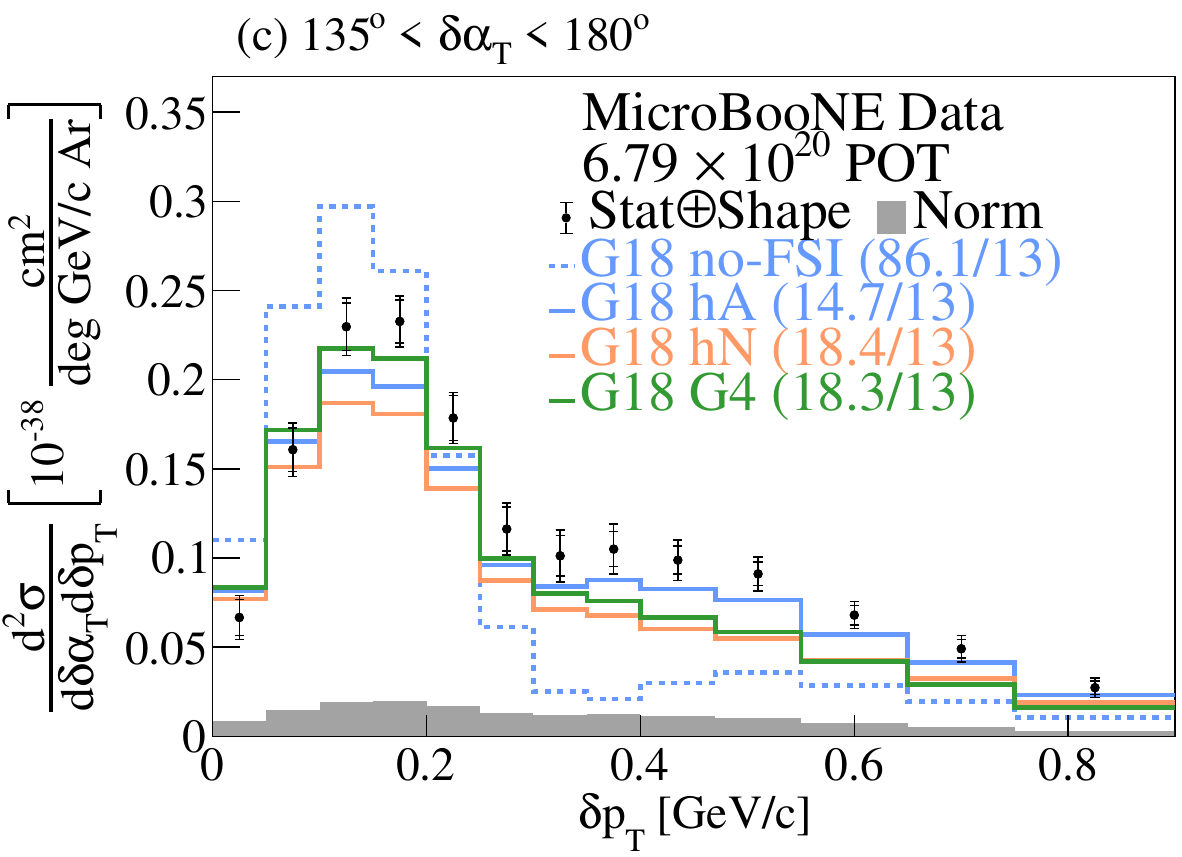}
\caption{
The flux-integrated (a) single- and (b-c) double- (in $\delta\alpha_{T}$ bins) differential \CCIpOpi\ cross sections as a function of the transverse missing momentum, $\delta p_{T}$. 
Inner and outer error bars show the statistical and total (statistical and shape systematic) uncertainty at the 1$\sigma$, or 68\%, confidence level. 
The gray band shows the separate normalization systematic uncertainty.
Colored lines show the results of theoretical cross section calculations with a number of $\texttt{G21}$ FSI modeling variations.
}
\label{SuppMatDeltaPTInDeltaAlphaT}
\end{figure*}

To quantify the FSI-modeling impact on $\delta p_{T}$, Fig.~\ref{SuppMatDeltaPTInDeltaAlphaT} presents the comparison of the single- and double-differential data cross sections against $\texttt{G18}$ FSI-modeling simulation variations.
More details on the $\texttt{G18}$ modeling configurations are included in the main text. 
We investigated the effect of the FSI-modeling choice by comparing the $\texttt{G18}$ results, which uses the hA FSI model ($\texttt{G18\,hA}$), to the ones obtained with $\texttt{G18\,hN}$, where the hN2018 FSI model was used instead~\cite{hN2018}, and to $\texttt{G18\,G4}$ with the recently coupled $\texttt{GEANT4}$ FSI framework~\cite{Wright:2015xia}.
The no-FSI prediction is disfavored across both the single- and the double-differential cross section comparisons.
The single-differential result as a function of $\delta p_{T}$ is shown in Fig.~\ref{SuppMatDeltaPTInDeltaAlphaT}a, where all the $\texttt{G18}$ variations with FSI effects yield $\chi^{2}$/bins ratios below one.
Figures~\ref{SuppMatDeltaPTInDeltaAlphaT}b and~\ref{SuppMatDeltaPTInDeltaAlphaT}c show the double-differential cross sections for $\delta\alpha_{T} <$ 45$^{\circ}$ and 135$^{\circ}$ $< \delta\alpha_{T} <$ 180$^{\circ}$ values, respectively.
All the predictions with FSI effects in Fig.~\ref{SuppMatDeltaPTInDeltaAlphaT}b result in comparable $\chi^{2}$/bins ratios, supporting the expectation that events with $\delta\alpha_{T} <$ 45$^{\circ}$ are weakly affected by FSI.
Figure~\ref{SuppMatDeltaPTInDeltaAlphaT}c shows the equivalent double-differential result for 135$^{\circ}$ $< \delta\alpha_{T} <$ 180$^{\circ}$, where the $\texttt{G18}\,hA$ prediction illustrates the lowest $\chi^{2}$/bins ratio amongst the predictions including FSI effects. 
The $\texttt{G18\,hN}$ and $\texttt{G18\,G4}$ predictions yield a mildly higher $\chi^{2}$/nbins ratio, supporting the observation in Fig.~1 of the main text related to the importance of adding FSI effects in simulation predictions. 

\clearpage
\section{QE-modeling Impact on \texorpdfstring{$\delta\alpha_{T}$}{deltaalphaT}}\label{Fig2QE}

\begin{figure*}[htb!]
\centering 
\includegraphics[width=0.32\linewidth]{\figures 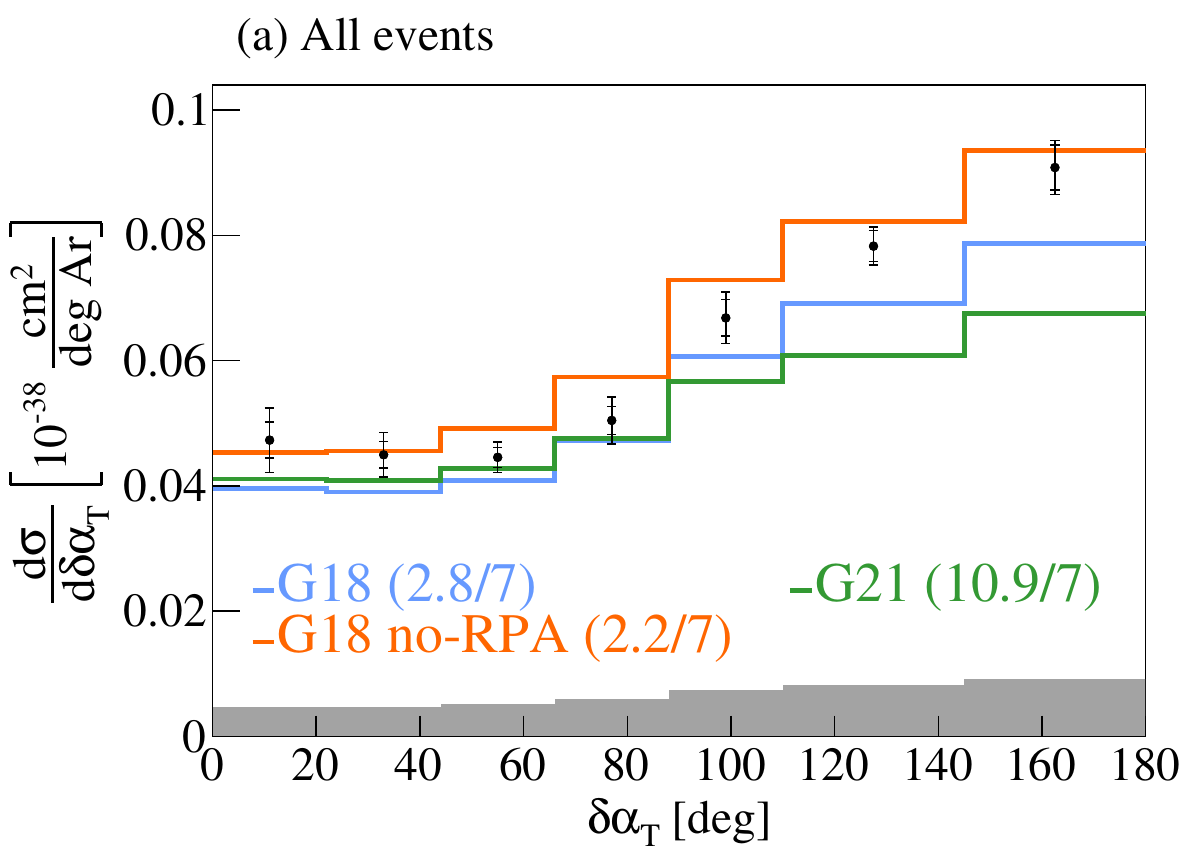}
\includegraphics[width=0.32\linewidth]{\figures 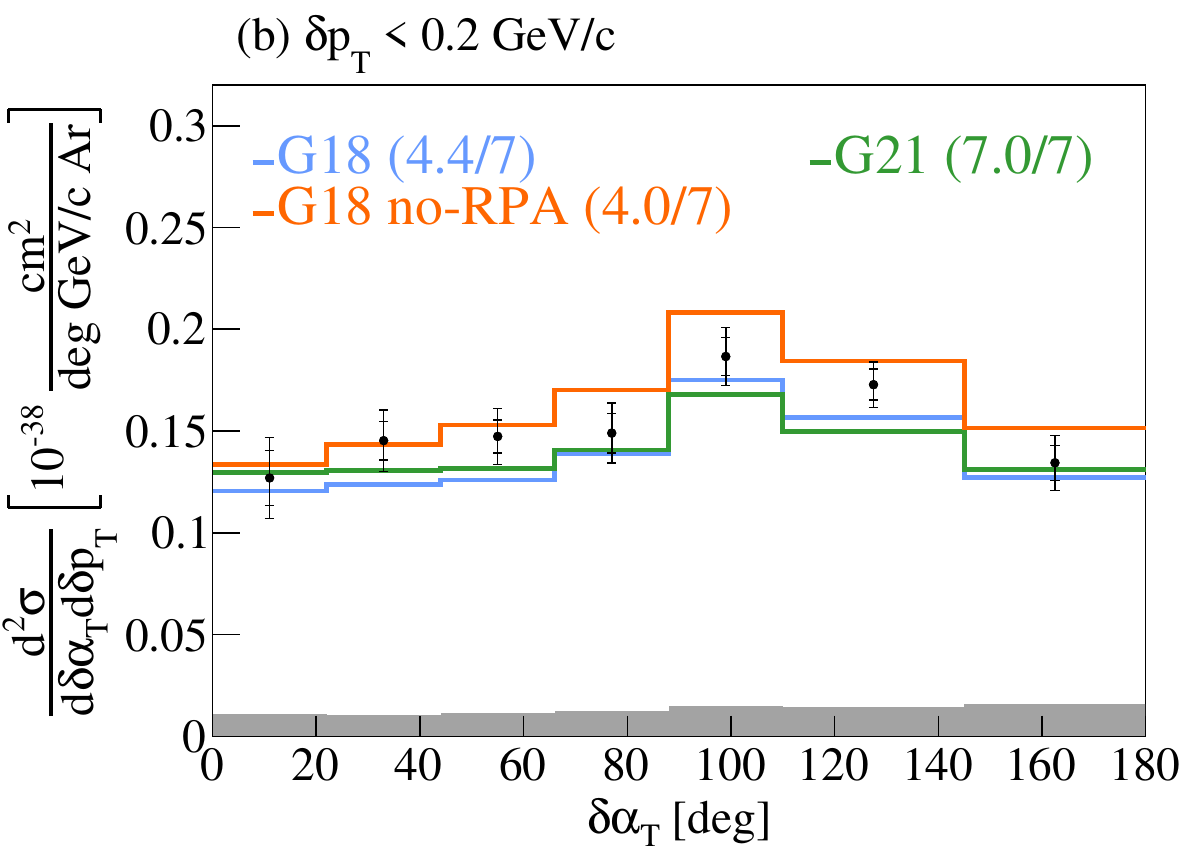}
\includegraphics[width=0.32\linewidth]{\figures 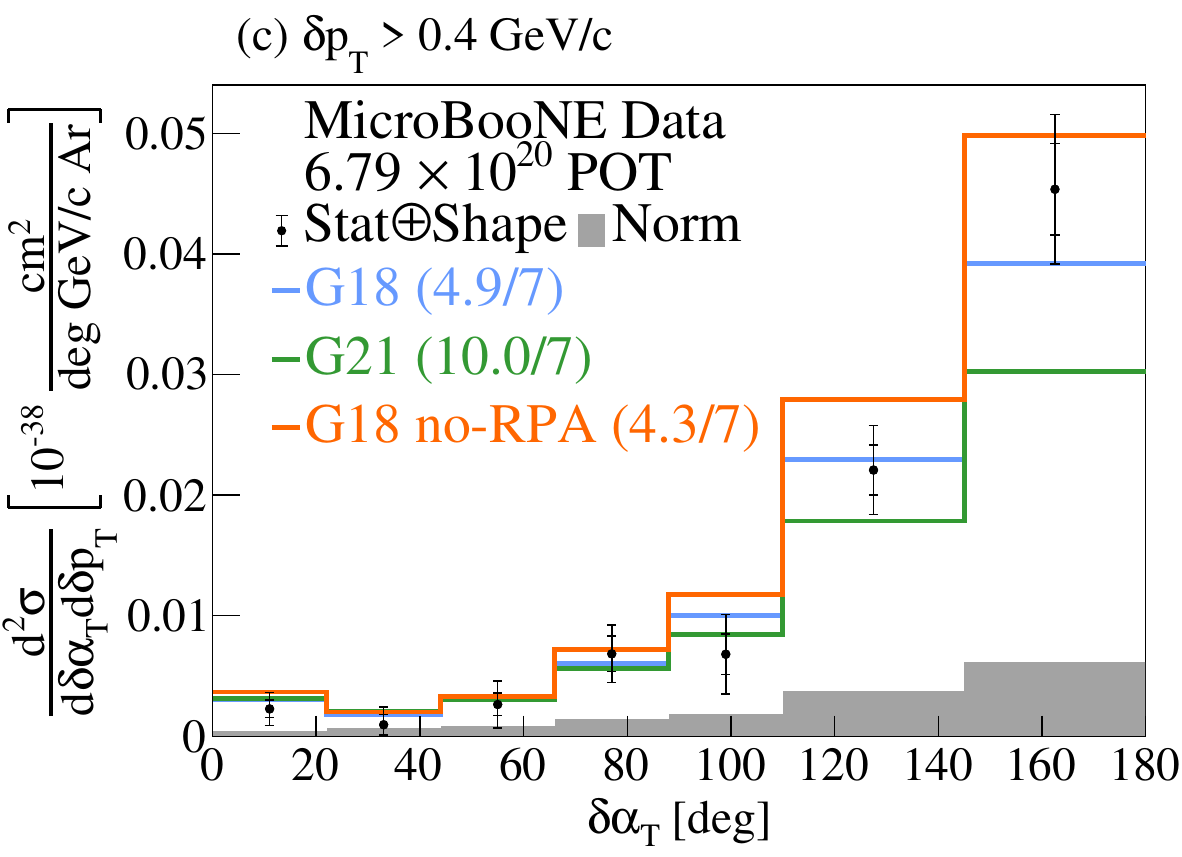}
\caption{The flux-integrated (a) single- and (b-c) double- (in $\delta p_{T}$ bins) differential \CCIpOpi\ cross sections as a function of the angle $\delta\alpha_{T}$. 
Inner and outer error bars show the statistical and total (statistical and shape systematic) uncertainty at the 1$\sigma$, or 68\%, confidence level. 
The gray band shows the separate normalization systematic uncertainty.
Colored lines show the results of theoretical cross section calculations with a number of QE-modeling choices based on the $\texttt{GENIE}$ event generator.}
\label{SuppMatDeltaAlphaTInDeltaPT}
\end{figure*}

To quantify the QE-modeling impact on $\delta\alpha_{T}$, Fig.~\ref{SuppMatDeltaAlphaTInDeltaPT} presents the comparison of the single- and double-differential data cross sections against QE-modeling simulation variations.
Namely, the $\texttt{G18}$ prediction (Nieves QE) is shown, along with the $\texttt{G21}$ (SuSAv2 QE) and $\texttt{G18 no-RPA}$ (Nieves QE without RPA effects) predictions.
Details on the modeling configuration of the aforementioned predictions are included in the main text.
Note that, though our selection is QE-dominated, there are still some more complex events, such as MEC events, that satisfy the CC1p0$\pi$ topological signal definition and that the $\texttt{G21}$ prediction uses the SuSAv2 MEC prediction, unlike the Nieves MEC prediction used by the $\texttt{G18}$ family.
Furthermore, note that the $\texttt{G18 no-RPA}$ prediction, which uses the Nieves QE prediction~\cite{Nieves:2012yz}, effectively corresponds to the Llewellyn Smith QE scattering prescription~\cite{LlewellynSmith:1971uhs} up to the Coulomb corrections, which are expected to be negligible. 
The single-differential result as a function of $\delta\alpha_{T}$ is shown in Fig.~\ref{SuppMatDeltaAlphaTInDeltaPT}a with the $\texttt{G21}$ prediction, which is based on the super-scaling treatment, illustrating the highest $\chi^{2}$/bins ratio.
Figure~\ref{SuppMatDeltaAlphaTInDeltaPT}b shows the double-differential cross section for low $\delta p_{T}$ values, where all the predictions result in a $\chi^{2}$/bins ratio equal to or less than unity.
Figure~\ref{SuppMatDeltaAlphaTInDeltaPT}c presents the double-differential cross section for higher $\delta p_{T}$ values, with $\texttt{G21}$ prediction being the one with the highest $\chi^{2}$/bins ratio.
Compared to the FSI variations presented in Fig.~2 of the main text, the choice of the QE model seems to be less impactful.

\section{FSI-modeling Impact on \texorpdfstring{$\delta p_{T,x}$}{deltaptx}}\label{Fig3FSI}

\begin{figure*}[htb!]
\centering 
\includegraphics[width=0.32\linewidth]{\figures 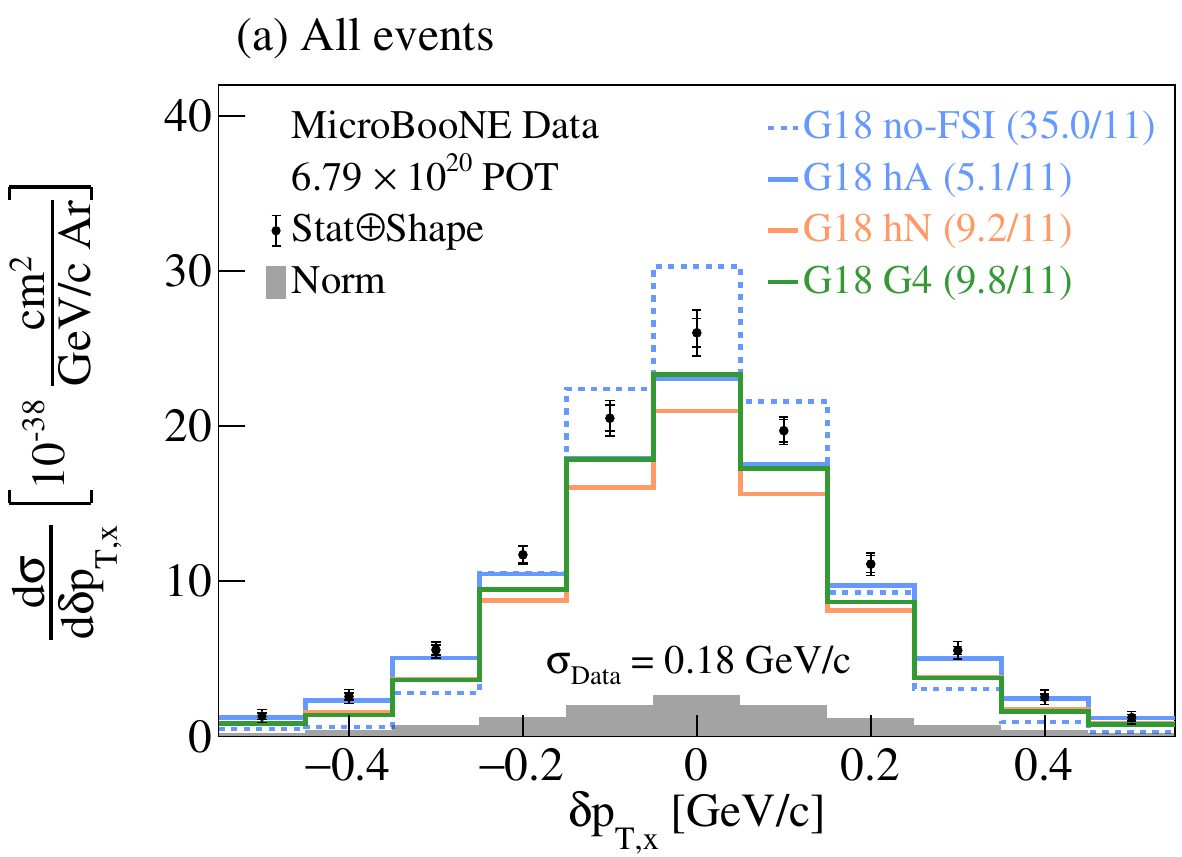}
\includegraphics[width=0.32\linewidth]{\figures 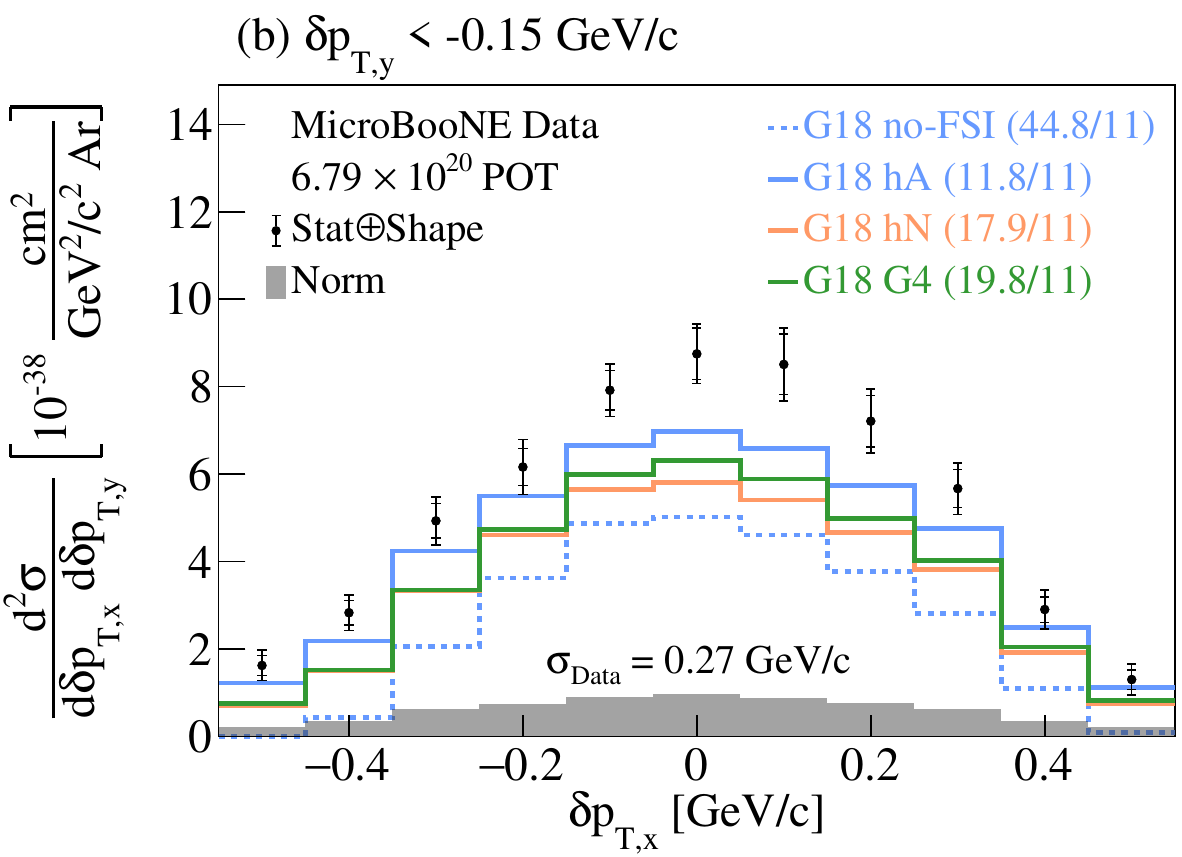}
\includegraphics[width=0.32\linewidth]{\figures 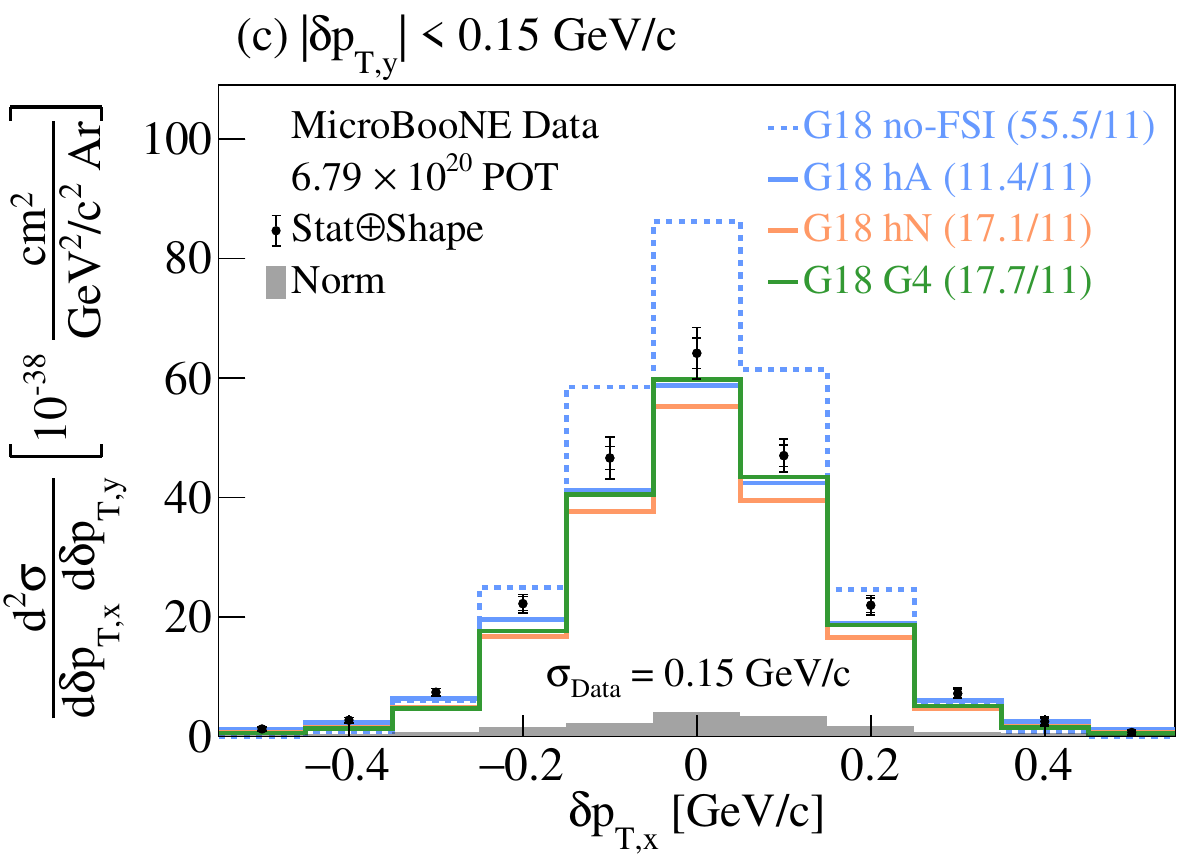}
\caption{
The flux-integrated (a) single- and (b-c) double- (in $\delta p_{T,y}$ bins) differential \CCIpOpi\ cross sections as a function of the transverse three-momentum transfer component, $\delta p_{T,x}$. 
Inner and outer error bars show the statistical and total (statistical and shape systematic) uncertainty at the 1$\sigma$, or 68\%, confidence level. 
The gray band shows the separate normalization systematic uncertainty.
Colored lines show the results of theoretical cross section calculations with a number of $\texttt{G18}$ FSI modeling variations.
The standard deviation ($\sigma_\mathrm{Data}$) of a Gaussian fit to the data is shown on each panel. 
}
\label{SuppMatDeltaPtxInDeltaPty}
\end{figure*}

To quantify the FSI-modeling impact on $\delta p_{T,x}$, Fig.~\ref{SuppMatDeltaPtxInDeltaPty} presents the comparison of the single- and double-differential data cross sections against $\texttt{G18}$ FSI-modeling simulation variations.
More details on the included modeling configurations are included in the main text and in the previous sections. 
The single-differential result as a function of $\delta p_{T,x}$ is shown in Fig.~\ref{SuppMatDeltaPtxInDeltaPty}a, with the no-FSI prediction being disfavored.
On the other hand, all the $\texttt{G18}$ variations with FSI effects yield comparable $\chi^{2}$/bins ratios below unity.
Figures~\ref{SuppMatDeltaPtxInDeltaPty}b and~\ref{SuppMatDeltaPtxInDeltaPty}c show the double-differential cross sections for $\delta p_{T,y} <$ -0.15\,GeV/c and $|\delta p_{T,y}| <$ 0.15\,GeV/c values, respectively.
All the predictions with FSI result in $\chi^{2}$/bins ratios of less than 2, while the no-FSI ratios as significantly higher.
Compared to the QE variations presented in Fig.~3 of the main text, the choice of the FSI model seems to be less impactful.

\clearpage
\bibliography{supp}
